\documentclass[onecolumn,showpacs,nofootinbib,preprintnumbers,prd]{revtex4-1}

\usepackage{amsmath}
\usepackage{amsfonts}
\usepackage{amssymb}
\usepackage{graphicx}
\usepackage[titletoc]{appendix}
\usepackage{color}
\usepackage{hyperref}
\usepackage{cleveref}
\usepackage[rightcaption]{sidecap}
\usepackage{subcaption}
\usepackage{comment}
\usepackage{slashed}
\usepackage{mathtools}
\usepackage{dcolumn}
\usepackage{array}
\usepackage{ctable}
\usepackage{multirow}
\usepackage{siunitx}
\usepackage{tabularx}
\usepackage{booktabs}
\usepackage{supertabular}
\usepackage{verbatim}
\usepackage{cancel}
\usepackage{graphicx}
\usepackage{dcolumn}
\usepackage{bm}
\usepackage{braket}

\begin{document}

\title{Vector-field spontaneous baryogenesis with Lorentz invariance violation}

\author{Mattia Dubbini}
\email{mattia.dubbini@unicam.it}
\affiliation{Universit\`a di Camerino, Via Madonna delle Carceri, Camerino, 62032, Italy.}

\author{Orlando Luongo}
\email{orlando.luongo@unicam.it}
\affiliation{Universit\`a di Camerino, Via Madonna delle Carceri, Camerino, 62032, Italy.}
\affiliation{Department of Nanoscale Science and Engineering, University at Albany-SUNY, Albany, New York 12222, USA.}
\affiliation{Istituto Nazionale di Astrofisica (INAF), Osservatorio Astronomico di Brera, 20121 Milano, Italy.}
\affiliation{Istituto Nazionale di Fisica Nucleare (INFN), Sezione di Perugia, Perugia, 06123, Italy,}
\affiliation{Al-Farabi Kazakh National University, Al-Farabi av. 71, 050040 Almaty, Kazakhstan.}

\author{Aniello Quaranta}
\email{aniello.quaranta@unicam.it}
\affiliation{Universit\`a di Camerino, Via Madonna delle Carceri, Camerino, 62032, Italy. }

\date{\today}

\begin{abstract}
We extend spontaneous baryogenesis by considering the spontaneous breaking of $U(1)_B$ through a complex vector field. This field interacts with baryons and leptons via a vector-current coupling and, by construction, acquires a nonzero vacuum expectation value. Accordingly, the theory also exhibits a spontaneous violation of Lorentz invariance, effectively realizing a Bumblebee model. In this picture, the pseudo-Nambu-Goldstone boson arising from spontaneous breaking of the $U(1)_B$ global symmetry is the global phase of the Bumblebee vector and, in the broken phase, it results minimally coupled with the baryonic current, guaranteeing the violation of the baryon number. Consequently, we assume that the pseudo-Nambu-Goldstone, arising from spontaneous breaking of $U(1)_B$, plays the role of the inflaton, leading to baryogenesis across the entire inflationary stage, up to when the inflaton decays into baryon-antilepton and antibaryon-lepton pairs through a CP-violating interaction that also violates the Lorentz symmetry. Afterwards, we address the issue of flavor oscillations among baryon and lepton fields, including the oscillation probability in the calculation of the baryon asymmetry. Remarkably, our framework predicts a non-null mixing factor even for massless fermions. This mixing acts on the spatial momenta rather than on the masses of the produced fermions, allowing larger values of the coupling constant even guaranteeing the production of light fermions. The net baryon asymmetry results accordingly modified, and may also reproduce the experimental data for allowed values of the coupling constant.
\end{abstract}

\maketitle
\tableofcontents

\section{Introduction}\label{sec:level1}

Baryogenesis is responsible for baryon/antibaryon asymmetry and for their production in the early universe \cite{Cline:2006ts, Dimopoulos:1978kv, Dimopoulos:1978pw, Weinberg:1979bt}. This asymmetry is commonly expressed in terms of the baryon-to-entropy ratio, $n_B/s$, whose experimental value turns out to be  $n_B/s=\left(8.59\pm0.10\right)\cdot 10^{-11}$. Within the standard Big Bang paradigm \cite{Riotto:1999yt}, there is no known mechanism capable of reproducing such a large baryon asymmetry, unless something by hand is introduced. Nonetheless, according to the well-established \emph{CPT theorem} \cite{Cordero:2024hjr, Barenboim:2017vlc, Wang:2017igw, Zhao:2015mqa}, the universe should have been symmetric in baryons and antibaryons immediately after the Big Bang\footnote{The CPT theorem simultaneously guarantees the charge (C), parity (P) and time reversal (T) symmetries, accounting for the observed matter-antimatter asymmetry, known as baryogenesis  \cite{Balazs:2014eba, Riotto:1998bt}.}. 

According to baryogenesis, since the seminal Sakharov's proposal \cite{Sakharov:1967dj} was presented in the late 60s, three fundamental criteria were introduced and, accordingly, every baryogenesis scheme might satisfy what follows.
\begin{itemize}
    \item[-] The Baryon number is  violated through some sort of dynamical mechanism, i.e., generating a net baryon asymmetry.
    \item[-] A $CP$ violation may occur, ensuring that the processes involving baryons and antibaryons evolve differently and preventing mutual compensation.
    \item[-] Out-of-equilibrium decays of heavy particles and, accordingly, departure from thermal equilibrium, as due to the universe's expansion are expected. 
\end{itemize}

In this respect, relevant examples turn out to be the Grand Unified Theory (GUT) baryogenesis, proposed in the 1970s \cite{Chung:1998rq}, predicting extremely massive particles whose decays can violate both baryon number and CP symmetry; electroweak baryogenesis \cite{Cohen:1990it, Farrar:1993hn, Morrissey:2012db, Bodeker:2020ghk} and baryogenesis via leptogenesis \cite{Fong:2012buy, Davidson:2008bu}, where  baryogenesis occurs during the electroweak phase transition, in the first case, while 
leptogenesis through a CP-violating decay of heavy Majorana neutrinos firstly produces a lepton asymmetry, from which recovering baryogenesis, in the second case.

Additional and more recent schemes also include  sphaleron processes, 
\cite{Buchmuller:2003gz}, as well as 
the Affleck-Dine mechanism \cite{Bettoni:2018pbl, Bettoni:2018utf, Bettoni:2019dcw, Bettoni:2021zhq, Bettoni:2024ixe, Laverda:2023uqv, Cline:2019fxx, Lloyd-Stubbs:2020sed}, proposed in 1985, within supersymmetric  frameworks, extending the standard model (SM) of particle physics. 

Almost purely geometric scenarios are even possible, as e.g. provided by the so-called  gravitational baryogenesis \cite{Davoudiasl:2004gf, Goodarzi:2023ltp, Arbuzova:2017zby, Mojahed:2024mvb, Sadjadi:2007dx, Arbuzova:2018hcj}, where  the gravitational field couples to a matter current via the derivative of the Ricci scalar, effectively generating a chemical potential for baryon number. 

Among the \emph{plethora} of possible paradigms, one of the most appealing scenario remains the \emph{spontaneous baryogenesis} \cite{Dolgov:1994zq, Dolgov:1996qq, Luongo:2021gho, DeSimone:2016ofp}, which assumes that the asymmetry originates from the spontaneous breaking of the $U(1)$ global symmetry associated with baryon number conservation in the early universe. The resulting pseudo-Nambu-Goldstone boson acts as the inflaton and becomes minimally coupled to the baryonic current, which is then no longer conserved. During reheating, the inflaton decays into baryon-antilepton and antibaryon-lepton pairs through CP-violating interactions, thereby producing a net baryon asymmetry consistent with Sakharov’s conditions.

The model has received great attention and has been extended in several ways, for example, by introducing a non-minimal coupling between the inflaton and gravity \cite{Dubbini:2025jjz}, or adding a complex scalar spectator field \cite{Dubbini:2025hcw}, showing how to significantly enhance the predicted asymmetry.

All these approaches share a common feature: \emph{the preservation of Lorentz invariance}, as a standard and consolidate fact. However, deviations from Lorentz invariance at energies approaching the Planck scale are predicted in several theoretical frameworks \cite{Jejelava:2025ijk}, including certain approaches to quantum gravity \cite{Collins:2004bp, Alfaro:2004aa}, string theory \cite{Mavromatos:2007xe, Li:2025yvq}, and other beyond-Standard Model scenarios \cite{Kepuladze:2025czd, Williams:2024woi}. 

Even though Lorentz invariance violation (LIV) is expected to be extremely small at experimentally accessible energies, \emph{the corresponding  effects can grow with energy and accumulate to potentially detectable levels over cosmological distances} \cite{Piran:2023xfg}. Astrophysical observations involving high-energy emissions and long baselines therefore provide exceptionally sensitive tests of Lorentz invariance \cite{Xi:2025ruv, Yang:2025kfr, Desai:2023rkd, Xiao:2009xe}. A relevant example is offered by gamma-ray bursts, widely-used to explore possible LIV effects. 

Within baryogenesis, it appears licit to conjecture a plausible LIV by introducing background vector fields minimally coupled to the baryonic current \cite{Carroll:2005dj, Shu:2007wi, Alfaro:2009xc}. When these vector fields acquire a vacuum expectation value (VEV), a Lorentz-violating term appears in the Lagrangian, acting as an effective chemical potential associated with baryon number. Once such a term appears in the Lagrangian, commonly the baryon asymmetry is computed as the number density of particles minus antiparticles subjected to an effective chemical potential \cite{Carroll:2005dj}. Pure (quantum) field theory-based approaches to LIV baryogenesis are therefore not extensively addressed in the scientific literature.

Motivated by the above considerations, in this work, we develop spontaneous baryogenesis with a LIV correction, leading to a LIV baryogenesis in which the baryon asymmetry is induced by the presence of the LIV term. From a physical standpoint, our starting point is the spontaneous baryogenesis framework, as our strategy for producing the baryon asymmetry follows the same underlying principles. To introduce a LIV, however, we break the $U(1)_B$ global symmetry in a different manner, specifically by employing a background complex vector field instead of a scalar one. In this way, besides ensuring baryon number violation in the broken phase, the vector field acquires a VEV with a preferred spacetime direction, which naturally generates a Lorentz-violating term in the Lagrangian. Since the background field is now vectorial in nature, the form of its interaction with fermions is accordingly modified compared to the spontaneous baryogenesis case. This new interaction structure has significant implications, particularly regarding the role of flavor oscillations between baryons and leptons in generating baryon asymmetry. Indeed, unlike the model of spontaneous baryogenesis, we find a non-null mixing factor even in the limit of massless fermions. Moreover, we do not obtain a pure mass-mixing but rather a mixing of the momenta, specifically consisting in a shift of the along-VEV component of the momentum proportional to the coupling constant of the vector-current interaction. In this scenario, such a coupling constant is not forced to be extremely small to guarantee small masses of fermions, essentially because it does not affect the fermions' masses but their momentum. Therefore, we can consider values of the coupling constant higher than those allowed within the framework of spontaneous baryogenesis, and this enables us to reproduce the experimental value of the baryon asymmetry.

The paper is outlined as follows. In Sect. \ref{sec:level2}, we briefly outline the spontaneous baryogenesis paradigm, presenting the underlying physics and the main results. In Sect. \ref{sec:level3}, we formulate the Lagrangian of our model and address the spontaneous symmetry breaking by the background vector field. We thus parametrize the VEV and write the broken phase Lagrangian. In Sect. \ref{sec:level4}, we find the equations of motion for the fields in question. In particular, we solve the equation of motion for the inflaton and then compute the inflationary dynamics. In Sect. \ref{sec:level5}, we compute the baryon asymmetry produced, ignoring the effect of lepton-baryon oscillations. In Sect. \ref{sec:level6}, we address the question of the flavor oscillations between baryons and leptons, deriving the oscillation probability. We then compute the corresponding correction to the baryon asymmetry and show the main differences between our model and scalar spontaneous baryogenesis. Finally, in Sect. \ref{sec:level7} we provide a brief discussion of Lorentz symmetry restoration. Sect. \ref{sec:level8} is dedicated to conclusions and perspectives.

\section{The spontaneous baryogenesis}\label{sec:level2}

The original model of spontaneous baryogenesis \cite{Dolgov:1994zq, Dolgov:1996qq}, hereafter referred to as scalar model of spontaneous baryogenesis (SSB), assumes the existence of a background complex scalar field subject to a symmetry breaking potential and interacting with fermion fields $Q$ and $L$, respectively carrying baryon number and not. The unbroken phase Lagrangian is thus
\begin{equation}
\begin{split}
\mathcal{L}&=\left(\partial^{\mu}\Phi\right)\left(\partial_{\mu}\Phi^*\right)-V(\Phi^*\Phi)+\overline{Q}\left(i\slashed{\partial}-m_Q\right)Q+\overline{L}\left(i\slashed{\partial}-m_L\right)L+g\Phi\overline{Q}L+g\Phi^*\overline{L}Q,
\end{split}
\label{eq1}
\end{equation}
where the symmetry breaking potential is given by
\begin{equation}
V(\Phi^*\Phi)=\lambda\left(\Phi^*\Phi-\frac{f^2}{2}\right)^2.
\label{eq2}
\end{equation}
In the model, the only quantized fields are the fermions $Q$ and $L$, while all the other fields are treated in a classical way, following \emph{de facto} a semiclassical approach.
The initial symmetry enjoyed by the Lagrangian in Eq. (\ref{eq1}) is the $U(1)$ global symmetry
\begin{equation}
\Phi\to\Phi e^{i\alpha},\quad Q\to Qe^{i\alpha},\quad L\to L,
\label{eq3}
\end{equation}
corresponding to the conservation of baryon number. When the symmetry is spontaneously broken at the energy scale $f$, presumably of the order of the Planck's scale, the complex scalar field acquires the VEV $\braket{\Phi}=(f/\sqrt{2})e^{i\theta}$, where $\theta$ is a dimensionless real scalar field representing the angular mode of the initial complex scalar field, that is not constrained by the choice of the VEV. Freezing the radial mode to $\rho=f/\sqrt{2}$, the broken phase Lagrangian results
\begin{equation}
\begin{split}
\mathcal{L}&=\frac{f^2}{2}\left(\partial^{\mu}\theta\right)\left(\partial_{\mu}\theta\right)+\overline{Q}\left(i\slashed{\partial}-m_Q\right)Q+\overline{L}\left(i\slashed{\partial}-m_L\right)L+\frac{gf}{\sqrt{2}}\left(\overline{Q}Le^{i\theta}+\overline{L}Qe^{-i\theta}\right).
\end{split}
\label{eq5}
\end{equation}
Finally, since $\theta$ is physically interpreted as the inflaton, it acquires a mass and becomes a pseudo-Nambu-Goldstone boson rather than a pure Goldstone one. Consequently, $\theta$ is subject to an inflationary potential $V(\theta)$, that should be included within the Lagrangian in Eq. (\ref{eq5}) yielding
\begin{equation}
\begin{split}
\mathcal{L}&=\frac{f^2}{2}\left(\partial^{\mu}\theta\right)\left(\partial_{\mu}\theta\right)-V(\theta)+\overline{Q}\left(i\slashed{\partial}-m_Q\right)Q+\overline{L}\left(i\slashed{\partial}-m_L\right)L+\frac{gf}{\sqrt{2}}\left(\overline{Q}Le^{i\theta}+\overline{L}Qe^{-i\theta}\right).
\end{split}
\label{eq6}
\end{equation}
The interaction term between fermions and the inflaton in Eq. (\ref{eq6}) violates the CP symmetry. This means that the inflaton decays into baryon-antilepton and lepton-antibaryon pairs with different decay rates, producing thus a net difference between number densities of baryons and antibaryons produced. In particular, the model assumes these decays to happen during the reheating epoch, when in fact the inflaton decays into fermion-antifermion pairs while performs small oscillations about the minimum of the inflationary potential. Therefore, the particular shape of the inflationary potential is irrelevant for the formulation of the model, since the latter needs merely to account for its second order Taylor expansion, providing the inflaton's mass.

In order to see the non-conservation of the baryon number in the broken phase, it is helpful to make the rotation $Q\to Qe^{i\theta}$, yielding
\begin{equation}
\begin{split}
\mathcal{L}&=\frac{f^2}{2}\left(\partial^{\mu}\theta\right)\left(\partial_{\mu}\theta\right)-V(\theta)+\overline{Q}\left(i\slashed{\partial}-m_Q\right)Q+\overline{L}\left(i\slashed{\partial}-m_L\right)L-\left(\partial_{\mu}\theta\right)J^{\mu}+\frac{gf}{\sqrt{2}}\left(\overline{Q}L+\overline{L}Q\right),
\end{split}
\label{eq8}
\end{equation}
where $J^{\mu}$ is the baryonic current, that from the Lagrangian in Eq. (\ref{eq8}) results clearly non-conserved in the broken phase. The Lagrangian in Eq. (\ref{eq8}) also tells us that the fields $Q$ and $L$ are not mass eigenstates.
Using Eq. (\ref{eq8}), the equation of motion for the inflaton results
\begin{equation}
\ddot{\theta}+\Gamma\dot{\theta}+\Omega^2\theta=0,
\label{eq9}
\end{equation}
where $\Gamma=(g^2\Omega)/(8\pi)$ is the decay rate of the inflaton and $\Omega$ is its renormalized mass, constrained by observations \cite{Adams:1992bn} to be\footnote{Actually, experimentally we have an allowed range of values, that is $f\sim[10^6-10^{12}]\Omega$. Choosing a value rather than another does not affect the building-up of the model, but precisely changes the values of the free parameter $g$ compatible with the experimental baryon asymmetry.} $\Omega\sim 10^{-6}f$.
Eq. (\ref{eq9}) admits a simple solution in the limit $g\ll1$, that we refer to as $\theta_0(t)$, namely
\begin{equation}
\theta_0(t)=\theta_Ie^{-\frac{\Gamma t}{2}}\cos(\Omega t).
\label{eq12}
\end{equation}
The solution in Eq. (\ref{eq12}) is then used for computing the baryon asymmetry. More precisely, the number density $n$ of particle-antiparticle pairs produced by the decay of a homogeneous classical scalar field is computed via
\begin{equation}
n=\frac{1}{\mathcal{V}}\sum_{s1,s2}\int\frac{d^3p_1}{(2\pi)^32p_1^0}\frac{d^3p_2}{(2\pi)^32p_2^0}|\mathcal{A}|^2,
\label{eq13}
\end{equation}
where $\mathcal{A}$ is the single pair production amplitude and $\mathcal{V}$ the volume element. Using this relation, the number density of baryons is computed by considering the decay of the inflaton into $Q-\overline{L}$ pairs, whereas the number density of antibaryons by the decay of the inflaton into $\overline{Q}-L$ pairs. The corresponding decay amplitudes are thus computed. The baryon asymmetry, that we label with $n_B^{(0)}$, is finally figured out by taking the difference between the number density of baryons and antibaryons, namely
\begin{equation}
n_B^{(0)}=\frac{1}{16\pi}\Omega g^2f^2\theta_I^3.
\label{eq14}
\end{equation}
Eq. (\ref{eq14}) takes into account neither the expansion of the universe nor the mass-mixing. In particular, the mass-mixing is addressed by diagonalizing the mass-matrix
\begin{equation}
\hat{M}=\begin{pmatrix}
m_Q & -\frac{gf}{\sqrt{2}} \\ -\frac{gf}{\sqrt{2}} & m_L
\end{pmatrix},
\label{eq15}
\end{equation}
finding the mass-eigenvalues and the corresponding mass-eigenvectors, resulting linear combinations of the quantum states $\ket{Q}$ and $\ket{L}$. Finally, computing the probability that the mass-eigenstates rotate into quantum states $\ket{Q}$ and $\ket{L}$, it is found that the baryon asymmetry in Eq. (\ref{eq14}) is modified by a multiplicative factor $[(1-\epsilon^2)/(1+\epsilon^2)]^2$,
with
\begin{equation}
\epsilon=\frac{\sqrt{2}gf}{\Delta m+\sqrt{\Delta m^2+2g^2f^2}}\quad \text{and} \quad \Delta m=m_Q-m_L.
\label{eq17}
\end{equation}
Finally, it is found that introducing the cosmic expansion increases the baryon asymmetry by a factor $8\pi/g^2$.

\section{The spontaneous baryogenesis with a background complex vector field}\label{sec:level3}

In this work, we formulate a model of spontaneous baryogenesis, hereafter dubbed vector-field spontaneous baryogenesis (VFSB), that spontaneously breaks the $U(1)_B$ symmetry with a complex vector field rather than with a scalar. Due to the vectorial nature of the background field, Lorentz symmetry is also broken. We thus obtain a LIV interaction term between baryons and inflaton in the Lagrangian, that is inspired by the well-known LIV interaction $\sim a_{\mu}J^{\mu}$ \cite{Kost98}. In particular, our LIV interaction term is the same that also violates the CP symmetry and baryon number conservation in the broken phase, and is thus able to generate a net baryon asymmetry according to the Sakharov criteria.

Moreover, the LIV interaction considered in this work shows a substantial difference with respect to that considered within SSB. In fact, there the interaction between the fields $Q$ and $L$ produces a pure mass-mixing between fermions, where the mixing term is proportional to the coupling constant, roughly $\sim gf$. This constrains the coupling constant to be extremely small in order to guarantee the production of fermions much lighter than the inflaton. However, taking such a small coupling constant does not allow the model to reproduce the experimental value of the baryon asymmetry.

Conversely, considering an initial vector field rather than a scalar would change the interaction from a Yukawa-like coupling to a vector-current one. In so doing, we expect that the mixing would affect the momenta of the produced fermions and not their masses. Consequently, we would obtain a much weaker constraint on the coupling constant\footnote{In this case, we no longer need to impose $g f \ll \Omega$, albeit  we still require $g \ll 1$ to ensure the validity of the perturbative regime.}, allowing us to reproduce the observed baryon-to-entropy ratio.

\subsection{The formulation of the theory}\label{subsec:level3-1}

We consider a complex vector field $A_{\mu}$ minimally coupled to the fermion fields $Q$ and $L$, respectively carrying baryon number and not. As well as in SSB, we treat fermions as quantized fields and the vector field in a classical way. We assume the vector field $A_{\mu}$ to be subject to the symmetry breaking potential
\begin{equation}
V(A^{\mu}A_{\mu}^*)=\pm\mu^2A^{\mu}A_{\mu}^*+\frac{\lambda}{2}(A^{\mu}A_{\mu}^*)^2,
\label{eq18}
\end{equation}
containing the mass term and the self-interaction, with coupling constant $\lambda$ taken positive to guarantee the potential to have a minimum. One may wonder about the $\pm$ in front of the mass term, since usually, e.g. in a complex scalar field theory, to be the symmetry spontaneously broken the signs of the mass term and the self-interaction should be opposite. However, in that case, the argument of the potential is the square modulus of the field, that is always positive-definite. Conversely, here, the sign of the argument of the potential depends on the nature of the VEV that we choose for the complex vector field. In particular, considering the Minkowski metric, $
\eta_{\mu\nu}=\text{diag}\left(+1,-1,-1,-1\right)$, a timelike VEV $\braket{A_{\mu}}$ reproduces the situation of the complex scalar field, since $\braket{A^{\mu}A_{\mu}^*}>0$. In this case, therefore, the mass term should have a flipped sign with respect to the self-interaction, thus the $-$ sign might be chosen. Conversely, for a spacelike VEV $\braket{A^{\mu}A_{\mu}^*}<0$, thus the sign in front of the mass term is fixed to be a $+$, like the sign of the quartic term. At this stage, we prefer not to specify the nature of the VEV, thus we leave the $\pm$ in front of the mass term, where the $+$ is correct for a spacelike VEV and the $-$ for a timelike one. The Lagrangian density that we consider results thus
\begin{equation}
\begin{split}
\mathcal{L}&=-\frac{1}{2}F^{\mu\nu}F_{\mu\nu}^*\mp V(A^{\mu}A_{\mu}^*)+\overline{Q}(i\slashed{\partial}-m_Q)Q+\overline{L}(i\slashed{\partial}-m_L)L+g(A_{\mu}\overline{Q}\gamma^{\mu}L+A_{\mu}^*\overline{L}\gamma^{\mu}Q),
\end{split}
\label{eq20}
\end{equation}
where $\mp$ in front of the potential refer respectively to a spacelike and a timelike VEV, guaranteeing the mass term to have a $-$ sign in front in both of the cases within the full Lagrangian. This is necessary in order to have non-tachyonic fields after the symmetry breaking but real particles with square mass positive-definite. Indeed, before the symmetry breaking, one usually assumes the fields to be tachyonic, with a square mass negative-definite, in order to obtain after the symmetry breaking real particles. In this case, for a timelike VEV the mass term already has a $-$ in front within the shape of the potential, therefore it is not necessary to add an extra $-$ sign in front of the potential itself. Conversely, for a spacelike VEV, one needs to put by hand a $-$ sign in front of the potential in order the fields to be tachyonic before of the symmetry breaking and acquire positive-definite mass after it.

The Lagrangian in Eq. (\ref{eq20}) is invariant under the following global-phase transformations of the fields,
\begin{equation}
Q\rightarrow Qe^{i\alpha},\quad A_{\mu}\rightarrow A_{\mu}e^{i\alpha},\quad L\rightarrow L,
\label{eq21}
\end{equation}
enjoying thus the corresponding $U(1)$ global symmetry. The shape of the potential in Eq. (\ref{eq18}) allows the symmetry to be spontaneously broken when the field $A_{\mu}$ arbitrarily chooses a minimum and acquires a VEV. In particular, we have
\begin{equation}
\braket{A^{\mu}A_{\mu}^*}=\mp\frac{\mu^2}{\lambda},
\label{eq22}
\end{equation}
constraining the modulus of the vector field in the minimum. It is important to remark the fact that the $\mp$ that appears in Eq. (\ref{eq22}) is due to not have chosen the nature of the VEV yet, i.e., the exact form of the potential. Once chosen the $+$ or the $-$ sign in the potential in Eq. (\ref{eq18}), the $\mp$ in Eq. (\ref{eq22}) becomes $-$ or $+$ respectively. The $\mp$ in Eq. (\ref{eq22}) has thus nothing to do with the arbitrariness in choosing a VEV when the symmetry is spontaneously broken. The modulus of the vector field is perfectly constrained to a precise value once the potential is well defined. We are thus allowed to freely choose a VEV of $A_{\mu}$, with the precaution that its modulus square respects Eq. (\ref{eq22}). In particular, we take a VEV pointing along a precise spacetime direction, that is
\begin{equation}
\braket{A_{\mu}}=v_{\mu}e^{i\theta},
\label{eq23}
\end{equation}
where $v_{\mu}$ is a real constant vector having simply one non-null component and such that
\begin{equation}
\braket{v_{\mu}v^{\mu}}=\mp\frac{\mu^2}{\lambda}.
\label{eq24}
\end{equation}
We do not yet specify which one of the four components of $v_{\mu}$ is non-null, adopting in this way the most general approach. Consequently, spacelike or timelike nature of the VEV is not yet specified too.

The dynamics of the vector field $A_{\mu}$ can be now studied in terms of its small oscillations about the VEV. In particular, we can write
\begin{equation}
A_{\mu}=(v_{\mu}+a_{\mu})e^{i\theta},
\label{eq25}
\end{equation}
where $a_{\mu}$ is a complex vector field describing the small oscillations about the VEV by the initial vector field. One may wonder that the expression in Eq. (\ref{eq25}) is overcounting the degrees of freedom of our theory, that seem $6$ associated to the complex vector field $a_{\mu}$, as usual for a Proca-like complex vector field, and $1$ to the scalar field $\theta$. However, writing the VEV as in Eq. (\ref{eq24}), we are already leaving freedom to rotate within the complex plane to the mode of $A_{\mu}$ pointing along the spacetime direction of the VEV. Therefore, among the four complex modes of $a_{\mu}$, $\theta$ represents the global phase of that pointing along the direction of the VEV. Therefore, this mode of $a_{\mu}$  carries one physical degree of freedom, associated to its modulus, since the phase has already been factorized as $e^{i\theta}$. Reasoning in terms of real and imaginary parts of $a_{\mu}$, say $\alpha_{\mu}$ and $\beta_{\mu}$ respectively, this means that the component of $\beta_{\mu}$ directed along the VEV does not carry a physical degree of freedom. From a mathematical point of view, this can be addressed by considering the component of $\beta_{\mu}$ directed along the VEV to be identically null.

\subsection{The broken phase Lagrangian}\label{subsec:level3-2}

Substituting $A_{\mu}$ as in Eq. (\ref{eq25}) within the Lagrangian in Eq. (\ref{eq20}), the kinetic term for the vector field $A_{\mu}$ becomes
\begin{equation}
\begin{split}
-\frac{1}{2}F^{\mu\nu}F_{\mu\nu}^{*}&=-\frac{1}{2}f^{\mu\nu}f^*_{\mu\nu}-\frac{1}{2}K^{\mu\nu}(v)K_{\mu\nu}(v)-\frac{1}{2}K^{\mu\nu}(a)K_{\mu\nu}(a^*)-\frac{1}{2}K^{\mu\nu}(v)K_{\mu\nu}(a+a^*)+\\&+\frac{i}{2}K^{\mu\nu}(v)\left(f_{\mu\nu}-f_{\mu\nu}^*\right)+\frac{i}{2}\left[K^{\mu\nu}(a^*)f_{\mu\nu}-K^{\mu\nu}(a)f_{\mu\nu}^*\right],
\end{split}
\label{eq26}
\end{equation}
where $f_{\mu\nu}$ is the field strength tensor for $a_{\mu}$ and $
K_{\mu\nu}(w)=w_{\nu}\left(\partial_{\mu}\theta\right)-w_{\mu}\left(\partial_{\nu}\theta\right)$, 
is an antisymmetric tensor satisfying the following properties,
\begin{equation}
\begin{split}
&K_{\mu\nu}(w_1+w_2)=K_{\mu\nu}(w_1)+K_{\mu\nu}(w_2),\quad \quad \text{(linearity)}\\& K_{\mu\nu}^*(w)=K_{\mu\nu}(w^*).
\end{split}
\label{eq28}
\end{equation}
As expected, the kinetic term for the vector field $a_{\mu}$ appears and, moreover, Eq. (\ref{eq28}) contains various interaction terms between the field $\theta$ and the vector field $a_{\mu}$. Finally, the kinetic term for the field $\theta$ is hidden inside $K^2(v)$, i.e.,
\begin{equation}
\begin{split}
&K^{\mu\nu}(v)K_{\mu\nu}(v)=\mp\frac{2\mu^2}{\lambda}\left(\partial\theta\right)^2-2\left(v\cdot\partial\theta\right)^2.
\end{split}
\label{eq29}
\end{equation}
We consider a spatially flat homogeneous and isotropic universe, where thus we can consider classical fields depending only on cosmic time. Thus, the second term among the two in Eq. (\ref{eq29}) can be written as $2(v^0\dot{\theta})^2$. If we consider a timelike VEV, viz., 
\begin{equation}
v_{\mu}=\left(\sqrt{\frac{\mu^2}{\lambda}},0,0,0\right),
\label{eq31}
\end{equation}
then it is immediate to see that Eq. (\ref{eq29}) would vanish, thus the kinetic term of the classical field $\theta$ would be canceled out from the Lagrangian. This means that the dynamics of $\theta$ would not be independent of the other fields and thus $\theta$ could not be considered as a separate degree of freedom for our theory. In so doing, we would be left with exactly $5$ degrees of freedom, that is inconsistent with our approach. Therefore, we necessarily have to choose a spacelike VEV, in a generic direction $i$, that is
\begin{equation}
v_{\mu}=v^i\eta_{\mu i}, \quad \text{with} \quad v^i=\sqrt{\frac{\mu^2}{\lambda}}.
\label{eq32}
\end{equation}
So, since the VEV is spacelike,
\begin{itemize}

\item[-] the $+$ sign within the expression for the potential, in Eq. (\ref{eq18}), is the right one and consequently the negative sign in Eq. (\ref{eq24}) is fixed; 
\item[-] the negative sign in front of the potential is fixed. 

\end{itemize}
Eq. (\ref{eq29}) is thus simplified to $
K^{\mu\nu}(v)K_{\mu\nu}(v)=-f^2\left(\partial_{\mu}\theta\right)\left(\partial^{\mu}\theta\right)$, where we define $f=\mu\sqrt{2/\lambda}$ as the energy scale proper of the symmetry breaking. Following the SSB, we assume $f$ to be of the order of the Planck's scale. 

The $-$ sign appearing in the expression of $K^2(v)$ plays a fundamental role for the physical meaning of the theory. Indeed, since $K^2(v)$ in the Lagrangian is multiplied by $-1/2$, the term in question finally results $(f^2/2)(\partial\theta)^2$, representing a valid kinetic term for the field $\theta$. In so doing, our broken phase Lagrangian contains the kinetic terms for both $a_{\mu}$ and $\theta$, further than for fermions, so that all the fields in question have independent dynamics, and the degrees of freedom are all preserved.

Analogously, we have to see how the potential $V$ in Eq. (\ref{eq18}) transforms when substituting for $A_{\mu}$ Eq. (\ref{eq25}). In particular, since $a_{\mu}$ strictly represents  a small displacement from the minimum, we consider terms up to quadratic order in $a_{\mu}$, yielding
\begin{equation}
V(A^{\mu}A^*_{\mu})=\frac{\lambda}{2}v^{\mu}v^{\nu}\left(a_{\mu}+a_{\mu}^*\right)\left(a_{\nu}+a_{\nu}^*\right).
\label{eq36}
\end{equation}
The potential appearing in the broken phase Lagrangian resembles somehow a mass term for the vector field $a_{\mu}$. However, this point can be clarified if we write $a_{\mu}$ in terms of its real and imaginary part, respectively $\alpha_{\mu}$ and $\beta_{\mu}$, as $a_{\mu}=(\alpha_{\mu}+i\beta_{\mu})/\sqrt{2}$. Notice that along the direction of the VEV $v_i$, only the real part $\alpha_i \neq 0$ and $\beta_i =0$, that is, $a_i \equiv \alpha_i.$ This is because the phase degree of freedom is already contained in $e^{i\theta}$: the field along the broken symmetry direction is $A_i = (v_i + \alpha_i)e^{i\theta}$. A nonvanishing $\beta_i$ would indeed lead to an overcounting of the degrees of freedom. 

With respect to the two real vector fields $\alpha_{\mu}$ and $\beta_{\mu}$, Eq. (\ref{eq36}) becomes
\begin{equation}
V(A^{\mu}A_{\mu}^*)=\lambda v^{\mu}v^{\nu}\alpha_{\mu}\alpha_{\nu}.
\label{eq38}
\end{equation}
Physically, therefore, after the symmetry breaking the real component of the initial vector field pointing along the direction of the VEV acquires mass $m_{\alpha}=\sqrt{2}\mu=f\sqrt{\lambda}$, whereas all the other modes remain massless.

We prefer writing also the kinetic term in Eq. (\ref{eq26}) in terms of the real vector fields $\alpha_{\mu}$ and $\beta_{\mu}$, obtaining
\begin{equation}
\begin{split}
-\frac{1}{2}F^{\mu\nu}F_{\mu\nu}^*=&-\frac{1}{4}\alpha^{\mu\nu}\alpha_{\mu\nu}-\frac{1}{4}\beta^{\mu\nu}\beta_{\mu\nu}+\frac{f^2}{2}\left(\partial_{\mu}\theta\right)\left(\partial^{\mu}\theta\right)-\frac{1}{4}\left[K^{\mu\nu}(\alpha)K_{\mu\nu}(\alpha)+K^{\mu\nu}(\beta)K_{\mu\nu}(\beta)\right]+\\&-\frac{1}{\sqrt{2}}K^{\mu\nu}(v)\left[K_{\mu\nu}(\alpha)+\beta_{\mu\nu}\right]+\frac{1}{2}\left[K^{\mu\nu}(\beta)\alpha_{\mu\nu}-K^{\mu\nu}(\alpha)\beta_{\mu\nu}\right],
\end{split}
\label{eq39}
\end{equation}
where $\alpha_{\mu\nu}$ and $\beta_{\mu\nu}$ are the field strength tensors respectively for $\alpha_{\mu}$ and $\beta_{\mu}$.

Finally, the interaction term in Eq. (\ref{eq20}) after the symmetry breaking becomes
\begin{equation}
\begin{split}
g\left(A_{\mu}\overline{Q}\gamma^{\mu}L+A_{\mu}^*\overline{L}\gamma^{\mu}Q\right)&=gv_{\mu}\left(e^{i\theta}\overline{Q}\gamma^{\mu}L+e^{-i\theta}\overline{L}\gamma^{\mu}Q\right)+\\&+\frac{g}{\sqrt{2}}\alpha_{\mu}\left(e^{i\theta}\overline{Q}\gamma^{\mu}L+e^{-i\theta}\overline{L}\gamma^{\mu}Q\right)+\frac{ig}{\sqrt{2}}\beta_{\mu}\left(e^{i\theta}\overline{Q}\gamma^{\mu}L-e^{-i\theta}\overline{L}\gamma^{\mu}Q\right).
\end{split}
\label{eq40}
\end{equation}
Therefore, the full broken phase Lagrangian is
\begin{equation}
\begin{split}
\mathcal{L}=&-\frac{1}{4}\alpha^{\mu\nu}\alpha_{\mu\nu}-\frac{1}{4}\beta^{\mu\nu}\beta_{\mu\nu}+\frac{f^2}{2}\left(\partial_{\mu}\theta\right)\left(\partial^{\mu}\theta\right)-\frac{1}{4}\left[K^{\mu\nu}(\alpha)K_{\mu\nu}(\alpha)+K^{\mu\nu}(\beta)K_{\mu\nu}(\beta)\right]+\\&-\frac{1}{\sqrt{2}}K^{\mu\nu}(v)\left[K_{\mu\nu}(\alpha)+\beta_{\mu\nu}\right]+\frac{1}{2}\left[K^{\mu\nu}(\beta)\alpha_{\mu\nu}-K^{\mu\nu}(\alpha)\beta_{\mu\nu}\right]-\lambda v^{\mu}v^{\nu}\alpha_{\mu}\alpha_{\nu}+\\&+\overline{Q}\left(i\slashed{\partial}-m_Q\right)Q+\overline{L}\left(i\slashed{\partial}-m_L\right)L+gv_{\mu}\left(e^{i\theta}\overline{Q}\gamma^{\mu}L+e^{-i\theta}\overline{L}\gamma^{\mu}Q\right)+\\&+\frac{g}{\sqrt{2}}\alpha_{\mu}\left(e^{i\theta}\overline{Q}\gamma^{\mu}L+e^{-i\theta}\overline{L}\gamma^{\mu}Q\right)+\frac{ig}{\sqrt{2}}\beta_{\mu}\left(e^{i\theta}\overline{Q}\gamma^{\mu}L-e^{-i\theta}\overline{L}\gamma^{\mu}Q\right).
\end{split}
\label{eq41}
\end{equation}
The Lagrangian in Eq. (\ref{eq41}) contains the key ingredient to active the same mechanism as in SSB, that is a non-conservation of baryon number and a CP-violating interaction between inflaton and fermions. Therefore, the way in which we want to produce the baryon asymmetry is physically the same of SSB. \emph{However, using a vector field rather than a scalar as background field makes our theory LIV, in particular because of the constant vectorial VEV acquired by the vector field. Therefore, while spontaneously breaking the $U(1)$ global symmetry, we are also producing a LIV, obtaining a model that is physically based on that SSB but containing also a spontaneous violation of Lorentz invariance.}

As done in SSB, it is useful to redefine the fermion field $Q$ according to
\begin{equation}
Q\rightarrow Q'e^{i\theta},
\label{eq42}
\end{equation}
in order to make the non-conservation of the baryonic current manifest. Moreover, the broken phase Lagrangian of Eq. (\ref{eq41}) contains in essence the derivatives of the scalar field $\theta$, that is thus a Nambu-Goldstone boson. The symmetry that still survives is now
\begin{equation}
\theta\rightarrow\theta+\alpha,
\label{eq43}
\end{equation}
with $\alpha$ constant. As in SSB, we would like to assume the field $\theta$ to be the inflaton, thus massive. We thus need to require also an explicit breaking of the symmetry of Eq. (\ref{eq43}), providing a potential energy $V(\theta)$ for the field $\theta$. This, independently of the inflationary model adopted, can be approximated up to the second order in $\theta$, describing \emph{de facto} the small oscillations of the inflaton during reheating epoch. The final expression for the broken phase Lagrangian is thus
\begin{equation}
\begin{split}
\mathcal{L}=&-\frac{1}{4}\alpha^{\mu\nu}\alpha_{\mu\nu}-\lambda v^{\mu}v^{\nu}\alpha_{\mu}\alpha_{\nu}-\frac{1}{4}\beta^{\mu\nu}\beta_{\mu\nu}+\frac{f^2}{2}\left(\partial_{\mu}\theta\right)\left(\partial^{\mu}\theta\right)-V(\theta)+\\&-\frac{1}{4}\left[K^{\mu\nu}(\alpha)K_{\mu\nu}(\alpha)+K^{\mu\nu}(\beta)K_{\mu\nu}(\beta)\right]-\frac{1}{\sqrt{2}}K^{\mu\nu}(v)\left[K_{\mu\nu}(\alpha)+\beta_{\mu\nu}\right]+\frac{1}{2}\left[K^{\mu\nu}(\beta)\alpha_{\mu\nu}-K^{\mu\nu}(\alpha)\beta_{\mu\nu}\right]+\\&+\overline{Q'}\left(i\slashed{\partial}-m_Q\right)Q'+\overline{L}\left(i\slashed{\partial}-m_L\right)L-\left(\partial_{\mu}\theta\right)\overline{Q'}\gamma^{\mu}Q'+gv_{\mu}\left(\overline{Q'}\gamma^{\mu}L+\overline{L}\gamma^{\mu}Q'\right)+\\&+\frac{g}{\sqrt{2}}\alpha_{\mu}\left(\overline{Q'}\gamma^{\mu}L+\overline{L}\gamma^{\mu}Q'\right)+\frac{ig}{\sqrt{2}}\beta_{\mu}\left(\overline{Q'}\gamma^{\mu}L-\overline{L}\gamma^{\mu}Q'\right).
\end{split}
\label{eq44}
\end{equation}
In summary, we start with an unbroken phase Lagrangian containing $8$ massive modes, that are the $4$ massive complex modes of the initial Proca-like complex vector field. Then, after having broken the $U(1)$ global symmetry, only the real mode along the direction of the VEV has left with mass, whereas all the other modes become massless, some of them playing the role of Goldstone bosons. 

In particular, to well identify which massless modes are Goldstone bosons and which are not, we examine the situation from the standpoint of symmetry groups.

\begin{itemize}

\item[-] The broken phase Lagrangian is not Lorentz invariant, whereas the unbroken phase Lagrangian is. This means that, further than breaking the $U(1)$ global symmetry, \emph{when symmetry is spontaneously broken we are also generating a LIV in our theory}. In particular, breaking the symmetry and choosing a spacelike VEV, we break the invariance of the theory under the boosts along the direction of the VEV and rotations around all the axis but that along the direction of the VEV. In so doing, $3$ generators are broken and $3$ Goldstone bosons are thus expected to come out.

\item[-] Breaking $U(1)$ produces another Goldstone boson, for a total of $4$.

\item[-] In terms of symmetry groups, initially the Lagrangian is invariant under the symmetry group $U(1)\times SO(1,3)$, whereas after the symmetry breaking strictly $SO(1,2)$ is left. Actually we thus have $4$ Goldstone bosons produced as a consequence of the symmetry breaking, that are $\theta$ and the components of $\alpha_{\mu}$ not directed along the VEV.

\item[-] The physical, non-null and orthogonal-to-VEV components of $\beta_{\mu}$ do not correspond to a broken symmetry, since they represent flat directions of the potential; thus, they cannot be regarded as Goldstone bosons, even if they are massless. 

\end{itemize}

The total number of Goldstone bosons that we obtain is thus $4$, that is equal to the total numbers of generators broken in the spontaneous symmetry breaking
\begin{equation}
U(1)\times SO(1,3)\to SO(1,2).
\label{eq45}
\end{equation}

\section{The equations of motion}\label{sec:level4}

We have now our full broken phase Lagrangian to study, represented by Eq. (\ref{eq44}). The dynamics of the fields in question can be thus found by using the usual Euler-Lagrange formalism.

\subsection{The fermion fields}\label{subsec:level4_1}

Considering the Lagrangian in Eq. (\ref{eq44}), the equations of motion for $Q'$ result into
\begin{equation}
\left(i\slashed{\partial}-m_Q\right)Q'-\left(\partial_{\mu}\theta\right)\gamma^{\mu}Q'=-g\left(\slashed{v}+\frac{\slashed{\alpha}+i\slashed{\beta}}{\sqrt{2}}\right)L,
\label{eq46}
\end{equation}
that can be rewritten in terms of $Q$ as
\begin{equation}
\left(i\slashed{\partial}-m_Q\right)Q=-g\left(\slashed{v}+\frac{\slashed{\alpha}+i\slashed{\beta}}{\sqrt{2}}\right)Le^{i\theta}.
\label{eq47}
\end{equation}
Analogously, in terms of the original field $Q$, for $L$ we find
\begin{equation}
\left(i\slashed{\partial}-m_L\right)L=-g\left(\slashed{v}+\frac{\slashed{\alpha}-i\slashed{\beta}}{\sqrt{2}}\right)Qe^{-i\theta}.
\label{eq48}
\end{equation}
From the equations of motion for fermions, we derive the explicit expression of the four-divergence of the baryonic current $J^{\mu}_{Q'}$, namely
\begin{equation}
\partial_{\mu}J^{\mu}_{(Q')}=ig\left[\overline{Q'}\left(\slashed{v}+\slashed{a}\right)L-\overline{L}\left(\slashed{v}+\slashed{a}^*\right)Q'\right].
\label{eq49}
\end{equation}
Non-conservation of baryonic current can be now clearly interpreted as a difference between baryons and antibaryons. It is important to point out that, since $a_{\mu}$ is only a small perturbation with respect to the VEV $v_{\mu}$, presumably the leading terms in the right-hand side (r.h.s.) of Eq. (\ref{eq49}) are those containing $\slashed{v}$, so those that violate the Lorentz invariance. It is thus clear that the LIV interaction term is the key ingredient to obtain the production of baryon asymmetry within VFSB.

\subsection{The vector fields}\label{subsec:level4-2}

Considering the Lagrangian in Eq. (\ref{eq44}), the equation of motion for $\alpha_{\mu}$ results
\begin{equation}
\begin{split}
&\partial_{\mu}\alpha^{\mu\nu}-2v^{\nu}v^{\mu}\alpha_{\mu}-K^{\mu\nu}\left(\alpha\right)\left(\partial_{\mu}\theta\right)=\partial_{\mu}K^{\mu\nu}(\beta)+\sqrt{2}K^{\mu\nu}\left(v\right)\left(\partial_{\mu}\theta\right)+\beta^{\mu\nu}\left(\partial_{\mu}\theta\right)-\frac{g}{\sqrt{2}}\left(\overline{Q'}\gamma^{\nu}L+\overline{L}\gamma^{\nu}Q'\right).
\end{split}
\label{eq50}
\end{equation}
We consider a homogeneous and isotropic universe, characterized by the spatially-flat Friedmann-Lemaître-Robertson-Walker (FLRW) metric, and therefore we assume the classical fields to be only time-dependent. 

In so doing, we follow a semiclassical approach, where all the fields but fermions are treated as classical fields. Therefore, Eq. (\ref{eq50}) is simplified to
\begin{equation}
\begin{split}
\ddot{\alpha^{\nu}}-\eta^{\nu}_0\ddot{\alpha^0}-2v^{\mu}\alpha_{\mu}v^{\nu}-\dot{\theta}^2\left(\alpha^{\nu}-\eta^{\nu}_0\alpha^0\right)&=\ddot{\theta}\left(\beta^{\nu}-\eta^{\nu}_0\beta^0\right)+\dot{\theta}\left(\dot{\beta^{\nu}}-\eta^{\nu}_0\dot{\beta^0}\right)+\sqrt{2}\dot{\theta}^2v^{\nu}+\\&+\dot{\theta}\left(\dot{\beta^{\nu}}-\eta^{\nu}_0\dot{\beta^0}\right)-\frac{g}{\sqrt{2}}\left(\overline{Q'}\gamma^{\nu}L+\overline{L}\gamma^{\nu}Q'\right).
\end{split}
\label{eq51}
\end{equation}
From the zero-component of Eq. (\ref{eq51}) we simply obtain the constraint
\begin{equation}
\overline{Q'}\gamma^{0}L+\overline{L}\gamma^{0}Q'=0,
\label{eq52}
\end{equation}
so the mode $\alpha_0$ does not have a dynamics. Conversely, for the generic spatial component $j$, we have
\begin{equation}
\begin{split}
&\ddot{\alpha^{j}}-2v^{\mu}\alpha_{\mu}v^{j}-\dot{\theta}^2\alpha^{j}=\ddot{\theta}\beta^{j}+2\dot{\theta}\dot{\beta^{j}}+\sqrt{2}\dot{\theta}^2v^{j}-\frac{g}{\sqrt{2}}\left(\overline{Q'}\gamma^{j}L+\overline{L}\gamma^{j}Q'\right).
\end{split}
\label{eq53}
\end{equation}
Since our theory violates the Lorentz invariance, we expect the three spatial components not to have the same dynamics. In particular, we expect the component along the VEV direction to have a different equation of motion with respect to the other two spatial ones. Indeed, considering the component along the VEV direction, labeled for simplicity with $i$, we get
\begin{equation}
\begin{split}
&\ddot{\alpha^{i}}+f^2\alpha^i-\dot{\theta}^2\alpha^{i}=f\dot{\theta}^2-\frac{g}{\sqrt{2}}\left(\overline{Q'}\gamma^{i}L+\overline{L}\gamma^{i}Q'\right),
\end{split}
\label{eq54}
\end{equation}
whereas for the other two, both labeled with the same index $k$, we obtain
\begin{equation}
\begin{split}
&\ddot{\alpha^{k}}-\dot{\theta}^2\alpha^{k}=\ddot{\theta}\beta^{k}+2\dot{\theta}\dot{\beta^{k}}-\frac{g}{\sqrt{2}}\left(\overline{Q'}\gamma^{k}L+\overline{L}\gamma^{k}Q'\right).
\end{split}
\label{eq55}
\end{equation}
Proceeding analogously, the equation of motion for $\beta_{\mu}$ results
\begin{equation}
\begin{split}
&\partial_{\mu}\beta^{\mu\nu}-K^{\mu\nu}\left(\beta\right)\left(\partial_{\mu}\theta\right)=-\partial_{\mu}K^{\mu\nu}\left(\alpha\right)-\sqrt{2}\partial_{\mu}K^{\mu\nu}\left(v\right)-\alpha^{\mu\nu}\left(\partial_{\mu}\theta\right)-\frac{ig}{\sqrt{2}}\left(\overline{Q'}\gamma^{\nu}L-\overline{L}\gamma^{\nu}Q'\right),
\end{split}
\label{eq56}
\end{equation}
that in a FLRW universe becomes
\begin{equation}
\begin{split}
\ddot{\beta^{\nu}}-\eta^{\nu}_0\ddot{\beta^0}-\dot{\theta}^2\left(\beta^{\nu}-\eta^{\nu}_0\beta^0\right)&=-\ddot{\theta}\left(\alpha^{\nu}-\eta^{\nu}_0\alpha^0\right)-\dot{\theta}\left(\dot{\alpha^{\nu}}-\eta^{\nu}_0\dot{\alpha^0}\right)-\sqrt{2}\ddot{\theta}v^{\nu}+\\&-\dot{\theta}\left(\dot{\alpha^{\nu}}-\eta^{\nu}_0\dot{\alpha}^0\right)-\frac{ig}{\sqrt{2}}\left(\overline{Q'}\gamma^{\nu}L-\overline{L}\gamma^{\nu}Q'\right).
\end{split}
\label{eq57}
\end{equation}
Once again, from the zero-component we immediately obtain the constraint
\begin{equation}
\overline{Q'}\gamma^{0}L-\overline{L}\gamma^{0}Q'=0,
\label{eq58}
\end{equation}
whereas for the generic spatial component $j$ we get
\begin{equation}
\ddot{\beta^{j}}-\dot{\theta}^2\beta^{j}=-\ddot{\theta}\alpha^{j}-2\dot{\theta}\dot{\alpha^{j}}-\sqrt{2}\ddot{\theta}v^{j}-\frac{ig}{\sqrt{2}}\left(\overline{Q'}\gamma^{j}L-\overline{L}\gamma^{j}Q'\right).
\label{eq59}
\end{equation}
We now distinguish between along-VEV direction and orthogonal ones. In particular, we want to remark that the component $\beta^i$ parallel to the VEV is assumed to be identically null in order not to overcount the physical degrees of freedom of the theory, since the corresponding one is already represented by $\theta$. Therefore, for $j=i$, Eq. (\ref{eq59}) becomes
\begin{equation}
\ddot{\theta}\alpha^i+\ddot{\theta}f+2\dot{\theta}\dot{\alpha^i}=-\frac{ig}{\sqrt{2}}\left(\overline{Q'}\gamma^{i}L-\overline{L}\gamma^{i}Q'\right).
\label{eq60}
\end{equation}
More precisely, Eq. (\ref{eq60}) is essentially a constraint, as $\beta^i$ is not a physical degree of freedom, as we previously argued. Therefore, we expect that Eq. (\ref{eq60}) is not independent of the equations of motion of the other fields. 
Conversely, the spatial components of $\beta^{\mu}$ orthogonal to the direction of the VEV, say $\beta^k$, have an independent dynamics. For them, we have
\begin{equation}
\ddot{\beta^{k}}-\dot{\theta}^2\beta^{k}=-\ddot{\theta}\alpha^{k}-2\dot{\theta}\dot{\alpha^{k}}-\frac{ig}{\sqrt{2}}\left(\overline{Q'}\gamma^{k}L-\overline{L}\gamma^{k}Q'\right).
\label{eq61}
\end{equation}

\subsection{The inflaton field}\label{subsec:level4_3}

Considering the Lagrangian in Eq. (\ref{eq44}), the inflaton equations of motion read
\begin{equation}
\begin{split}
&\partial_{\nu}\left[f^2\partial^{\nu}\theta+K^{\mu\nu}\left(\alpha\right)\alpha_{\mu}+K^{\mu\nu}\left(\beta\right)\beta_{\mu}\right]+m^2_{\theta}f^2\theta+\\&+\partial_{\nu}\bigg{[}+\sqrt{2}K^{\mu\nu}(v)\alpha_{\mu}+\sqrt{2}K^{\mu\nu}(\alpha)v_{\mu}+\sqrt{2}\beta^{\mu\nu}v_{\mu}-\alpha^{\mu\nu}\beta_{\mu}+\beta^{\mu\nu}\alpha_{\mu}-\overline{Q'}\gamma^{\nu}Q'\bigg{]}=0,
\end{split}
\label{eq62}
\end{equation}
that in a FLRW universe become
\begin{equation}
\begin{split}
&f^2\left(\ddot{\theta}+m^2_{\theta}\theta\right)-\partial_0\bigg{[}\dot{\theta}\alpha^j\alpha_j+\dot{\theta}\beta^k\beta_k-2f\dot{\theta}\alpha^{i}-\dot{\alpha^k}\beta_k+\dot{\beta^k}\alpha_k\bigg{]}=\partial_{\mu}J^{\mu}_{Q'}.
\end{split}
\label{eq63}
\end{equation}
The notation adopted in Eq. \eqref{eq63} and in the subsequent equations is as follows. The index $i$ denotes the spatial direction determined by the spacelike VEV $v^i$; the index $k$ runs over the remaining orthogonal spatial components, while $j$ runs over all the three spatial directions $j=i,k$. The Einstein convention is understood on repeated $j$ and $k$ indices.

This distinction is necessary since the modes parallel to the direction of the VEV have a different dynamics with respect to those orthogonal.

Eq. (\ref{eq63}) can be further simplified using the equations of motion for the other fields and the various constraint equations. In particular, Eqs. (\ref{eq55}) and (\ref{eq61}), for the spatial and orthogonal-to-VEV components of $\alpha^{\mu}$ and $\beta^{\mu}$, respectively, are helpful to let on the left-hand side (l.h.s.) only terms dependent on $\theta$ and $\alpha^i$, whereas on the r.h.s. only quantities depending on fermions, acting thus as a source-like term. Moreover, the obtained r.h.s. can be further simplified using Eq. (\ref{eq49}) for the four-divergence of the baryonic current and the constraints in Eqs. (\ref{eq52}) and (\ref{eq58}), yielding\footnote{The full calculations can be found in Appendix \ref{app1}.}
\begin{equation}
\begin{split}
&f^2\left(\ddot{\theta}+m^2_{\theta}\theta\right)-\ddot{\theta}\alpha^i\alpha_i-2\dot{\theta}\alpha_i\dot{\alpha^i}+2f\ddot{\theta}\alpha^i+2f\dot{\theta}\dot{\alpha^i}=ig\left(v_i+\frac{\alpha_i}{\sqrt{2}}\right)\left(\overline{Q'}\gamma^iL-\overline{L}\gamma^iQ'\right).
\end{split}
\label{eq64}
\end{equation}
Finally, using Eq. (\ref{eq60}) as a constraint equation within Eq. (\ref{eq64}), we obtain the final equation of motion for the inflaton field:
\begin{equation}
\ddot{\theta}\left(1+\frac{\alpha^i}{f}\right)+2\dot{\theta}\left(\frac{\dot{\alpha^i}}{f}\right)+m_{\theta}^2\theta=\frac{ig}{f^2}v_i\left(\overline{Q'}\gamma^iL-\overline{L}\gamma^iQ'\right).
\label{eq65}
\end{equation}
Passing from Eq. (\ref{eq64}) to Eq. (\ref{eq65}), we cancel the redundancy due to the fact that our theory formally contains an extra degree of freedom represented by $\beta^i$. The corresponding equation, Eq. (\ref{eq60}), is thus used to simplify the actual equation of motion for $\theta$ and to prevent an overcounting of degrees of freedom. 

Once simplified, we expect Eq. (\ref{eq65}) not to be independent of Eq. (\ref{eq60}), as they describe a single degree of freedom, $\theta$. However, these two seem incompatible, due to the presence of the mass term in the latter. We remarkably point out that this extra term is due to the explicit symmetry breaking under which $\theta$ undergoes to play the role of the inflaton. Conversely, for $\beta^i$ we do not introduce any additional potential term, as it is not a physical degree of freedom and we actually put it equal to zero to avoid any possible redundancy.

Eqs. (\ref{eq54}) and (\ref{eq65}) respectively for $\alpha^i$ and $\theta$ constitute a system of second-order differential equations, both of them containing source-like terms depending on fermions. Since the latter are quantized, the corresponding terms appearing within the equations of motion of the classical fields $\alpha^i$ and $\theta$ are understood as the respective VEVs. In particular, since we are not accounting for the expansion of the universe for the moment, following the same approach adopted in SSB, the vacuum that we consider is the usual Minkowski vacuum.

In order to compute the VEVs it is necessary to solve the equations of motion for the fermions $Q'$ and $L$. In particular, it is convenient to solve the equations in terms of the fields $Q$ and $L$, namely Eqs. (\ref{eq47}) and (\ref{eq48}). To do so, we adopt perturbation method over the small coupling constant $g\ll1$. As in SSB, zero-order solutions give vanishing VEVs, thus we need to compute first-order solutions. Labeling with $Q_{\text{free}}$ and $L_{\text{free}}$ the zero-order solutions of Eqs. (\ref{eq47}) and (\ref{eq48}) respectively, representing free fermions, the first-order solutions, $Q_1$ and $L_1$, can be written as follows:
\begin{subequations}
\begin{equation}
Q_1(x)=Q_{\text{free}}(x)-g\int d^4y G_Q(x,y)\slashed{w}(y)L_{\text{free}}(y)e^{i\theta(y)},
\label{eq66}
\end{equation}
\vspace{-1.5em}
\begin{equation}
L_1(x)=L_{\text{free}}(x)-g\int d^4yG_L(x,y)\slashed{w^*}(y)Q_{\text{free}}(y)e^{-i\theta(y)},
\label{eq67}
\end{equation}
\end{subequations}
where $w_{\mu}=v_{\mu}+a_{\mu}$ and $G$ is the usual fermion propagator. As we are working at first order in $g$, corresponding to keeping terms up to second order in the equations of motion, we set $w_{\mu}=v_{\mu}$ and neglect the perturbation $a_{\mu}$, which would otherwise give rise to third-order contributions. In so doing, following the same procedure as in SSB, the VEV contained in the source term of Eq. (\ref{eq65}) becomes\footnote{The full calculations can be found in Appendix \ref{app2}.}
\begin{equation}
\braket{\overline{Q'}\gamma^iL-\overline{L}\gamma^iQ'}=-\frac{igf\Omega}{3\pi\sqrt{2}}\left[\frac{\dot{\theta}}{2}-\Omega\theta\lim_{\omega\to+\infty}\ln\left(\frac{2\omega}{\Omega}\right)\right],
\label{eq68}
\end{equation}
where $\Omega$ is understood as the renormalized mass of the inflaton and $\omega$ has the dimension of an energy. Consequently, Eq. (\ref{eq65}) becomes
\begin{equation}
\ddot{\theta}\left(1+\frac{\alpha^i}{f}\right)+\dot{\theta}\left(\Gamma+\frac{2\dot{\alpha^i}}{f}\right)+\Omega^2\theta=0,
\label{eq69}
\end{equation}
where
\begin{equation}
\Gamma=\frac{g^2\Omega}{12\pi}
\label{eq70}
\end{equation}
and the renormalized mass is defined in terms of the bare mass $m_{\theta}$ as
\begin{equation}
m_{\theta}^2=\Omega^2\left[1+\frac{g^2}{6\pi}\lim_{\omega\to+\infty}\ln\left(\frac{2\omega}{\Omega}\right)\right].
\label{eq71}
\end{equation}
Comparing Eqs. (\ref{eq70}) and (\ref{eq71}) with the expressions for $\Gamma$ and $\Omega$ found within SSB, we observe that our expressions for them are exactly $2/3$ of those obtained in SSB.

In principle, in order to find the complete equation of motion for $\alpha^i$, we would need to compute $\braket{\overline{Q'}\gamma^iL+\overline{L}\gamma^iQ'}$, proceeding similarly to what already done for the other VEV. However, in Appendix \ref{app2} we show that the whole VEV $\braket{\overline{Q'}\gamma^iL+\overline{L}\gamma^iQ'}$ results of the order $\theta^2$. The corresponding term appearing in r.h.s. of Eq. (\ref{eq54}) would be thus a pure source-like term of the order of $g^2f\theta^2$, that is definitely negligible with respect to the other one, that is of the order of $\theta^2 f$. Therefore, we neglect that term and finally obtain
\begin{equation}
\begin{cases}
\ddot{\alpha^{i}}+\left(f^2-\dot{\theta}^2\right)\alpha^{i}=f\dot{\theta}^2, \\ \ddot{\theta}\left(1+\frac{\alpha^i}{f}\right)+\dot{\theta}\left(\Gamma+\frac{2\dot{\alpha^i}}{f}\right)+\Omega^2\theta=0.
\end{cases}
\label{eq72}
\end{equation}
Since we take $\alpha^i$ essentially representing a small perturbation with respect to the VEV acquired by the vector field, we have $\alpha^i\ll v^i\sim f$. Moreover, $\dot{\theta}^2\ll f^2$, since at leading order we expect $\dot{\theta}\sim \Omega \theta\ll\Omega$. Therefore, Eq. (\ref{eq72}) is simplified to
\begin{equation}
\begin{cases}
\ddot{\alpha^{i}}+f^2\alpha^{i}=f\dot{\theta}^2, \\ \ddot{\theta}+\dot{\theta}\left(\Gamma+\frac{2\dot{\alpha^i}}{f}\right)+\Omega^2\theta=0.
\end{cases}
\label{eq73}
\end{equation}
We chose $\alpha^i(0)=\dot{\alpha}^i(0)=0$ as initial conditions for $\alpha^i$, assuming the vector field initially to be fixed at the minimum of the potential and receiving the impulse to oscillate by the inflaton. In so doing, the equation of motion of the inflaton becomes
\begin{equation}
\ddot{\theta}+\Gamma\dot{\theta}+\Omega^2\theta+2\dot{\theta}\int_0^{t}\cos\left[f\left(t-s\right)\right]\dot{\theta}^2(s)ds=0,
\label{eq74}
\end{equation}
representing an integro-differential non-linear oscillator equation with memory kernel. In particular, the last term is a non-local and non-linear damping term. Clearly, non-linearity is due to the presence of $\dot{\theta}^2$ within the integral, whereas non-locality in time is due to the presence of the integral itself over $s\in[0,t]$. This means that the system has memory, i.e., the present dynamics depends on the full history of $\dot{\theta}$. Finally, the kernel is $\cos\left[f\left(t-s\right)\right]$, which oscillates at frequency $f$. 

This kernel modulates how strongly past values of $\dot{\theta}^2$ influence the current damping. Since the non-linear term is of the order of $\theta^3$, we are allowed to consider it as  a perturbation. In so doing, we can apply perturbation method, writing $\theta$ as $
\theta(t)=\theta_0(t)+\theta_1(t)$,
where $\theta_0$ and $\theta_1$ satisfy respectively
\begin{subequations}
\begin{equation}
\ddot{\theta}_0+\Gamma\dot{\theta}_0+\Omega^2\theta_0=0,
\label{eq76}
\end{equation}
\vspace{-2em}
\begin{equation}
\ddot{\theta}_1+\Gamma\dot{\theta}_1+\Omega^2\theta_1=-2\dot{\theta}_0\int_0^t\cos\left[f\left(t-s\right)\right]\dot{\theta}_0^2(s)ds.
\label{eq77}
\end{equation}
\end{subequations}
In particular, we can see that Eq. (\ref{eq76}) formally recovers Eq. (\ref{eq9}), i.e., the SSB, and thus admits the solution in Eq. (\ref{eq12}), where $\Gamma$ and $\Omega$ are defined according to Eqs. (\ref{eq70}) and (\ref{eq71}). When $\theta_0(t)$ is inserted within Eq. (\ref{eq77}), in the limit $f\gg\Omega\gg\Gamma$ we find the first-order solution
\begin{equation}
\theta_1(t)=-\frac{\theta_I^3\Omega^3}{2\Gamma f^2}e^{-\frac{\Gamma t}{2}}\left(1-e^{-\Gamma t}\right)\sin(\Omega t),
\label{eq78}
\end{equation}
imposing the initial conditions $\theta_1(0)=\dot{\theta}_1(0)=0$, since the true initial conditions for $\theta$ are already respected by $\theta_0(t)$. 

The full solution for $\theta(t)$ is thus
\begin{equation}
\theta(t)=\theta_Ie^{-\frac{\Gamma t}{2}}\left[\cos(\Omega t)-\frac{6\pi\theta_I^2\Omega^2}{g^2 f^2}\left(1-e^{-\Gamma t}\right)\sin(\Omega t)\right].
\label{eq79}
\end{equation}
The magnitude of our correction to the background solution due to the interaction between $\theta$ and $\alpha^i$ can be quantified by the dimensionless parameter $
\xi=\frac{6\pi\theta_I^2\Omega^2}{g^2f^2}$. Here $\theta_I = \theta (0)$ is the value of the field at the reheating temperature. This is defined in correspondence with $H=\Gamma$, so that, analogously to SSB
\begin{equation}
\theta_I=\sqrt{\frac{3}{4\pi}}\frac{\Gamma M_{\text{Pl}}}{f\Omega}\sim g^2 \ .
\label{eq81}
\end{equation}
Then, we obtain $\xi\sim 10^{-12}g^2\ll1$, coherently with perturbative regime.

\section{The produced baryon asymmetry}\label{sec:level5}

Once found the time-dependent expression for the inflaton, we compute the baryon asymmetry through calculating the number density of baryons and antibaryons produced. 

In particular, the average number density of fermion-antifermion pairs produced by the decays of a classical scalar field is computed via Eq. (\ref{eq13}), where $\mathcal{A}$ is the single pair production amplitude. In particular, neglecting the small perturbations $\alpha^{\mu}$ and $\beta^{\mu}$ with respect to the VEV $v^{\mu}$ and considering the constraints in Eqs. (\ref{eq52}) and (\ref{eq58}), the interaction term between inflaton and fermions is given by
\begin{equation}
\mathcal{L}_I=gv_i\left\{e^{i\theta}\overline{Q}\gamma^iL+e^{-i\theta}\overline{L}\gamma^iQ\right\}.
\label{eq82}
\end{equation}
This interaction term allows the production of $Q-\overline{L}$ and $\overline{Q}-L$ pairs by starting from a classical background vectorial VEV, the first associated to baryons and the second to antibaryons. Since the interaction contains complex phases, it violates the CP symmetry, thus we expect baryons and antibaryons to be produced with different decay amplitudes, finally yielding a net difference between matter and antimatter.

We use Eq. (\ref{eq13}) once considering the $Q-\overline{L}$ and $\overline{Q}-L$ pair amplitudes and, then, we find respectively the number densities of baryons and antibaryons produced. 

To this end, up to  second order Taylor expansion of $e^{\pm i\theta}$ for $\theta\ll 1$ as performed in Appendix \ref{app3},  we find the following relation for the baryon shift
\begin{equation}
n_B=\frac{g^2f^2}{3\pi^2}\int_0^{+\infty}d\omega\omega^2\mathcal{F}(\omega),
\label{eq84}
\end{equation}
where $
\mathcal{F}(\omega)=2\text{Re}\left\{\frac{\tilde{\theta}(2\omega)\tilde{\theta}^{2*}(2\omega)}{i}\right\}$, and $
\tilde{\theta}(2\omega)=\int_{-\infty}^{+\infty}\theta(t)e^{2i\omega t}dt,\quad 
\tilde{\theta}^{2*}(2\omega)=\int_{-\infty}^{+\infty}\theta^2(t)e^{-2i\omega t}dt$. 
Since our time-dependent expression is $\theta_0$ plus a small correction of the order of $\xi$, we have  $\tilde{\theta}(2\omega) = \tilde{\theta}_0 (2\omega) +\xi \tilde{\theta}_1 (2\omega)$, and something similar holds for $\tilde{\theta}^{2*}(2\omega)$. Accordingly, expanding $\mathcal{F(\omega)}$ up to first order in $\xi$, we find 
\begin{equation}
n_B=\frac{\Omega g^2f^2\theta_I^3}{24\pi}\left(1+\frac{\xi g^2}{12\pi}\right)=\frac{\Omega g^2f^2\theta_I^3}{24\pi}\left(1+\frac{\theta_I^2\Omega^2}{2f^2}\right) = \frac{2}{3}n^{(0)}_B \left(1+\frac{\theta_I^2\Omega^2}{2f^2}\right),
\label{eq87}
\end{equation}
where $n_B^{(0)}$ is given by Eq. (\ref{eq14}). Since $\theta_I\ll1$ and $\Omega/f\sim 10^{-6}$, the correction to the leading order results negligible, reason for which from now on we consider essentially the leading order contribution. This means that, up to this point, the VFSB predicts a baryon asymmetry exactly $2/3$ of that obtained in the SSB framework.

Eq. (\ref{eq87}) is obtained ignoring the impact of lepton-baryon oscillations. We now proceed to estimate their impact on the final baryon asymmetry.

\section{Lorentz-violating flavor oscillations and baryon asymmetry}\label{sec:level6}

If we look at the Lagrangian in Eq. (\ref{eq41}) and focus on the fermionic sector, we notice that the fields $Q$ and $L$ do not evolve freely in time, since they are not eigenstates of the Hamiltonian. In order to diagonalize the Hamiltonian, it is helpful to rewrite the fermionic sector of the Lagrangian in Eq. (\ref{eq41}). In particular, we can neglect the perturbations $\alpha^{\mu}$ and $\beta^{\mu}$ with respect to the VEV $v^{\mu}$ and consider $\theta\ll 1$, so to be allowed to use the zero-order expansion $e^{i\theta}\simeq 1$. Moreover, we also account for the constraints in Eqs. (\ref{eq52}) and (\ref{eq58}), finally obtaining
\begin{equation}
\begin{split}
\mathcal{L}_{\text{fermions}}&=\overline{Q}\left(i\slashed{\partial}-m_Q\right)Q+\overline{L}\left(i\slashed{\partial}-m_L\right)L+gv_{i}\left(\overline{Q}\gamma^{i}L+\overline{L}\gamma^{i}Q\right).
\end{split}
\label{eq88}
\end{equation}

The $Q \rightarrow L$ transition probabilities can be computed, starting from Eq. \eqref{eq88}, following either of the two approaches discussed in Appendix \ref{app4}. We will stick to the massless limit $m_L, m_Q \rightarrow 0$ for simplicity. For a fixed momentum $\vec{p}$ the result is (Eq. \eqref{eq111} and Eq. \eqref{eq126})
\begin{equation}\label{OscEq}
    P_{Q_{\vec{p}} \ \rightarrow L_{\vec{p}}}(t)= \sin^2 \left[t\left(\lambda_{++}(\vec{p})-\lambda_{+-}(\vec{p})\right)\right/2 ] \ ,
\end{equation}
where the eigenvalues are given by $\lambda_{+ \pm} (\vec{p}) = |\vec{p}\pm g\vec{v}|$. Note that the same result holds for the inverse transition $L \rightarrow Q$ and the antiparticle transitions ($\bar{Q} \rightarrow \bar{L}$ and $\bar{L} \rightarrow \bar{Q}$, see also the Appendix \ref{app4}).

\subsection{The effects of the flavor oscillations on the baryon asymmetry}\label{subsec:level6-3}

Once obtained the result in Eq. (\ref{OscEq}), we can compute the number densities of baryons and antibaryons produced per momentum at the generic time $t$. We call them $n_b$ and $n_{\overline{b}}$ respectively. Since the probabilities of oscillation are all equal, we find
\begin{equation}
n_b-n_{\overline{b}}=\left[n(Q,\overline{L})-n(\overline{Q},L)\right]\left[1-2P_{Q\to L}(t)\right],
\label{eq127}
\end{equation}
where $\left[n(Q,\overline{L})-n(\overline{Q},L)\right]$ is the baryon asymmetry per momentum when the flavor oscillations are neglected, namely when $P=0$. In this case, therefore, we recover Eq. (\ref{eq84}) for the baryon asymmetry. Differently, when $P\neq 0$, the integrand in Eq. (\ref{eq84}) should be multiplied by the time-average of the quantity $\left[1-2P_{Q\to L}(t)\right]$ on a time $\tau=2/\Gamma$, that is the characteristic time during which baryogenesis happens. Actually, in this case, the integral should be performed on the whole $3D$ momenta space, since the probability of oscillation is not strictly speaking a function of the modulus of the momentum but also depends explicitly to its along-VEV component. In formulas, we have
\begin{equation}
n_B=\frac{g^2f^2}{12\pi^3}\int d^3p\mathcal{F}\left(\omega\right)\left[1-2\braket{P_{Q\to L}(t)}_{\tau}\right],
\label{eq128}
\end{equation}
where $\omega=|\vec{p}|$ and $\braket{P_{Q\to L}(t)}_{\tau}$ is the time-average of Eq. (\ref{eq126}) on the time $\tau$, that is in general a function of $\omega$ and the along-VEV component of the momentum. We can arbitrarily chose to write this component as $\omega\cos(\theta)$, where $\theta$ is the polar angle, and perform thus trivially the integral over the azimuthal angle $\phi$. Moreover, for simplicity of notation we define\footnote{The dimensionless quantities $a$, $b$ and $c$ defined in this context should not be confused with the homonymous parameters defined in Appendix \ref{app4} to simplify the form of the coefficients $c_j$.} the two dimensionless parameters $a$ and $b$ as
\begin{equation}
a=\frac{\Omega}{gv^i}=10^{-6}\sqrt{2}g^{-1},\quad
b=\frac{\Gamma}{gv^i}=\frac{\sqrt{2}}{12\pi}10^{-6}g,
\label{eq129}
\end{equation}
and the dimensionless integration variable $c=\omega/(gv^i)$. In so doing, the baryon asymmetry becomes
\begin{equation}
n_B=\frac{g^2f^2\left(gv^i\right)^3}{3\pi^2}\int_0^{+\infty}c^2\mathcal{F}(c)\mathcal{P}(c)dc,
\label{eq131}
\end{equation}
where the function $\mathcal{P}(c)$ is defined as
\begin{equation}
\mathcal{P}(c)=\frac{b}{4}\int_{-1}^1dx\frac{\sin\left[\frac{2}{b}\left(\sqrt{c^2+1-2cx}-\sqrt{c^2+1+2cx}\right)\right]}{\sqrt{c^2+1-2cx}-\sqrt{c^2+1+2cx}},
\label{eq132}
\end{equation}
and describes the effect of the flavor oscillations. In particular, we want to point out that for $\mathcal{P}(c)=1$ we recover exactly Eq. (\ref{eq84}) written in terms of the dimensionless integration variable $c$. Moreover, from now on we consider the function $\mathcal{F}(c)$ at leading order, thus neglecting all the corrections proportional to $\xi$. Indeed, Eq. (\ref{eq87}) shows that this contribution results in a negligible correction to the baryon asymmetry. Therefore, in the limit $\mathcal{P}=1$, Eq. (\ref{eq87}) should be recovered at leading order.

\begin{figure}
\centering
\includegraphics[width=0.85\textwidth]{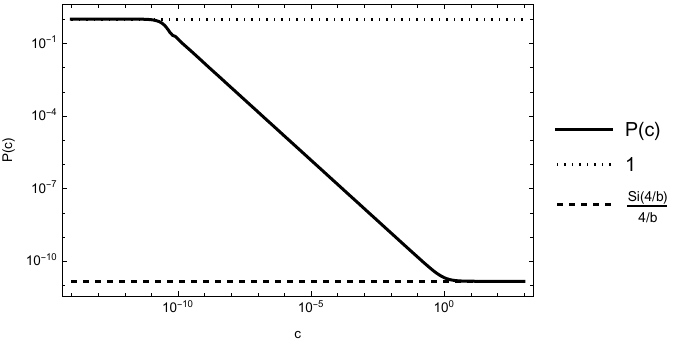}
\captionsetup{justification=raggedright,singlelinecheck=false}
\caption{Continuous line: function $\mathcal{P}(c)$ numerically computed with respect to the normalized momentum $c$, obtained for $g=10^{-3}$. Dotted line: limit value of the $\mathcal{P}(c)$ for low momenta, $\lim_{c\to 0}\mathcal{P}(c)=1$. Dashed line: limit value of the $\mathcal{P}(c)$ for high momenta, $\lim_{c\to+\infty}\mathcal{P}(c)=[\text{Si}(4/b)]/[4/b]$.}
\label{fig1}
\end{figure}

In order understand the effects of the flavor oscillations for a given momentum we show in Fig. (\ref{fig1}) the shape of the function $\mathcal{P}(c)$, from Eq. \eqref{eq132}. We see that for low momenta the flavor oscillations almost do not affect the baryon asymmetry, whereas they suppress the latter for high momenta.

It is convenient to introduce the dimensionless function
\begin{equation}
\mathcal{I}(g)=\frac{8\left(gv^i\right)^2}{\theta_I^3b}\int_0^{+\infty}c^2\mathcal{F}(c)\mathcal{P}(c)dc,
\label{eq133}
\end{equation}
such that $n_B=\frac{g^2f^2\Gamma\theta_I^3}{24\pi^2}\mathcal{I}(g)$.
When $\mathcal{P}(c)=1$, the function $\mathcal{I}(g)$ can be analytically computed and its explicit expression, labeled with $\mathcal{I}_0(g)$, results 
$\mathcal{I}_0(g)=\frac{12\pi^2}{g^2}$.

\begin{figure}
\centering
\includegraphics[width=0.85\textwidth]{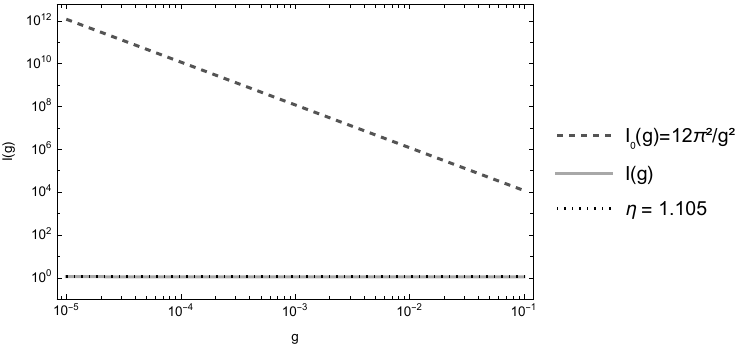}
\captionsetup{justification=raggedright,singlelinecheck=false}
\caption{Continuous line: function $\mathcal{I}(g)$ numerically computed accounting for the $Q-L$ oscillations. Dashed line: theoretical curve $\mathcal{I}_0(g)=12\pi^2/g^2$, obtained for $\mathcal{P}(c)=1$. Dotted line: best fit of $\mathcal{I}(g)$ through the model of the constant function, that provides the value $\eta=1.105$.}
\label{fig2}
\end{figure}

Obtaining an exact analytical expression for Eq. \eqref{eq133} turns out to be challenging. Performing the integration numerically, we find that $\mathcal{I}(g)$ is essentially a constant, to a good approximation equal to $\mathcal{I}(g) = \eta  \simeq 1.105$. The numerical results are shown in Fig. (\ref{fig2}).

The baryon asymmetry, when flavor oscillations are involved, can be thus written as
\begin{equation}
n_B=\left(\frac{\Omega g^2f^2\theta_I^3}{24\pi}\right)\frac{g^2\eta}{12\pi^2},
\label{eq136}
\end{equation}
representing the leading order term of Eq. (\ref{eq87}) multiplied by a factor $
r(g)=\frac{g^2\eta}{12\pi^2}$, that shows up the ratio between baryon asymmetry with and without considering the flavor oscillations. 

We can see that, for small $g$, the reduction to the baryon asymmetry caused by considering the $Q-L$ oscillations becomes noticeably large. Nevertheless, even working in the limit of exactly massless fermions, we obtain a non-null baryon asymmetry when considering the flavor oscillations. 

This result marks a clear distinction with respect to the scalar model of spontaneous baryogenesis \cite{Dolgov:1994zq,Dolgov:1996qq}, where the oscillations among massless fermions would lead to a complete disappearance of the baryon asymmetry.

\subsection{The baryon-to-entropy ratio}\label{subsec6-4}

\begin{figure}
\centering
\includegraphics[width=0.85\textwidth]{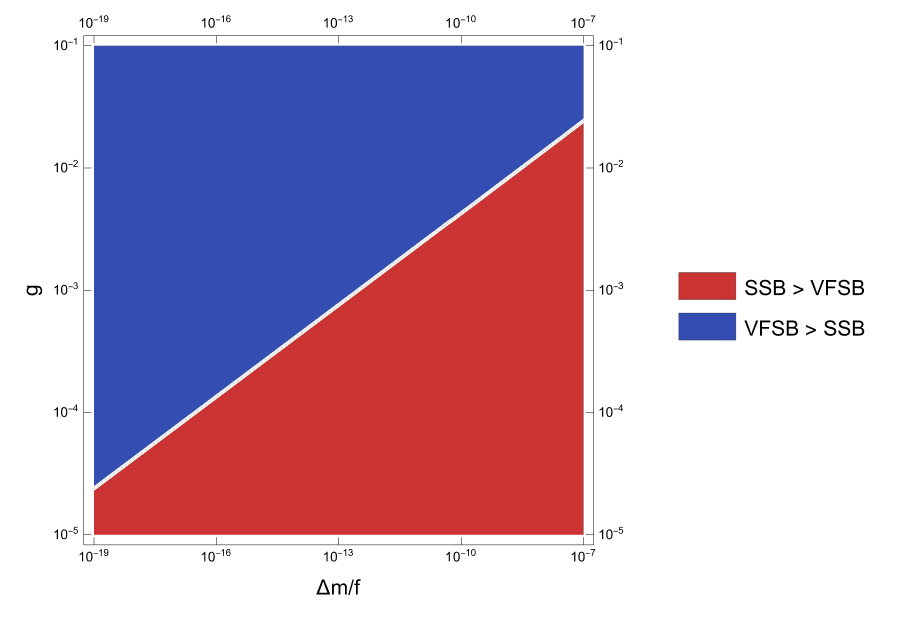}
\captionsetup{justification=raggedright,singlelinecheck=false}
\caption{Comparison between the baryon-to-entropy ratio obtained within SSB, by Eq. (\ref{eq141}), and VFSB, by Eq. (\ref{eq140}). In particular, the first one is obtained in the limit $\Delta m\ll gf$, considering light fermions but not totally massless, whereas the second in the purely massless fermions limit. The dark (light) colored region is the region of parameter space where VFSB predicts a baryon asymmetry higher (lower) than SSB.}
\label{fig3}
\end{figure}

The baryon asymmetry from Eq. \eqref{eq136} can be written as 
\begin{equation}
n_B=\frac{2}{3}n_B^{(0)}r(g),
\label{eq138}
\end{equation}
where $n^{(0)}_B$, introduced in Eq. \eqref{eq14}, is the baryon asymmetry, in absence of oscillations, as resulting from the SSB model. 
In order to incorporate the effect of the universe expansion we may proceed analogously to SSB \cite{Dolgov:1994zq, Dolgov:1996qq}. This amounts to a redefinition of $\theta_I$ as the value of $\theta(t)$ when $H=\Omega$, instead of its value at reheating (Eq. \eqref{eq81}), effectively enhancing the baryon asymmetry by a factor $(\Omega/\Gamma)$. The resulting baryon to entropy ratio, including the effect of the expansion of the universe, is

\begin{equation}
\frac{n_B}{s}= \frac{2g(r)}{3}\left(\frac{n_B^{(0)}}{s}\right)\simeq 2\times 10^{-3}\frac{g^3}{g_*^{\frac{1}{4}}}\biggl(\frac{M_{\text{Pl}}}{f}\biggl)^{\frac{3}{2}}\sqrt{\frac{f}{\Omega}}\times r(g).
\label{eq139}
\end{equation}
Since we take roughly $f\sim 10^{6}\Omega$, $f\sim M_{\text{Pl}}$ and $g_*\sim 100$, Eq. (\ref{eq139}) becomes
\begin{equation}
\left(\frac{n_B}{s}\right)_{\text{VFSB}}\simeq 5.9\times 10^{-3}g^5.
\label{eq140}
\end{equation}
Comparing Eq. (\ref{eq140}) with the baryon-to-entropy ratio obtained within SSB, namely
\begin{equation}
\left(\frac{n_B}{s}\right)_{\text{SSB}}\simeq g^3\left(\frac{1-\epsilon^2}{1+\epsilon^2}\right)^2\sim \frac{g}{2}\left(\frac{\Delta m}{f}\right)^2,
\label{eq141}
\end{equation}
we obtain the result displayed in Fig. (\ref{fig3}). 

By looking at Fig. (\ref{fig3}), we find a remarkable improvement in using the VFSB paradigm and in particular,

\begin{itemize}
    \item[-] the VFSB performs quite well  if compared with the standard SSB, for lighter fermions, as expected since in the purely massless case SSB returns zero;
    \item[-]  on the other hand, SSB works better than VFSB for smaller $g$. Actually, when we compare Eq. (\ref{eq140}) with the experimental value $n_B/s=\left(8.59\pm0.10\right)\times10^{-11}$, we find that VFSB is able to reproduce the observations for $g\in\left[2,3\right]\times10^{-2}$. Within the SSB's framework, such values of $g$ would be too large to be acceptable. There, indeed, $g$ is proportional to the mass of the produced fermions, that is roughly $\sim gf$, thus it should be as small as possible in order to facilitate the decays of the inflaton into fermion-antifermion pairs. Conversely, in the framework of VFSB, $gf$, or analogously $gv^i$, does not play the role of a mass, but rather it is merely a shift of one of the three components of the spatial momentum.
\end{itemize}

According to the above considerations, we physically have no reason to fine-tune $g$ to turn into very small values, within the context of VFSB, allowing also values of the order of $10^{-2}$, that is also compatible with the perturbative regime. 

Consequently, we can reproduce the experimental value of the baryon asymmetry and state that the VFSB appears favored in describing such an asymmetry, suggesting that at so large scales the LIV processes may be more relevant than what expected. 

\section{Restoring the Lorentz symmetry at late times}\label{sec:level7}

Challengingly, the Lorentz invariance might be restored immediately after the overall baryogenesis process occurs.

Rephrasing it differently, ensuring a LIV phase, it turns out to be finite, in order to fulfill cosmic observations at late and early times \cite{LIGOScientific:2017zic,Ellis:2018ogq,Maccione:2009ju}. 

To do so, we may invoke a similar mechanism prompted in those models named Bumblebee paradigms, see e.g. \cite{SPLV1,SPLV2, SPLV3}, where analogously with our prescription a proper $v^\mu \rightarrow 0$ limit appears somehow tricky. 

Nevertheless, it turns plausible to recover massless photon modes as the Nambu-Goldstone bosons in the spontaneously broken theory \cite{SPLV1, SPLV3} and to roughly mimic a Lorentz invariant theory. 

The symmetry restoration \emph{tout court}, that is, a phase transition from $v^\mu \neq 0$ to $v^\mu = 0$, may occur through an \emph{inverse symmetry breaking} mechanism \cite{ WeinbergISB, ISB0, ISB1, ISB2, ISBX, ISBB1, ISBB2, ISBB3} as basic demand, here invoked to make our paradigm work.

One coarse-grained possibility is represented by the appearance of parametric resonant field modes, dubbed $\alpha^i$.
More precisely, to understand how such a phenomenon may take place, let us consider Eq. (\ref{eq73}) for $\alpha^i(t)$, using the leading-order solution for the inflaton, $\theta$, say $\theta_0(t)$, given by Eq. (\ref{eq12}). Hence, for $\alpha^i(t)$ we have

\begin{equation}
\ddot{\alpha^i}+f^2\alpha^i=f\theta_I^2\Omega^2e^{-\Gamma t}\sin^2\left(\Omega t\right),
\label{eq142}
\end{equation}

where we have already considered, when computing the inflatonic solution, the initial conditions $\alpha^i(0)=\dot{\alpha}^i(0)=0$. Solving Eq. (\ref{eq142}) in the limit $f\gg\Omega\gg\Gamma$, we find\footnote{The solution in Eq. (\ref{eq143}) does not effectively respect our imposed initial conditions. More practically, this occurrence bids  $f\gg\Omega\gg\Gamma$, yielding the initial values of $\alpha^i$ and $\dot{\alpha}^i$ to be much smaller than unity, albeit not exactly zero. However, because of the linearity of the equation, we do not expect a significant change of the result when slightly deviating from the exact initial conditions.}

\begin{equation}
\begin{split}
\frac{\alpha^i(t)}{f}&=\frac{\theta_I^2\Omega^2}{f^2}\bigg{\{}e^{-\Gamma t}\sin^2\left(\Omega t\right)+\frac{4\Omega^2}{f^2}\cos\left(ft\right)\bigg{\}}.
\end{split}
\label{eq143}
\end{equation}

Estimating the order of magnitude of Eq. (\ref{eq143}), it is straightforward to see that $\alpha^i(t)\ll f\quad \forall t$. This means that the small perturbations of the background vector field about the VEV remain roughly small, at least in the limit of $f\gg\Omega$. 

Otherwise, indeed, we could not go from Eq. (\ref{eq72}) to Eq. (\ref{eq73}). In this case, in contrast, the equation of motion for $\alpha^i$ would become a forced, damped Mathieu-like equation, possibly leading to the phenomenon of parametric resonance when $f$ and $\Omega$ are comparable. 

From a theoretical perspective, this mechanism could induce exponentially growing oscillations, that can overcome the energy difference between minima and the symmetric state potentially driving the field out of the potential well \cite{ISBPR}. The system thus settles into a symmetric configuration once the oscillations redistribute energy evenly across the potential. 

Another viable scheme  for the strict restoration of Lorentz symmetry lies in the nature of the symmetry breaking potential $V(A^2 )=\pm \mu^2 A^2 + \frac{\lambda}{2} A^4$, where $\mu^2>0 $ for a spacelike VEV. 

In general, the mass parameter $\mu^2$ depends on the background temperature $\mu^2 = \mu^2 (T)$, as certified by the fact that external spectator fields may play significant roles throughout the overall dynamics. 

The study of the temperature dependence of the symmetry breaking potential can be traced back to the seminal paper by Weinberg \cite{WeinbergISB}, whereas theories with multiple scalar/vector fields \cite{WeinbergISB, ISB0, ISB1, ISB2} display a phase transition at a certain critical temperature $T_c$, below which the symmetry is naturally restored. They provide a physical realization of \emph{inverse symmetry breaking} in that the broken symmetry phase appears at high, rather than low, temperature, i.e., associated with more disordered than ordered systems \cite{WeinbergISB, ISB0, ISB1, ISB2, ISBX, ISBPR, ISBB1, ISBB2, ISBB3, PhysRevD.54.7153}
. A temperature dependence for the mass parameter also appears in fermion models that dynamically generate a bumblebee model \cite{ISBB1, ISBB2, ISBB3}, allowing for a phase transition. The exact behavior of $\mu^2(T)$ obviously depends on the details of the underlying fundamental theory. 
On the other hand, it is legitimate to expect that Lorentz violation, if present, characterizes only the very early stages of the universe. The mass parameter should therefore be positive up to a certain critical temperature $T_c$, corresponding to some early epoch of the universe, and drop down to zero, eventually flipping its sign, for cosmic temperature $T < T_c$, i.e.
\begin{equation}
\mu^2=\mu^2(T)\begin{cases}
>0 & T>T_c \\ \leq 0 & T \leq  T_c
\end{cases} \ .
\label{eq144}
\end{equation}

In practice $T_c$ should lie close to the reheating temperature, in order for the LIV to affect baryogenesis. In a very simple model, in which the reheating temperature is estimated assuming an instantaneous conversion of the matter energy density into radiation, the critical temperature $T_c$ should coincide with the reheating temperature $T_{\text{RH}}$. This is in fact the approximation followed in both SSB and VFSB that finds direct justification into our baryogenesis landscape. 

Notice, importantly, that, as it happens in the various inverse symmetry breaking scenarios \cite{WeinbergISB,ISB0,ISB1,ISB2, ISBB1, ISBB2, ISBB3}, a temperature dependence of the form of Eq. \eqref{eq144} requires some extra field coupling to the Bumblebee vector. The Lagrangian of Eq. \eqref{eq20} might necessarily be understood as an \emph{effective Lagrangian} arising from some, more involved, fundamental theory.
In assuming Eq. \eqref{eq144}, we are saying that, when baryogenesis comes to an end, the potential shape is modified, eliminating the non-trivial minimum configuration in favor of a trivial, Lorentz-invariant vacuum state. Alternatively speaking, the previously acquired VEV would be thus removed and consequently the Lorentz invariance restored, as direct consequence of the aforementioned mechanism.

\section{Final remarks}\label{sec:level8}

In this work, we explored the consequences of a plausible baryogenesis mechanism \emph{induced} by the presence of a LIV term in the Lagrangian. 

We motivated our choice remarking that, across the very early universe, energies, temperatures and pressures are expected to be compatible with the breakdown of special relativity prescription, implying possible fundamental/spacetime symmetry breaking, such as the evidence for a LIV and for a violation of the $CPT$ theorem that may eventually be spoiled by quantum gravity or other speculative high energy effects, see e.g. \cite{Hor1, ELV1, ELV2, ELV3, ELV4, Hor2}. 

Scenarios in which the early universe dynamics leads to a breakdown of Lorentz symmetry have been explored since the eighties, in connection with inflation \cite{Gasp85}, and spontaneously arising from string theory \cite{Kost89}. Later developments have led to the systematization of the (Lorentz-violating) extension of the Standard Model of particle physics \cite{Kost98,Kost04,Kost11} and to the appearance of Bumblebee models \cite{ BB2, BB3, BB4, BB5, BB6, ISBB1, ISBB2, ISBB3}, which provide a concrete and prototypical model of \emph{spontaneous} LIV.

Motivated by such schemes, we employed a primordial baryogenesis scenario, based on modifying the spontaneous baryogenesis, taking  place at primordial times. In this framework, we wondered the consequences of breaking Lorentz invariance by means of a vector field and the impact of such a paradigm on baryon asymmetry generation.

We remarked that a thorough analysis of baryogenesis mechanisms cannot ignore the impact of a Lorentz-violating background, as exemplified, for instance, by Bumblebee models.

Accordingly, while the original spontaneous baryogenesis approach involves a scalar field, we have here considered a complex vector field carrying baryon number, where this field simultaneously plays the role of a Bumblebee vector, selecting a preferred spacetime direction after spontaneous symmetry breaking, and breaks $U(1)_B$ by coupling to baryons and leptons.  

The (pseudo) Nambu-Goldstone boson $\theta$, corresponding to the global phase of the Bumblebee vector, decays asymmetrically, favoring the creation of baryons over antibaryons. The same mechanism that leads to LIV has been here explored to generate the aforementioned baryon asymmetry.

Consequently, we demonstrated that our treatment represents a natural extension of spontaneous baryogenesis by the inclusion of Lorentz-violating effects, with deep influence on the baryon production.

We remarked throughout the text that the Bumblebee models come with an inherent ambiguity within the symmetry breaking potential, absent in the scalar case. The reason is simple, i.e., the scalar norm $|\phi|^2$ is always positive, whereas the vector (pseudo-)norm $A^2 =A^\mu A_\mu$ admits both signs. As a consequence, the sign of the mass term $\pm \mu^2 A^2$ in the symmetry breaking potential depends on whether the VEV is timelike or spacelike. 

In principle, both the options are possible. Nevertheless, in an isotropic and homogeneous cosmological context, where fields are assumed to depend only on cosmic time $t$, we found that only a spacelike VEV turns out to be viable to have a nontrivial dynamics of the Bumblebee phase $\theta$.

As its decay is what determines the baryon asymmetry, we have found that only a spacelike bumblebee vector can allow for such a scheme of spontaneous baryogenesis.

In addition, the pseudo-Nambu-Goldstone boson $\theta$ is deemed to play the role of the inflaton, which would be impossible without an appropriate kinetic term. 

Within our framework, we have then analyzed the CP-violating coupling of the Bumblebee field to leptons and baryons. Such vector interaction, reminiscent of the Yukawa CP-violating Lagrangian that characterizes the scalar model of spontaneous baryogenesis, yields two main effects. First, by computing the production amplitude for lepton-antibaryon and antilepton-baryon pairs, we have shown that the decay of $\theta$ induces a net excess of baryons over antibaryons, in analogy with scalar spontaneous baryogenesis. 
Conversely, the same interaction acts as a sort of mixing between the lepton and the baryon fields. It is well known, in the scalar case, that lepton-baryon oscillations tend to wash out the baryon asymmetry, impairing the baryogenesis mechanism. In the limiting case of massless fermions, the scalar model predicts a vanishing asymmetry. 
Here we have conducted an in-depth analysis of the lepton-baryon oscillations triggered by the vector interaction. Hence, as second byproduct, we have found that, although diminished, the baryon asymmetry survives after taking into account the flavor oscillations, even in the case of massless fermions. 

We showed that the above circumstance represents a major advantage of the vector model over the scalar model, as the mechanism remains efficient even at energies high enough that the fermion fields are still massless, e.g. above the electroweak scale.

We then compared our results with those of the well-known scalar spontaneous baryogenesis model and with the experimental value of the baryon asymmetry. Our vector-field framework provides higher values for the baryon asymmetry than the scalar model for lighter fermions and, in general, can reproduce the observed asymmetry for values of the coupling constant of the vector-current interaction that are not extremely small. 

Additionally, we found one more advantage: In the scalar spontaneous baryogenesis paradigm, having such large values of the coupling constant appeared quite problematic, since the latter was tied to fermion masses. Conversely, in our model, the coupling constant is roughly proportional to a shift of the spatial momentum along the broken symmetry direction. 

Consequently, the vector model does not suffer from the same limitation of its scalar counterpart, allowing for larger coupling constants and, in turn, to reproduce the experimental value of the baryon asymmetry.

As for any Bumblebee model, a full-fledged restoration of Lorentz symmetry at later times could lead to caveats. For the sake of completeness, the way it could be effectively realized has been naively explored, albeit leaving open other mechanisms that, for the sake of brevity, have been omitted in our discussions.

Nonetheless, the possibility of mimicking a Lorentz invariant theory by the Goldstone modes is well established \cite{SPLV1, SPLV3} and, once the temperature dependency of the symmetry breaking potential is taken into account, even a full restoration of the Lorentz invariance was shown to be possible. 

Clearly this comes at the price of abandoning the idea of the Bumblebee model as fundamental, and rather recognize it as an effective field theory. We stress, in each case, that the symmetry restoration issue is not unique to the model considered in this paper, affecting all the known schemes of spontaneous breakdown of Lorentz symmetry. 

Overall, we intended to show that the vector model provides a viable alternative mechanism for baryogenesis, and candidates to represent a different view toward defining the baryon asymmetry as observed. 

Moreover, linking the baryon asymmetry of the universe with LIV exhibits robust advantages, as above stressed, although it turns out to be plagued by the model-dependence on the kind of LIV produced, requiring moreover a fully-restoration at later times.

For these reasons, future works could shed light on the fundamental machinery behind Lorentz restoration and the baryon asymmetry. To do so, we will include explicit LIV terms in the Lagrangian of SSB and study their impact on baryon asymmetry. Within the VFSB framework, the model can also be extended in multiple ways, similar to the approaches already considered for SSB. For example, non-minimal couplings with gravity have been shown to induce significant corrections to baryon asymmetry in the context of SSB \cite{Dubbini:2025jjz,Dubbini:2025hcw,Luongo:2021gho} and, then, incorporating such couplings could show additional effects. 

Additional findings may be expected by exploring well-established LIV models, such as the Horava scenario \cite{Hor1, Hor2,Luongo:2018oil}, or phenomenological and/or more theoretical schemes widely-used in several contexts of cosmology, see e.g. \cite{Biesiada:2007zzb,Kost98,Mattingly:2005re,Liberati:2013xla,Minami:2020odp}. Finally, as in the case of SSB, a more realistic treatment of VFSB would require the inclusion of the fundamental quark fields, rather than the composite baryon fields $Q$, potentially revealing additional mechanisms for asymmetry generation.

\section*{Acknowledgements} 
MD is thankful to Alessio Belfiglio for insightful discussions on particle production in cosmology. OL and AQ acknowledge support by the  Fondazione  ICSC, Spoke 3 Astrophysics and Cosmos Observations. National Recovery and Resilience Plan (Piano Nazionale di Ripresa e Resilienza, PNRR) Project ID $CN00000013$ ``Italian Research Center on  High-Performance Computing, Big Data and Quantum Computing" funded by MUR Missione 4 Componente 2 Investimento 1.4: Potenziamento strutture di ricerca e creazione di ``campioni nazionali di R\&S (M4C2-19)" - Next Generation EU (NGEU).

\newpage
\appendix

\section{Simplification of the equation of motion for the inflaton}\label{app1}

In this section, we show the steps to pass by Eq. (\ref{eq63}) to Eq. (\ref{eq64}). Simply computing the time-derivative of the quantity in square brackets in Eq. (\ref{eq63}), we obtain
\begin{equation}
\begin{split}
&f^2\left(\ddot{\theta}+m^2_{\theta}\theta\right)-\ddot{\theta}\alpha^i\alpha_i-2\dot{\theta}\alpha_i\dot{\alpha^i}+2f\ddot{\theta}\alpha^i+2f\dot{\theta}\dot{\alpha^i}-\left[\ddot{\beta^k}\alpha_k-\ddot{\alpha^k}\beta_k+\ddot{\theta}\left(\alpha^k\alpha_k+\beta^k\beta_k\right)+2\dot{\theta}\left(\alpha_k\dot{\alpha^k}+\beta_k\dot{\beta^k}\right)\right]=\partial_{\mu}J^{\mu}_{Q'}.
\end{split}
\label{eqAPP1}
\end{equation}
Considering Eqs. (\ref{eq55}) and (\ref{eq61}), we can widely simplify Eq. (\ref{eqAPP1}). In particular, we take Eq. (\ref{eq61}) multiplied by $\alpha_k$ and subtract Eq. (\ref{eq55}) multiplied by $\beta_k$. In so doing, we obtain
\begin{equation}
\begin{split}
&\ddot{\beta^k}\alpha_k-\ddot{\alpha^k}\beta_k+\ddot{\theta}\left[\alpha^k\alpha_k+\beta^k\beta_k\right]+2\dot{\theta}\left(\alpha^k\dot{\alpha_k}+\beta^k\dot{\beta_k}\right)=\frac{g}{\sqrt{2}}\left[\left(\overline{Q'}\gamma^{k}L+\overline{L}\gamma^{k}Q'\right)\beta_{k}-i\left(\overline{Q'}\gamma^{k}L-\overline{L}\gamma^{k}Q'\right)\alpha_k\right].
\end{split}
\label{eqAPP2}
\end{equation}
If we substitute Eq. (\ref{eqAPP2}) within Eq. (\ref{eqAPP1}) we get
\begin{equation}
f^2\left(\ddot{\theta}+m^2_{\theta}\theta\right)-\ddot{\theta}\alpha^i\alpha_i-2\dot{\theta}\alpha_i\dot{\alpha^i}+2f\ddot{\theta}\alpha^i+2f\dot{\theta}\dot{\alpha^i}=\partial_{\mu}J^{\mu}_{(Q')}+\frac{g}{\sqrt{2}}\left[\left(\overline{Q'}\gamma^{k}L+\overline{L}\gamma^{k}Q'\right)\beta_{k}-i\left(\overline{Q'}\gamma^{k}L-\overline{L}\gamma^{k}Q'\right)\alpha_k\right].
\label{eqAPP3}
\end{equation}
Using Eq. (\ref{eq49}) for the divergence of $J^{\mu}_{Q'}$, the r.h.s. of Eq. (\ref{eqAPP3}) becomes
\begin{equation}
\begin{split}
&ig\left(v_i+\frac{\alpha_i}{\sqrt{2}}\right)\left(\overline{Q'}\gamma^iL-\overline{L}\gamma^iQ'\right)+\frac{ig}{\sqrt{2}}\alpha_0\left(\overline{Q'}\gamma^0L-\overline{L}\gamma^0Q'\right)-\frac{g}{\sqrt{2}}\beta_0\left(\overline{Q'}\gamma^0L+\overline{L}\gamma^0Q'\right)-\frac{g}{\sqrt{2}}\beta_i\left(\overline{Q'}\gamma^iL+\overline{L}\gamma^iQ'\right),
\end{split}
\label{eqAPP4}
\end{equation}
that can be further simplified considering $\beta_i=0$ and using the constraints in Eqs. (\ref{eq52}) and (\ref{eq58}), obtaining
\begin{equation}
ig\left(v_i+\frac{\alpha_i}{\sqrt{2}}\right)\left(\overline{Q'}\gamma^iL-\overline{L}\gamma^iQ'\right).
\label{eqAPP5}
\end{equation}
Finally, if we substitute Eq. (\ref{eqAPP5}) in lieu of the r.h.s. of Eq. (\ref{eqAPP3}), we find Eq. (\ref{eq64}).\\

\section{Calculation of $\braket{\overline{Q'}\gamma^iL\pm\overline{L}\gamma^iQ'}$}\label{app2}

In this section, we prove the result reported in Eq. (\ref{eq68}) and show that $\braket{\overline{Q'}\gamma^iL+\overline{L}\gamma^iQ'}$ is of the order of $\theta^2$. In particular, we start by Eqs. (\ref{eq66}) and (\ref{eq67}), where we take $w_{\mu}=v_{\mu}$. In so doing, at first order in $g$ it is immediate to find
\begin{equation}
\begin{split}
&\braket{\overline{Q'}\gamma^iL}\pm h.c.=\frac{gf}{\sqrt{2}}\int d^4y\left[\left\langle \overline{L}_{\text{free}}(y)\gamma^iG_Q(y,x)\gamma^iL_{\text{free}}(x)+\overline{Q}_{\text{free}}(x)\gamma^iG_L(x,y)\gamma^iQ_{\text{free}}(y)\right\rangle e^{-i\theta(y)+i\theta(x)}\pm h.c.\right].
\end{split}
\label{eqAPP6}
\end{equation}
In particular, the quantity in square brackets can be rewritten as
\begin{equation}
\left[\braket{\overline{Q}_{\text{free}}(x)\gamma^iG_L(x,y)\gamma^iQ_{\text{free}}(y)}e^{-i\theta(y)+i\theta(x)}\pm\braket{\overline{L}_{\text{free}}(x)\gamma^iG_Q(x,y)\gamma^iL_{\text{free}}(y)}e^{-i\theta(x)+i\theta(y)}\right]\pm h.c.
\label{eqAPP7}
\end{equation}
In the massless fermions limit, each of the two VEVS in Eq. (\ref{eqAPP7}) results equal to
\begin{equation}
\begin{split}
&2\int \frac{d^3p}{(2\pi)^3}\frac{1}{|\vec{p}|}\int \frac{d^4l}{(2\pi)^4}\frac{e^{-il(x-y)}e^{-ip(x-y)}}{E_l^2-|\vec{l}|^2+i\epsilon}\left[2l^ip^i+E_l|\vec{p}|-\vec{l}\cdot\vec{p}\right],
\end{split}
\label{eqAPP8}
\end{equation}
therefore Eq. (\ref{eqAPP7}) becomes
\begin{equation}
\begin{split}
&2\int \frac{d^3p}{(2\pi)^3}\frac{1}{|\vec{p}|}\int \frac{d^4l}{(2\pi)^4}\frac{e^{-i(l+p)(x-y)}}{E_l^2-|\vec{l}|^2+i\epsilon}\left[2l^ip^i+E_l|\vec{p}|-\vec{l}\cdot\vec{p}\right]\left[e^{i\theta(x)-i\theta(y)}\pm e^{-i\theta(x)+i\theta(y)}\right]\pm h.c.
\end{split}
\label{eqAPP9}
\end{equation}
When the integral over $d^3y$ is performed, we find
\begin{equation}
\begin{split}
&2\left\{\left[e^{i\theta(t_x)-i\theta(t_y)}\pm e^{-i\theta(t_x)+i\theta(t_y)}\right]\int \frac{d^3p}{(2\pi)^3}\frac{1}{|\vec{p}|}\left[\int_{-\infty}^{+\infty} \frac{dE_l}{2\pi}\frac{E_l|\vec{p}|+|\vec{p}|^2-2(p^i)^2}{E_l^2-|\vec{p}|^2+i\epsilon}e^{-i(E_l+|\vec{p}|)(t_x-t_y)}\right]\right\}\pm h.c.
\end{split}
\label{eqAPP10}
\end{equation}
The integral in square brackets can be evaluated throughout the integration over the complex plane. In so doing, Eq. (\ref{eqAPP10}) becomes
\begin{equation}
\begin{split}
&\frac{i}{3\pi^2}\left\{\left[e^{i\theta(t_x)-i\theta(t_y)}\pm e^{-i\theta(t_x)+i\theta(t_y)}\right]\int_0^{+\infty} dp p^2\left[\Theta(t_y-t_x)-2e^{-2ip(t_x-t_y)}\Theta(t_x-t_y)\right]\right\}\pm h.c.,
\end{split}
\label{eqAPP12}
\end{equation}
where $\Theta$ is the Heaviside function. When either the $+$ sign or the $-$ sign is chosen, respectively for $\braket{\overline{Q'}\gamma^iL+\overline{L}\gamma^iQ}$ and $\braket{\overline{Q'}\gamma^iL-\overline{L}\gamma^iQ}$, Eq. (\ref{eqAPP12}) becomes
\begin{subequations}
\begin{equation}
-\frac{8}{3\pi^2}\Theta(t_x-t_y)\cos\left[\theta(t_x)-\theta(t_y)\right]\int_0^{+\infty}d\omega \omega^2\sin\left[2\omega(t_x-t_y)\right],
\end{equation}
\begin{equation}
-\frac{8i}{3\pi^2}\Theta(t_x-t_y)\sin\left[\theta(t_x)-\theta(t_y)\right]\int_0^{+\infty}d\omega\omega^2\sin\left[2\omega(t_x-t_y)\right].
\end{equation}
\label{eqAPP13}
\end{subequations}
Consequently, we obtain
\begin{subequations}
\begin{equation}
\braket{\overline{Q'}\gamma^iL+\overline{L}\gamma^iQ'}=\frac{8 gf}{3\pi^2\sqrt{2}}\int_{-\infty}^{0}dt_y\int_0^{+\infty} d\omega\omega^2\cos\left[\theta(t_y+t_x)-\theta(t_x)\right]\sin\left[2\omega t_y\right],
\label{eqAPP14a}
\end{equation}
\begin{equation}
\braket{\overline{Q'}\gamma^iL-\overline{L}\gamma^iQ'}=-\frac{8igf}{3\pi^2\sqrt{2}}\int_{-\infty}^{0}dt_y\int_0^{+\infty} d\omega\omega^2\sin\left[\theta(t_y+t_x)-\theta(t_x)\right]\sin\left[2\omega t_y\right].
\label{eqAPP14b}
\end{equation}
\end{subequations}
In particular, the double integral within the expression for $\braket{\overline{Q'}\gamma^iL-\overline{L}\gamma^iQ'}$ is exactly the same appearing also in SSB, resulting
\begin{equation}
\frac{\pi\Omega\dot{\theta}}{16}-\frac{\pi\Omega^2\theta}{8}\lim_{\omega\to+\infty}\ln\left(\frac{2\omega}{\Omega}\right).
\label{eqAPP15}
\end{equation}
If we substitute Eq. (\ref{eqAPP15}) into Eq. (\ref{eqAPP14b}) for $\braket{\overline{Q'}\gamma^iL-\overline{L}\gamma^iQ'}$ we find exactly Eq. (\ref{eq68}).
About $\braket{\overline{Q'}\gamma^iL+\overline{L}\gamma^iQ'}$, we start from its expression in Eq. (\ref{eqAPP14a}) and compute the respective double integral following the same approach used in SSB to compute $\braket{\overline{Q'}\gamma^iL-\overline{L}\gamma^iQ'}$. In so doing, we obtain
\begin{equation}
\frac{1}{8}\lim_{\omega\to+\infty}\int_{-\infty}^0dt_y\cos(\Delta\theta)\frac{\partial^2}{\partial t_y^2}\left(\frac{\cos\left(2\omega t_y\right)-1}{t_y}\right)=\frac{1}{8}\lim_{\omega\to+\infty}\int_{-\infty}^0dt_y\left(\frac{1-\cos\left(2\omega t_y\right)}{t_y}\right)\left[\ddot{\theta}(t_y+t_x)\sin(\Delta\theta)+\dot{\theta}^2\cos(\Delta\theta)\right],
\label{eqAPP16}
\end{equation}
where $\Delta\theta=\theta(t_y+t_x)-\theta(t_x)$. From Eq. (\ref{eqAPP16}), it is clear that for small $\theta$ the leading contribution is of the order $\theta^2$, that is our initial claim.

\section{Calculation of the baryon asymmetry}\label{app3}

In this section, we want to prove the result obtained in Eq. (\ref{eq84}). At this purpose, we start by Eq. (\ref{eq13}) and use it to compute the number density of produced baryons and antibaryons. In particular, the number density of baryons is
\begin{equation}
n_b=\frac{1}{V}\sum_{s_Q,s_L}\int\frac{d^3p_Q}{(2\pi)^32p_Q^0}\frac{d^3p_L}{(2\pi)^32p_L^0}|\mathcal{A}(\overline{L}-Q)|^2,
\label{eqAPP17}
\end{equation}
where
\begin{equation}
\begin{split}
\mathcal{A}(\overline{L}-Q)&=i\int d^4x\braket{\overline{L}(p_L,s_L),Q(p_Q,s_Q)|\hat{T}\left\{\mathcal{L}_I(x)\right\}|0}=\\&=igv_i\overline{u}(\vec{p}_Q,s_Q)\gamma^iv(\vec{p}_L,s_L)(2\pi)^3\delta^3(\vec{p}_Q+\vec{p}_L)\int_{-\infty}^{+\infty}dt_xe^{i\theta(t_x)}e^{i(p^0+q^0)t_x}.
\end{split}
\label{eqAPP18}
\end{equation}
Summing over the spins, considering massless fermions and assuming $p_i^2=|\vec{p}|^2/3$, it is not difficult to find
\begin{equation}
\begin{split}
&n_b=\frac{g^2f^2}{3\pi^2}\int_0^{+\infty}d\omega\omega^2\left|\int_{-\infty}^{+\infty}dte^{i\theta(t)}e^{2i\omega t}\right|^2.
\end{split}
\label{eqAPP19}
\end{equation}
Similarly, for the number density of antibaryons we obtain the Eq. (\ref{eqAPP19}) but with $-i\theta(t)$ in lieu of $+i\theta(t)$ at the exponent. Therefore, formally, our expressions for the number density of baryons and antibaryons are analogous to that found within SSB, but with a factor $2/3$ in front. The same is definitely true for the baryon asymmetry, that results thus that in Eq. (\ref{eq84}).

\section{Oscillation probabilities}\label{app4}

In this Appendix we work out the $Q \rightarrow L$ transition probabilities, following two alternative approaches. The first approach deals directly with the spectral resolution of the field Hamiltonian $\mathcal{H}(p)$, diagonalizing, for a fixed momentum $p$, the $8 \times 8$ matrix on the original $\lbrace \ket{Q}, \ket{L} \rbrace$ basis. The analysis then proceeds in a standard fashion, with the determination of the eigenvalues and of the eigenvectors and the identification of the physical (oscillating) $Q$ and $L$ states. The second approach acts directly at the level of the field Lagrangian, leveraging on a clever choice of basis to identify the eigenstates as particles with definite energy, chirality and helicity (in the massless limit).

\subsection{The first approach}\label{subsec:level6-1}

In this approach, we directly write the Hamiltonian starting by the Lagrangian in Eq. (\ref{eq88}) and diagonalize it. The Hamiltonian in question is
\begin{equation}
\begin{split}
\mathcal{H}&=-\overline{Q}\left(i\gamma^j\partial_j-m_Q\right)Q-\overline{L}\left(i\gamma^j\partial_j-m_L\right)L-gv_i\left(\overline{Q}\gamma^{i}L+\overline{L}\gamma^{i}Q\right),
\end{split}
\label{eq90}
\end{equation}
whose matrix form in Fourier space can be written in Dirac representation as
\begin{equation}
\begin{split}
&\hat{\mathcal{H}}(p)=\begin{pmatrix}
0 & \vec{p}\cdot\vec{\sigma} & 0 & -gv_i\sigma^i \\ \vec{p}\cdot\vec{\sigma} & 0 & -gv_i\sigma^i & 0 \\ 0 & -gv_i\sigma^i & 0 & \vec{p}\cdot\vec{\sigma} \\ -gv_i\sigma^i & 0 & \vec{p}\cdot\vec{\sigma} & 0
\end{pmatrix},
\end{split}
\label{eq91}
\end{equation}
in the limit of massless fermions. In order to find eigenvalues and eigenvectors of the matrix $\hat{\mathcal{H}}(p)$, we need to consider the latter as an $8\times 8$ matrix, thus we have to expand all the $2\times 2$ blocks containing the Pauli matrices. Therefore, we have to consider an $8$-dimensional Hilbert space, whose basis is $B_{\text{old}}=\{\ket{q_1},\ket{q_2},\ket{q_3},\ket{q_4},\ket{l_1},\ket{l_2},\ket{l_3},\ket{l_4}\}$, representing the basis over which the matrix $\hat{\mathcal{H}}(p)$ is written in Eq. (\ref{eq91}). Therefore, eigenvalues and eigenvectors can be found only once a precise spatial direction is chosen for the component $i$. This is necessary because $\hat{\mathcal{H}}(p)$ is in general an $8\times 8$ matrix and we need to consider the whole $8$-dimensional Hilbert space in order to obtain the new basis of $8$ eigenvectors and the corresponding eigenvalues. At this purpose, we arbitrarily choose $i=z$. In so doing, we find the following four different eigenvalues:
\begin{equation}
\lambda_{\pm\pm}=\pm\sqrt{\left(p_x^2+p_y^2\right)+\left(p_z\pm gv_i\right)^2},
\label{eq93}
\end{equation}
where the signs should be all taken independently. Each of these $4$ eigenvalues has multiplicity $\mu=2$, therefore we expect for each eigenvalue to find a $2$-dimensional associated eigenspace.

The generic eigenvector can be written as
\begin{equation}
\begin{split}
\ket{\psi}&=\frac{1}{\sum_{j=1}^8|c_j|^2}\bigg{\{}c_1\ket{q_1}+c_2\ket{q_2}+c_3\ket{q_3}+c_4\ket{q_4}+c_5\ket{l_1}+c_6\ket{l_2}+c_7\ket{l_3}+c_8\ket{l_4}\bigg{\}},
\end{split}
\label{eq94}
\end{equation}
where the complex coefficients $c_j$ need to be found. Performing the calculation, we find that for each eigenvalue $\lambda$ we cannot associate merely a single eigenvector but a $2$-dimensional Hilbert space, as expected, whose basis vectors must be constructed in such a way that their scalar product vanishes. Given the generic eigenvalue $\lambda_{\pm\pm}$, we label with $\ket{\psi^{\uparrow}_{\lambda_{\pm\pm}}}$ and $\ket{\psi^{\downarrow}_{\lambda_{\pm\pm}}}$ the two basis vectors of the eigenspace associated to $\lambda_{\pm\pm}$ and with $c_j^{\uparrow\pm\pm}$ and $c_j^{\downarrow\pm\pm}$ the respective associated coefficients.

For simplicity of notation, introduce the complex variable $z$ as
\begin{equation}
z\equiv\sqrt{\frac{p_x+ip_y}{p_x-ip_y}}=\frac{p_x+ip_y}{\sqrt{p^2_x+p^2_y}},
\label{eq95}
\end{equation}
and define the four dimensionless real parameters $a$, $b$, $c$ and $d$ as
\begin{equation}
a\equiv\frac{\lambda_{++}+\sqrt{p_x^2+p^2_y}}{p_z+gv_i},\quad b\equiv\frac{\lambda_{+-}+\sqrt{p_x^2+p^2_y}}{p_z-gv_i},\quad c\equiv\frac{\lambda_{++}-\sqrt{p_x^2+p^2_y}}{p_z+gv_i},\quad d\equiv\frac{\lambda_{+-}-\sqrt{p_x^2+p^2_y}}{p_z-gv_i}.
\label{eq97}
\end{equation}
In so doing, the coefficients can be rewritten in a very simple form in terms of $a$, $b$, $c$, $d$ and $z$. We summarize them within Tab. \ref{tab1}. The corresponding full set of eigenvectors is a new basis for the Hilbert space in question, namely $B_{\text{new}}=\bigg{\{}\ket{\psi^{\uparrow++}},\ket{\psi^{\uparrow+-}},\ket{\psi^{\uparrow-+}},\ket{\psi^{\uparrow--}}, \ket{\psi^{\downarrow++}},\ket{\psi^{\downarrow+-}},\ket{\psi^{\downarrow-+}},\ket{\psi^{\downarrow--}}\bigg{\}}$.

\begin{table*}
\renewcommand{\arraystretch}{1.5}
\begin{tabular}{ |c|c|c|c|c|c|c|c| }
\hline
\hline
$c_1^{\uparrow ++}=a$ & $c_2^{\uparrow ++}=-z$ & $c_3^{\uparrow ++}=1$ & $c_4^{\uparrow ++}=az$ & $c_5^{\uparrow ++}=-a$ & $c_6^{\uparrow ++}=z$ & $c_7^{\uparrow ++}=-1$ & $c_8^{\uparrow ++}=-az$ \\
\hline
$c_1^{\uparrow +-}=-b$ & $c_2^{\uparrow +-}=z$ & $c_3^{\uparrow +-}=-1$ & $c_4^{\uparrow +-}=-bz$ & $c_5^{\uparrow+-}=-b$ & $c_6^{\uparrow +-}=z$ & $c_7^{\uparrow +-}=-1$ & $c_8^{\uparrow +-}=-bz$ \\
\hline
$c_1^{\uparrow -+}=-c$ & $c_2^{\uparrow -+}=-z$ & $c_3^{\uparrow -+}=1$ & $c_4^{\uparrow -+}=-cz$ & $c_5^{\uparrow -+}=c$ & $c_6^{\uparrow -+}=z$ & $c_7^{\uparrow -+}=-1$ & $c_8^{\uparrow -+}=cz$ \\
\hline
$c_1^{\uparrow --}=d$ & $c_2^{\uparrow --}=z$ & $c_3^{\uparrow --}=-1$ & $c_4^{\uparrow --}=dz$ & $c_5^{\uparrow--}=d$ & $c_6^{\uparrow --}=z$ & $c_7^{\uparrow --}=-1$ & $c_8^{\uparrow --}=dz$ \\
\hline
$c_1^{\downarrow++}=-c$ & $c_2^{\downarrow ++}=-z$ & $c_3^{\downarrow ++}=-1$ & $c_4^{\downarrow ++}=cz$ & $c_5^{\downarrow ++}=c$ & $c_6^{\downarrow++}=z$ & $c_7^{\downarrow++}=1$ & $c_8^{\downarrow ++}=-cz$ \\
\hline
$c_1^{\downarrow+-}=d$ & $c_2^{\downarrow +-}=z$ & $c_3^{\downarrow +-}=1$ & $c_4^{\downarrow +-}=-dz$ & $c_5^{\downarrow+-}=d$ & $c_6^{\downarrow+-}=z$ & $c_7^{\downarrow+-}=1$ & $c_8^{\downarrow+-}=-dz$ \\
\hline
$c_1^{\downarrow-+}=a$ & $c_2^{\downarrow -+}=-z$ & $c_3^{\downarrow -+}=-1$ & $c_4^{\downarrow -+}=-az$ & $c_5^{\downarrow -+}=-a$ & $c_6^{\downarrow-+}=z$ & $c_7^{\downarrow-+}=1$ & $c_8^{\downarrow -+}=az$ \\
\hline
$c_1^{\downarrow--}=-b$ & $c_2^{\downarrow --}=z$ & $c_3^{\downarrow --}=1$ & $c_4^{\downarrow --}=bz$ & $c_5^{\downarrow--}=-b$ & $c_6^{\downarrow--}=z$ & $c_7^{\downarrow--}=1$ & $c_8^{\downarrow--}=bz$ \\
\hline
\hline
\end{tabular}
\captionsetup{justification=raggedright,singlelinecheck=false}
\caption{Coefficients of the linear combinations expressing the eigenstates in terms of the old basis vectors. Each row corresponds to a specific eigenvector, and each column contains the coefficient of the corresponding old basis vector in the linear combination.}
\label{tab1}
\end{table*}

For convention, we take also the quantum states normalized to the unity. At this purpose, we have to consider the following squared normalization factors:
\begin{subequations}
\begin{align}
\left(N^{\uparrow++}\right)^2=\left(N^{\downarrow-+}\right)^2=4\left(1+a^2\right)\equiv A^2,\quad \left(N^{\uparrow+-}\right)^2=\left(N^{\downarrow--}\right)^2=4\left(1+b^2\right)\equiv B^2,\\
\left(N^{\uparrow-+}\right)^2=\left(N^{\downarrow++}\right)^2=4\left(1+c^2\right)\equiv C^2,\quad
\left(N^{\uparrow--}\right)^2=\left(N^{\downarrow+-}\right)^2=4\left(1+d^2\right)\equiv D^2.
\end{align}
\label{eq100}
\end{subequations}
In order to proceed with the calculation of the oscillation probability, we need to know how to write the old basis vectors in terms of the eigenvectors. In so doing, indeed, we can know how to write the time evolution of every old basis vector. At this purpose, we have to invert the relations $\ket{\psi}=\ket{\psi}(\ket{q_i},\ket{l_j})$ determined by the coefficients in Tab. \ref{tab1}. In this respect, it is helpful to define the $4$ real dimensionless parameters
\begin{equation}
\alpha\equiv\frac{A}{4\left(a+c\right)},\quad
\beta\equiv\frac{B}{4\left(b+d\right)},\quad 
\gamma\equiv\frac{C}{4\left(a+c\right)},\quad 
\delta\equiv\frac{D}{4\left(b+d\right)}.
\label{eq101}
\end{equation}
The coefficients used for writing the old basis vectors in terms of the eigenvectors are summarized in Tab. \ref{tab2}.

\begin{table}
\renewcommand{\arraystretch}{1.5}
\begin{tabular}{ |c|c|c|c|c|c|c|c| }
\hline\hline
$\alpha$ & $-\beta$ & $-\gamma$ & $\delta$ & $-\gamma$ & $\delta$ & $\alpha$ & $-\beta$ \\
\hline
$-z^*c\alpha$ & $z^*d\beta$ & $-z^*a\gamma$ & $z^*b\delta$ & $-z^*a\gamma$ & $z^*b\delta$ & $-z^*c\alpha$ & $z^*d\beta$ \\
\hline
$c\alpha$ & $-d\beta$ & $a\gamma$ & $-b\delta$ & $-a\gamma$ & $b\delta$ & $-c\alpha$ & $d\beta$ \\
\hline
$z^*\alpha$ & $-z^*\beta$ & $-z^*\gamma$ & $z^*\delta$ & $z^*\gamma$ & $-z^*\delta$ & $-z^*\alpha$ & $z^*\beta$ \\
\hline
$-\alpha$ & $-\beta$ & $\gamma$ & $\delta$ & $\gamma$ & $\delta$ & $-\alpha$ & $-\beta$ \\
\hline
$z^*c\alpha$ & $z^*d\beta$ & $z^*a\gamma$ & $z^*b\delta$ & $z^*a\gamma$ & $z^*b\delta$ & $z^*c\alpha$ & $z^*d\beta$ \\
\hline
$-c\alpha$ & $-d\beta$ & $-a\gamma$ & $-b\delta$ & $a\gamma$ & $b\delta$ & $c\alpha$ & $d\beta$ \\
\hline
$-z^*\alpha$ & $-z^*\beta$ & $z^*\gamma$ & $z^*\delta$ & $-z^*\gamma$ & $-z^*\delta$ & $z^*\alpha$ & $z^*\beta$ \\
\hline\hline
\end{tabular}
\captionsetup{justification=raggedright,singlelinecheck=false}
\caption{Coefficients of the linear combinations expressing the old basis vectors in terms of the eigenstates. Each row corresponds to a specific old basis vector, and each column contains the coefficient of the corresponding eigenstate in the linear combination.}
\label{tab2}
\end{table}

In the Schrödinger picture, each eigenstate $\ket{\psi^{\lambda\pm\pm}}$ of the Hamiltonian, associated to the eigenvalue $\lambda_{\pm\pm}$, evolves in time as $\ket{\psi^{\lambda\pm\pm}(t)}=e^{-i\lambda_{\pm\pm} t}\ket{\psi^{\lambda\pm\pm}(0)}$. In order to compute the time-dependent probability that a particle produced as $Q$ rotates into $L$ we have to find the particle states corresponding to $Q$ and $L$. At this purpose, we firstly define the quantum states
\begin{subequations}
\begin{align}
\ket{Q}=\frac{\ket{q_1}+\ket{q_2}+\ket{q_3}+\ket{q_4}}{2},\\
\ket{L}=\frac{\ket{l_1}+\ket{l_2}+\ket{l_3}+\ket{l_4}}{2}.
\end{align}
\label{eq104}
\end{subequations}
These two states, however, contain both positive and negative energy states. Therefore, we can interpret them as quantum states describing both particles and antiparticles. In order to find the corresponding particle states, we have to project the states $\ket{Q}$ and $\ket{L}$ into the Hilbert subspace associated exclusively to positive energy states. At this purpose, we define the projector on positive energy states as
\begin{equation}
\begin{split}
\hat{P}_+&=p_1\ket{\psi^{\uparrow++}}\bra{\psi^{\uparrow++}}+p_2\ket{\psi^{\uparrow+-}}\bra{\psi^{\uparrow+-}}+p_3\ket{\psi^{\downarrow++}}\bra{\psi^{\downarrow++}}+p_4\ket{\psi^{\downarrow+-}}\bra{\psi^{\downarrow+-}},
\end{split}
\label{eq105}
\end{equation}
where $p_j$ for $j=1,2,3,4$ have to be determined such that the projected states obtained by applying $\hat{P}_+$ to $\ket{Q}$ and $\ket{L}$ also result orthogonal at time $t=0$. The projector $\hat{P}_+$, when applied to both the states $\ket{Q}$ and $\ket{L}$, return respectively the $Q$-particle state and the $L$-particle state, to be then opportunely normalized.
To find the coefficients $p_j$, however, we have some freedom left even if imposing the orthogonality between $\hat{P}_+\ket{Q}$ and $\hat{P}_+\ket{L}$. Thus, we make the ansatz $|p_1|^2=|p_3|^2$, $|p_2|^2=|p_4|^2$, yielding $\frac{|p_2|^2}{|p_1|^2}=\frac{1+p_x/\lambda_{++}}{1+p_x/\lambda_{+-}}$. Accordingly, we arbitrarily take
\begin{equation}
p_1=p_3=\sqrt{1+\frac{p_x}{\lambda_{+-}}},\quad p_2=p_4=\sqrt{1+\frac{p_x}{\lambda_{++}}}.
\label{eq108}
\end{equation}
Finally, we normalize the particle states obtained by $\ket{Q}$ and $\ket{L}$, defined respectively as
\begin{equation}
\ket{Q_+}=\frac{\hat{P}_+\ket{Q}}{N_Q},\quad \ket{L_+}=\frac{\hat{P}_+\ket{L}}{N_L}.
\label{eq109}
\end{equation} 
We find for both of the particle states the same normalization factor, that is
\begin{equation}
N^2_Q=N_L^2=\frac{1}{2}\left(1+\frac{p_x}{\lambda_{++}}\right)\left(1+\frac{p_x}{\lambda_{+-}}\right).
\label{eq110}
\end{equation}
We are now ready to compute the probability that a particle produced in the quantum state $\ket{Q_+}$ at time $t=0$ rotates into $\ket{L_+}$ at the generic time $t$. We obtain
\begin{equation}
P_{Q\to L}(t)=\left|\braket{L_+|Q_+(t)}\right|^2=\sin^2\left[\left(\frac{\lambda_{++}-\lambda_{+-}}{2}\right)t\right].
\label{eq111}
\end{equation}
It is straightforward to see that the probability of the inverse process is the same. 

For the antiparticle states, we define a projector $\hat{P}_-$ formally analogous to $\hat{P}_+$ but projecting on negative energy states. In so doing, we obtain for the coefficients an equation similar to Eq. (\ref{eq108}) but substituting $\lambda_{++}\to-\lambda_{++}$ and $\lambda_{+-}\to-\lambda_{+-}$. The same should be done also for $N^2_Q$ and $N^2_L$ and consequently for the oscillation probability, obtaining for the latter the same result as in Eq. (\ref{eq111}).

Following this first approach, in summary, we obtain the same oscillation probability for a particle $Q$ into a particle $L$, an antiparticle $\overline{Q}$ into an antiparticle $\overline{L}$ and the respective inverse processes, given by Eq. (\ref{eq111}).

\subsection{The second approach}\label{subsec:level6-2}

In this subsection, we are going to see another method for computing the oscillation probability in the massless fermions limit that yields the same result essentially obtained following the first approach.

Looking at the fermion Lagrangian in Eq. (\ref{eq88}), here we want to consider a rotation of the fields $Q$ and $L$ into new fields $\Psi_1$ and $\Psi_2$ such to diagonalize the interaction term. In particular, we show that a rotation like
\begin{equation}
\begin{pmatrix}
\Psi_1 \\ \Psi_2
\end{pmatrix}=\begin{pmatrix}
\cos\left(\theta\right) & -\sin\left(\theta\right) \\ \sin\left(\theta\right) & \cos\left(\theta\right)
\end{pmatrix}\begin{pmatrix}
Q \\ L
\end{pmatrix}
\label{eq112}
\end{equation}
diagonalizes the interaction term for $\theta=\pi/4$. In so doing, in the limit of massless fermions, the fermion Lagrangian becomes
\begin{equation}
\mathcal{L}_{\text{fermions}}=\overline{\Psi}_1i\slashed{\partial}\Psi_1+\overline{\Psi}_2i\slashed{\partial}\Psi_2+gv_i\left[\overline{\Psi}_2\gamma^i\Psi_2-\overline{\Psi}_1\gamma^i\Psi_1\right].
\label{eq113}
\end{equation}
In the basis $\{\Psi_1,\Psi_2\}$ the effect of the interaction is to increase the $i$-th component of the spatial momentum for one of the two new fields, e.g. $\Psi_1$, and to reduce the same component by the same quantity for the other new field, e.g. $\Psi_2$. We thus expect the interaction to remove partially the degeneracy of the eigenvalues of the Hamiltonian, in particular discriminating between energies with $i$-th component of the spatial momentum equal to $p_i+gv_i$ and $p_i-gv_i$. In this respect, for simplicity we define
\begin{equation}
\vec{p}_{\pm}=\vec{p}\pm g\vec{v},\quad \vec{v}=v_i\left(\delta^{ji}\hat{j}\right),
\label{eq114}
\end{equation}
where the index $j$ runs over the three spatial components, whereas $i$ is fixed and coincides with the along-VEV direction, coherently with the notation always used in this work.

In this context, it is also useful to write the new fields $\Psi_1$ and $\Psi_2$ in terms of the chiral states, namely
\begin{equation}
\Psi_1=\begin{pmatrix}
\Psi_1^L \\ \Psi_1^R
\end{pmatrix},\quad \Psi_2=\begin{pmatrix}
\Psi_2^L \\ \Psi_2^R
\end{pmatrix}
\label{eq115}.
\end{equation}
Adopting the chiral representation for the $\gamma$-matrices, with respect to the chiral basis $\{\Psi_1^L, \Psi_1^R,\Psi_2^L,\Psi_2^R\}$ we obtain the following Hamiltonian written in Fourier space:
\begin{equation}
\hat{\mathcal{H}}(p)=\begin{pmatrix}
-h_+ & 0 & 0 & 0 \\ 0 & h_+ & 0 & 0 \\ 0 & 0 & -h_- & 0 \\ 0 & 0 & 0 & h_-
\end{pmatrix},
\label{eq116}
\end{equation}
where
\begin{equation}
h_{\pm}=\vec{\sigma}\cdot\vec{p}\pm gv_i\sigma^i=\vec{\sigma}\cdot\vec{p}_{\pm}.
\label{eq117}
\end{equation}
The chiral basis $\{\Psi_1^L, \Psi_1^R,\Psi_2^L,\Psi_2^R\}$ thus represents the basis of the eigenvectors for the Hamiltonian in Eq. (\ref{eq116}). Each of the $2\times 2$ blocks appearing in Eq. (\ref{eq116}) has $2$ associated eigenvalues, one with positive energy and the other with negative energy. In particular, the blocks $h_+$ and $h_-$ have respectively eigenvalues
\begin{subequations}
\begin{align}
 &   \lambda_{\pm +}=\pm|\vec{p}_+|=\pm\sqrt{\sum_{j\neq i}\left(p^j\right)^2+\left(p^i+gv_i\right)^2},\\
&\lambda_{\pm -}=\pm|\vec{p}_-|=\pm\sqrt{\sum_{j\neq i}\left(p^j\right)^2+\left(p^i-gv_i\right)^2}.
\label{eq119}
\end{align}
\end{subequations}
This means that each Weyl spinor contains both positive and negative energy states. In order to distinguish between positive and negative energy states, it is helpful to write each Weyl spinor as linear combination of helicity eigenstates. In particular, in this context, we define the helicity with respect to the shifted momentum $\vec{p}_{\pm}$, that is $\hat{h}_{\pm}=\frac{\vec{\sigma}\cdot\vec{p}_{\pm}}{|\vec{p}_{\pm}|}$, where the $+$ signs are used for $\Psi_1$ and the $-$ for $\Psi_2$, in correspondence of the blocks with $h_+$ and $h_-$ respectively. To see the correspondence between definite-helicity states and definite-energy states is rather simple when looking at the Weyl equations arising from the eigenvalue equation for the Hamiltonian in Eq. (\ref{eq116}). Indeed, the corresponding Weyl equations are
\begin{equation}
\begin{cases}
-\hat{h}_+\Psi_1^L=\pm \Psi_1^L, \\ \hat{h}_+\Psi_1^R=\pm\Psi_1^R, \\ -\hat{h}_-\Psi_2^L=\pm\Psi_2^L, \\ \hat{h}_-\Psi_2^R=\pm\Psi_2^R,
\end{cases}
\label{eq121}
\end{equation}
where the $\pm$ on the r.h.s. of all the four Weyl equations refer to either positive $(+)$ or negative $(-)$ energy. It is straightforward to see that, both for $\Psi_1$ and $\Psi_2$, we obtain
\begin{itemize}

\item[-] left-handed chirality, positive (negative) energy and negative (positive) helicity;

\item[-] right-handed chirality, positive (negative) energy and positive (negative) helicity.

\end{itemize}
What changes between $\Psi_1$ and $\Psi_2$ is only that $\Psi_1$ is associated to shifted momentum $\vec{p}_+$ whereas $\Psi_2$ to $\vec{p}_-$, thus the absolute values of the corresponding energies are respectively $|\vec{p}_+|$ and $|\vec{p}_-|$. With this in mind, we can define the generic eigenstate of the Hamiltonian as represented by four quantum numbers, in the form $\ket{L/R,\pm1,\pm,\pm}$, where
\begin{itemize}

\item[-] $L/R$ refers to either left-handed or right-handed chirality;

\item[-] $\pm1$ indicates the value of the helicity;

\item[-] the first $\pm$ defines the sign of the energy;

\item[-] the second $\pm$ indicates if the absolute value of the energy is $|\vec{p}_+|$ or $|\vec{p}_-|$ and consequently if the quantum state is associated to the field $\Psi_1$ or $\Psi_2$.

\end{itemize}
In this notation, the full set of eigenstates of the Hamiltonian with the corresponding helicity and chirality is reported in Tab. \ref{tab3}.

\begin{table*}
\renewcommand{\arraystretch}{1.5}
\begin{tabular}{ |c|c|c|c|c|c|c|c|c| }
\hline\hline
Eigenstate & $\ket{L,+1,-,+}$ & $\ket{L,-1,+,+}$ & $\ket{R,+1,+,+}$ & $\ket{R,-1,-,+}$ & $\ket{L,+1,-,-}$ & $\ket{L,-1,+,-}$ & $\ket{R,+1,+,-}$ & $\ket{R,-1,-,-}$ \\
\hline\hline
Helicity & $+1$ & $-1$ & $+1$ & $-1$ & $+1$ & $-1$ & $+1$ & $-1$ \\
\hline
Energy & $-|\vec{p}_+|$ & $+|\vec{p}_+|$ & $+|\vec{p}_+|$ & $-|\vec{p}_+|$ & $-|\vec{p}_-|$ & $+|\vec{p}_-|$ & $+|\vec{p}_-|$ & $-|\vec{p}_-|$ \\
\hline\hline
\end{tabular}
\captionsetup{justification=raggedright,singlelinecheck=false}
\caption{Full set of the eigenstates of the Hamiltonian with the corresponding helicity, chirality, energy and associated particle-type.}
\label{tab3}
\end{table*}
All the eigenstates in Tab. \ref{tab3} are considered normalized to the unity. Moreover, they are all orthogonal between each other, since there is no pair of states with all the four quantum numbers identical. We thus construct the $\Psi_1$-particle state, the $\Psi_2-$particle state and the corresponding antiparticle states by taking appropriate linear combinations of the quantum states in Tab. \ref{tab3}. In particular, for particle states we consider linear combinations of positive energy states. Looking at Tab. \ref{tab3}, we see that we have two positive energy states both for $\Psi_1$ and $\Psi_2$, one left-handed and the other right-handed. In particular, the $\Psi_1$-particle state can be written as
\begin{equation}
\ket{(\Psi_1)_+}=\cos\left(\phi\right)\ket{L,-1,+,+}+\sin\left(\phi\right)\ket{R,+1,+,+},
\label{eq123}
\end{equation}
whereas $\Psi_2$-particle state as
\begin{equation}
\ket{(\Psi_2)_+}=\cos\left(\phi\right)\ket{L,-1,+,-}+\sin\left(\phi\right)\ket{R,+1,+,-}.
\label{eq124}
\end{equation}
The angle $\phi$ is for now left generic. The same combination of chiral states, i.e., the same $\phi$, is chosen for $\ket{(\Psi_1)_+}$ and $\ket{(\Psi_2)_+}$ due to the symmetry between $\Psi_1$ and $\Psi_2$ within Eq. (\ref{eq121}).

By reversing Eq. (\ref{eq112}) with $\theta=\pi/4$, we can construct the particle states $\ket{Q_+}$ and $\ket{L_+}$ using $\ket{(\Psi_1)_+}$ and $\ket{(\Psi_2)_+}$. Then, we can proceed as in the first approach. Each of the eigenstates in Tab. \ref{tab3} evolves freely in time with a phase factor given by the respective eigenvalue. In particular, we have
\begin{equation}
\ket{L/R,\pm1,\pm,\pm}(t)=\ket{L/R,\pm1,\pm,\pm}(0)e^{-i\lambda_{\pm\pm}t},
\label{eq125}
\end{equation}
and consequently we can know the time evolution of the states $\ket{Q_+}$ and $\ket{L_+}$. Therefore, we can compute the probability that a particle produced in the state $\ket{Q_+}$ rotates at the generic time $t$ into $\ket{L_+}$. For it, we obtain exactly the result displayed in Eq. (\ref{eq111}), that we know to even coincide with the probability for the inverse process to happen. Finally, we can do the same treatment for the antiparticle states. In so doing, the probability of oscillation results the same for particles and antiparticles, as already found following the first method. The two approaches thus yield the same result for the oscillation probability, that can be summarized in
\begin{equation}
\begin{split}
P_{Q\to L}(t)&=P_{L\to Q}(t)=P_{\overline{Q}\to\overline{L}}(t)=P_{\overline{L}\to\overline{Q}}(t)=\sin^2\left[\left(\frac{\lambda_{++}-\lambda_{+-}}{2}\right)t\right].
\end{split}
\label{eq126}
\end{equation}


\begin{thebibliography}{93}%
\makeatletter
\providecommand \@ifxundefined [1]{%
 \@ifx{#1\undefined}
}%
\providecommand \@ifnum [1]{%
 \ifnum #1\expandafter \@firstoftwo
 \else \expandafter \@secondoftwo
 \fi
}%
\providecommand \@ifx [1]{%
 \ifx #1\expandafter \@firstoftwo
 \else \expandafter \@secondoftwo
 \fi
}%
\providecommand \natexlab [1]{#1}%
\providecommand \enquote  [1]{``#1''}%
\providecommand \bibnamefont  [1]{#1}%
\providecommand \bibfnamefont [1]{#1}%
\providecommand \citenamefont [1]{#1}%
\providecommand \href@noop [0]{\@secondoftwo}%
\providecommand \href [0]{\begingroup \@sanitize@url \@href}%
\providecommand \@href[1]{\@@startlink{#1}\@@href}%
\providecommand \@@href[1]{\endgroup#1\@@endlink}%
\providecommand \@sanitize@url [0]{\catcode `\\12\catcode `\$12\catcode `\&12\catcode `\#12\catcode `\^12\catcode `\_12\catcode `\%12\relax}%
\providecommand \@@startlink[1]{}%
\providecommand \@@endlink[0]{}%
\providecommand \url  [0]{\begingroup\@sanitize@url \@url }%
\providecommand \@url [1]{\endgroup\@href {#1}{\urlprefix }}%
\providecommand \urlprefix  [0]{URL }%
\providecommand \Eprint [0]{\href }%
\providecommand \doibase [0]{http://dx.doi.org/}%
\providecommand \selectlanguage [0]{\@gobble}%
\providecommand \bibinfo  [0]{\@secondoftwo}%
\providecommand \bibfield  [0]{\@secondoftwo}%
\providecommand \translation [1]{[#1]}%
\providecommand \BibitemOpen [0]{}%
\providecommand \bibitemStop [0]{}%
\providecommand \bibitemNoStop [0]{.\EOS\space}%
\providecommand \EOS [0]{\spacefactor3000\relax}%
\providecommand \BibitemShut  [1]{\csname bibitem#1\endcsname}%
\let\auto@bib@innerbib\@empty
\bibitem [{\citenamefont {Cline}(2006)}]{Cline:2006ts}%
  \BibitemOpen
  \bibfield  {author} {\bibinfo {author} {\bibfnamefont {J.~M.}\ \bibnamefont {Cline}},\ }in\ \href@noop {} {\emph {\bibinfo {booktitle} {{Les Houches Summer School - Session 86: Particle Physics and Cosmology: The Fabric of Spacetime}}}}\ (\bibinfo {year} {2006})\ \Eprint {http://arxiv.org/abs/hep-ph/0609145} {arXiv:hep-ph/0609145} \BibitemShut {NoStop}%
\bibitem [{\citenamefont {Dimopoulos}\ and\ \citenamefont {Susskind}(1978)}]{Dimopoulos:1978kv}%
  \BibitemOpen
  \bibfield  {author} {\bibinfo {author} {\bibfnamefont {S.}~\bibnamefont {Dimopoulos}}\ and\ \bibinfo {author} {\bibfnamefont {L.}~\bibnamefont {Susskind}},\ }\href {\doibase 10.1103/PhysRevD.18.4500} {\bibfield  {journal} {\bibinfo  {journal} {Phys. Rev. D}\ }\textbf {\bibinfo {volume} {18}},\ \bibinfo {pages} {4500} (\bibinfo {year} {1978})}\BibitemShut {NoStop}%
\bibitem [{\citenamefont {Dimopoulos}\ and\ \citenamefont {Susskind}(1979)}]{Dimopoulos:1978pw}%
  \BibitemOpen
  \bibfield  {author} {\bibinfo {author} {\bibfnamefont {S.}~\bibnamefont {Dimopoulos}}\ and\ \bibinfo {author} {\bibfnamefont {L.}~\bibnamefont {Susskind}},\ }\href {\doibase 10.1016/0370-2693(79)90366-6} {\bibfield  {journal} {\bibinfo  {journal} {Phys. Lett. B}\ }\textbf {\bibinfo {volume} {81}},\ \bibinfo {pages} {416} (\bibinfo {year} {1979})}\BibitemShut {NoStop}%
\bibitem [{\citenamefont {Weinberg}(1979)}]{Weinberg:1979bt}%
  \BibitemOpen
  \bibfield  {author} {\bibinfo {author} {\bibfnamefont {S.}~\bibnamefont {Weinberg}},\ }\href {\doibase 10.1103/PhysRevLett.42.850} {\bibfield  {journal} {\bibinfo  {journal} {Phys. Rev. Lett.}\ }\textbf {\bibinfo {volume} {42}},\ \bibinfo {pages} {850} (\bibinfo {year} {1979})}\BibitemShut {NoStop}%
\bibitem [{\citenamefont {Riotto}\ and\ \citenamefont {Trodden}(1999)}]{Riotto:1999yt}%
  \BibitemOpen
  \bibfield  {author} {\bibinfo {author} {\bibfnamefont {A.}~\bibnamefont {Riotto}}\ and\ \bibinfo {author} {\bibfnamefont {M.}~\bibnamefont {Trodden}},\ }\href {\doibase 10.1146/annurev.nucl.49.1.35} {\bibfield  {journal} {\bibinfo  {journal} {Ann. Rev. Nucl. Part. Sci.}\ }\textbf {\bibinfo {volume} {49}},\ \bibinfo {pages} {35} (\bibinfo {year} {1999})},\ \Eprint {http://arxiv.org/abs/hep-ph/9901362} {arXiv:hep-ph/9901362} \BibitemShut {NoStop}%
\bibitem [{\citenamefont {Cordero}\ \emph {et~al.}(2025)\citenamefont {Cordero}, \citenamefont {Delgadillo}, \citenamefont {Miranda},\ and\ \citenamefont {Moura}}]{Cordero:2024hjr}%
  \BibitemOpen
  \bibfield  {author} {\bibinfo {author} {\bibfnamefont {R.}~\bibnamefont {Cordero}}, \bibinfo {author} {\bibfnamefont {L.~A.}\ \bibnamefont {Delgadillo}}, \bibinfo {author} {\bibfnamefont {O.~G.}\ \bibnamefont {Miranda}}, \ and\ \bibinfo {author} {\bibfnamefont {C.~A.}\ \bibnamefont {Moura}},\ }\href {\doibase 10.1140/epjc/s10052-024-13719-0} {\bibfield  {journal} {\bibinfo  {journal} {Eur. Phys. J. C}\ }\textbf {\bibinfo {volume} {85}},\ \bibinfo {pages} {6} (\bibinfo {year} {2025})},\ \Eprint {http://arxiv.org/abs/2407.18513} {arXiv:2407.18513 [hep-ph]} \BibitemShut {NoStop}%
\bibitem [{\citenamefont {Barenboim}\ and\ \citenamefont {Salvado}(2017)}]{Barenboim:2017vlc}%
  \BibitemOpen
  \bibfield  {author} {\bibinfo {author} {\bibfnamefont {G.}~\bibnamefont {Barenboim}}\ and\ \bibinfo {author} {\bibfnamefont {J.}~\bibnamefont {Salvado}},\ }\href {\doibase 10.1140/epjc/s10052-017-5347-y} {\bibfield  {journal} {\bibinfo  {journal} {Eur. Phys. J. C}\ }\textbf {\bibinfo {volume} {77}},\ \bibinfo {pages} {766} (\bibinfo {year} {2017})},\ \Eprint {http://arxiv.org/abs/1707.08155} {arXiv:1707.08155 [hep-ph]} \BibitemShut {NoStop}%
\bibitem [{\citenamefont {Wang}(2020)}]{Wang:2017igw}%
  \BibitemOpen
  \bibfield  {author} {\bibinfo {author} {\bibfnamefont {S.}~\bibnamefont {Wang}},\ }\href {\doibase 10.1140/epjc/s10052-020-7812-2} {\bibfield  {journal} {\bibinfo  {journal} {Eur. Phys. J. C}\ }\textbf {\bibinfo {volume} {80}},\ \bibinfo {pages} {342} (\bibinfo {year} {2020})},\ \Eprint {http://arxiv.org/abs/1712.06072} {arXiv:1712.06072 [gr-qc]} \BibitemShut {NoStop}%
\bibitem [{\citenamefont {Zhao}\ \emph {et~al.}(2015)\citenamefont {Zhao}, \citenamefont {Wang}, \citenamefont {Xia}, \citenamefont {Li},\ and\ \citenamefont {Zhang}}]{Zhao:2015mqa}%
  \BibitemOpen
  \bibfield  {author} {\bibinfo {author} {\bibfnamefont {G.-B.}\ \bibnamefont {Zhao}}, \bibinfo {author} {\bibfnamefont {Y.}~\bibnamefont {Wang}}, \bibinfo {author} {\bibfnamefont {J.-Q.}\ \bibnamefont {Xia}}, \bibinfo {author} {\bibfnamefont {M.}~\bibnamefont {Li}}, \ and\ \bibinfo {author} {\bibfnamefont {X.}~\bibnamefont {Zhang}},\ }\href {\doibase 10.1088/1475-7516/2015/07/032} {\bibfield  {journal} {\bibinfo  {journal} {JCAP}\ }\textbf {\bibinfo {volume} {07}},\ \bibinfo {pages} {032} (\bibinfo {year} {2015})},\ \Eprint {http://arxiv.org/abs/1504.04507} {arXiv:1504.04507 [astro-ph.CO]} \BibitemShut {NoStop}%
\bibitem [{\citenamefont {Balazs}(2014)}]{Balazs:2014eba}%
  \BibitemOpen
  \bibfield  {author} {\bibinfo {author} {\bibfnamefont {C.}~\bibnamefont {Balazs}},\ }\href@noop {} {\  (\bibinfo {year} {2014})},\ \Eprint {http://arxiv.org/abs/1411.3398} {arXiv:1411.3398 [hep-ph]} \BibitemShut {NoStop}%
\bibitem [{\citenamefont {Riotto}(1998)}]{Riotto:1998bt}%
  \BibitemOpen
  \bibfield  {author} {\bibinfo {author} {\bibfnamefont {A.}~\bibnamefont {Riotto}},\ }in\ \href@noop {} {\emph {\bibinfo {booktitle} {{ICTP Summer School in High-Energy Physics and Cosmology}}}}\ (\bibinfo {year} {1998})\ pp.\ \bibinfo {pages} {326--436},\ \Eprint {http://arxiv.org/abs/hep-ph/9807454} {arXiv:hep-ph/9807454} \BibitemShut {NoStop}%
\bibitem [{\citenamefont {Sakharov}(1967)}]{Sakharov:1967dj}%
  \BibitemOpen
  \bibfield  {author} {\bibinfo {author} {\bibfnamefont {A.~D.}\ \bibnamefont {Sakharov}},\ }\href {\doibase 10.1070/PU1991v034n05ABEH002497} {\bibfield  {journal} {\bibinfo  {journal} {Pisma Zh. Eksp. Teor. Fiz.}\ }\textbf {\bibinfo {volume} {5}},\ \bibinfo {pages} {32} (\bibinfo {year} {1967})}\BibitemShut {NoStop}%
\bibitem [{\citenamefont {Chung}\ \emph {et~al.}(1999)\citenamefont {Chung}, \citenamefont {Kolb},\ and\ \citenamefont {Riotto}}]{Chung:1998rq}%
  \BibitemOpen
  \bibfield  {author} {\bibinfo {author} {\bibfnamefont {D.~J.~H.}\ \bibnamefont {Chung}}, \bibinfo {author} {\bibfnamefont {E.~W.}\ \bibnamefont {Kolb}}, \ and\ \bibinfo {author} {\bibfnamefont {A.}~\bibnamefont {Riotto}},\ }\href {\doibase 10.1103/PhysRevD.60.063504} {\bibfield  {journal} {\bibinfo  {journal} {Phys. Rev. D}\ }\textbf {\bibinfo {volume} {60}},\ \bibinfo {pages} {063504} (\bibinfo {year} {1999})},\ \Eprint {http://arxiv.org/abs/hep-ph/9809453} {arXiv:hep-ph/9809453} \BibitemShut {NoStop}%
\bibitem [{\citenamefont {Cohen}\ \emph {et~al.}(1991)\citenamefont {Cohen}, \citenamefont {Kaplan},\ and\ \citenamefont {Nelson}}]{Cohen:1990it}%
  \BibitemOpen
  \bibfield  {author} {\bibinfo {author} {\bibfnamefont {A.~G.}\ \bibnamefont {Cohen}}, \bibinfo {author} {\bibfnamefont {D.~B.}\ \bibnamefont {Kaplan}}, \ and\ \bibinfo {author} {\bibfnamefont {A.~E.}\ \bibnamefont {Nelson}},\ }\href {\doibase 10.1016/0550-3213(91)90395-E} {\bibfield  {journal} {\bibinfo  {journal} {Nucl. Phys. B}\ }\textbf {\bibinfo {volume} {349}},\ \bibinfo {pages} {727} (\bibinfo {year} {1991})}\BibitemShut {NoStop}%
\bibitem [{\citenamefont {Farrar}\ and\ \citenamefont {Shaposhnikov}(1994)}]{Farrar:1993hn}%
  \BibitemOpen
  \bibfield  {author} {\bibinfo {author} {\bibfnamefont {G.~R.}\ \bibnamefont {Farrar}}\ and\ \bibinfo {author} {\bibfnamefont {M.~E.}\ \bibnamefont {Shaposhnikov}},\ }\href {\doibase 10.1103/PhysRevD.50.774} {\bibfield  {journal} {\bibinfo  {journal} {Phys. Rev. D}\ }\textbf {\bibinfo {volume} {50}},\ \bibinfo {pages} {774} (\bibinfo {year} {1994})},\ \Eprint {http://arxiv.org/abs/hep-ph/9305275} {arXiv:hep-ph/9305275} \BibitemShut {NoStop}%
\bibitem [{\citenamefont {Morrissey}\ and\ \citenamefont {Ramsey-Musolf}(2012)}]{Morrissey:2012db}%
  \BibitemOpen
  \bibfield  {author} {\bibinfo {author} {\bibfnamefont {D.~E.}\ \bibnamefont {Morrissey}}\ and\ \bibinfo {author} {\bibfnamefont {M.~J.}\ \bibnamefont {Ramsey-Musolf}},\ }\href {\doibase 10.1088/1367-2630/14/12/125003} {\bibfield  {journal} {\bibinfo  {journal} {New J. Phys.}\ }\textbf {\bibinfo {volume} {14}},\ \bibinfo {pages} {125003} (\bibinfo {year} {2012})},\ \Eprint {http://arxiv.org/abs/1206.2942} {arXiv:1206.2942 [hep-ph]} \BibitemShut {NoStop}%
\bibitem [{\citenamefont {Bodeker}\ and\ \citenamefont {Buchmuller}(2021)}]{Bodeker:2020ghk}%
  \BibitemOpen
  \bibfield  {author} {\bibinfo {author} {\bibfnamefont {D.}~\bibnamefont {Bodeker}}\ and\ \bibinfo {author} {\bibfnamefont {W.}~\bibnamefont {Buchmuller}},\ }\href {\doibase 10.1103/RevModPhys.93.035004} {\bibfield  {journal} {\bibinfo  {journal} {Rev. Mod. Phys.}\ }\textbf {\bibinfo {volume} {93}},\ \bibinfo {pages} {035004} (\bibinfo {year} {2021})},\ \Eprint {http://arxiv.org/abs/2009.07294} {arXiv:2009.07294 [hep-ph]} \BibitemShut {NoStop}%
\bibitem [{\citenamefont {Fong}\ \emph {et~al.}(2012)\citenamefont {Fong}, \citenamefont {Nardi},\ and\ \citenamefont {Riotto}}]{Fong:2012buy}%
  \BibitemOpen
  \bibfield  {author} {\bibinfo {author} {\bibfnamefont {C.~S.}\ \bibnamefont {Fong}}, \bibinfo {author} {\bibfnamefont {E.}~\bibnamefont {Nardi}}, \ and\ \bibinfo {author} {\bibfnamefont {A.}~\bibnamefont {Riotto}},\ }\href {\doibase 10.1155/2012/158303} {\bibfield  {journal} {\bibinfo  {journal} {Adv. High Energy Phys.}\ }\textbf {\bibinfo {volume} {2012}},\ \bibinfo {pages} {158303} (\bibinfo {year} {2012})},\ \Eprint {http://arxiv.org/abs/1301.3062} {arXiv:1301.3062 [hep-ph]} \BibitemShut {NoStop}%
\bibitem [{\citenamefont {Davidson}\ \emph {et~al.}(2008)\citenamefont {Davidson}, \citenamefont {Nardi},\ and\ \citenamefont {Nir}}]{Davidson:2008bu}%
  \BibitemOpen
  \bibfield  {author} {\bibinfo {author} {\bibfnamefont {S.}~\bibnamefont {Davidson}}, \bibinfo {author} {\bibfnamefont {E.}~\bibnamefont {Nardi}}, \ and\ \bibinfo {author} {\bibfnamefont {Y.}~\bibnamefont {Nir}},\ }\href {\doibase 10.1016/j.physrep.2008.06.002} {\bibfield  {journal} {\bibinfo  {journal} {Phys. Rept.}\ }\textbf {\bibinfo {volume} {466}},\ \bibinfo {pages} {105} (\bibinfo {year} {2008})},\ \Eprint {http://arxiv.org/abs/0802.2962} {arXiv:0802.2962 [hep-ph]} \BibitemShut {NoStop}%
\bibitem [{\citenamefont {Buchmuller}\ \emph {et~al.}(2003)\citenamefont {Buchmuller}, \citenamefont {Di~Bari},\ and\ \citenamefont {Plumacher}}]{Buchmuller:2003gz}%
  \BibitemOpen
  \bibfield  {author} {\bibinfo {author} {\bibfnamefont {W.}~\bibnamefont {Buchmuller}}, \bibinfo {author} {\bibfnamefont {P.}~\bibnamefont {Di~Bari}}, \ and\ \bibinfo {author} {\bibfnamefont {M.}~\bibnamefont {Plumacher}},\ }\href {\doibase 10.1016/S0550-3213(03)00449-8} {\bibfield  {journal} {\bibinfo  {journal} {Nucl. Phys. B}\ }\textbf {\bibinfo {volume} {665}},\ \bibinfo {pages} {445} (\bibinfo {year} {2003})},\ \Eprint {http://arxiv.org/abs/hep-ph/0302092} {arXiv:hep-ph/0302092} \BibitemShut {NoStop}%
\bibitem [{\citenamefont {Bettoni}\ \emph {et~al.}(2019)\citenamefont {Bettoni}, \citenamefont {Dom\`enech},\ and\ \citenamefont {Rubio}}]{Bettoni:2018pbl}%
  \BibitemOpen
  \bibfield  {author} {\bibinfo {author} {\bibfnamefont {D.}~\bibnamefont {Bettoni}}, \bibinfo {author} {\bibfnamefont {G.}~\bibnamefont {Dom\`enech}}, \ and\ \bibinfo {author} {\bibfnamefont {J.}~\bibnamefont {Rubio}},\ }\href {\doibase 10.1088/1475-7516/2019/02/034} {\bibfield  {journal} {\bibinfo  {journal} {JCAP}\ }\textbf {\bibinfo {volume} {02}},\ \bibinfo {pages} {034} (\bibinfo {year} {2019})},\ \Eprint {http://arxiv.org/abs/1810.11117} {arXiv:1810.11117 [astro-ph.CO]} \BibitemShut {NoStop}%
\bibitem [{\citenamefont {Bettoni}\ and\ \citenamefont {Rubio}(2018)}]{Bettoni:2018utf}%
  \BibitemOpen
  \bibfield  {author} {\bibinfo {author} {\bibfnamefont {D.}~\bibnamefont {Bettoni}}\ and\ \bibinfo {author} {\bibfnamefont {J.}~\bibnamefont {Rubio}},\ }\href {\doibase 10.1016/j.physletb.2018.07.046} {\bibfield  {journal} {\bibinfo  {journal} {Phys. Lett. B}\ }\textbf {\bibinfo {volume} {784}},\ \bibinfo {pages} {122} (\bibinfo {year} {2018})},\ \Eprint {http://arxiv.org/abs/1805.02669} {arXiv:1805.02669 [astro-ph.CO]} \BibitemShut {NoStop}%
\bibitem [{\citenamefont {Bettoni}\ and\ \citenamefont {Rubio}(2020)}]{Bettoni:2019dcw}%
  \BibitemOpen
  \bibfield  {author} {\bibinfo {author} {\bibfnamefont {D.}~\bibnamefont {Bettoni}}\ and\ \bibinfo {author} {\bibfnamefont {J.}~\bibnamefont {Rubio}},\ }\href {\doibase 10.1088/1475-7516/2020/01/002} {\bibfield  {journal} {\bibinfo  {journal} {JCAP}\ }\textbf {\bibinfo {volume} {01}},\ \bibinfo {pages} {002} (\bibinfo {year} {2020})},\ \Eprint {http://arxiv.org/abs/1911.03484} {arXiv:1911.03484 [astro-ph.CO]} \BibitemShut {NoStop}%
\bibitem [{\citenamefont {Bettoni}\ \emph {et~al.}(2022)\citenamefont {Bettoni}, \citenamefont {Lopez-Eiguren},\ and\ \citenamefont {Rubio}}]{Bettoni:2021zhq}%
  \BibitemOpen
  \bibfield  {author} {\bibinfo {author} {\bibfnamefont {D.}~\bibnamefont {Bettoni}}, \bibinfo {author} {\bibfnamefont {A.}~\bibnamefont {Lopez-Eiguren}}, \ and\ \bibinfo {author} {\bibfnamefont {J.}~\bibnamefont {Rubio}},\ }\href {\doibase 10.1088/1475-7516/2022/01/002} {\bibfield  {journal} {\bibinfo  {journal} {JCAP}\ }\textbf {\bibinfo {volume} {01}},\ \bibinfo {pages} {002} (\bibinfo {year} {2022})},\ \Eprint {http://arxiv.org/abs/2107.09671} {arXiv:2107.09671 [hep-ph]} \BibitemShut {NoStop}%
\bibitem [{\citenamefont {Bettoni}\ \emph {et~al.}(2025)\citenamefont {Bettoni}, \citenamefont {Laverda}, \citenamefont {Eiguren},\ and\ \citenamefont {Rubio}}]{Bettoni:2024ixe}%
  \BibitemOpen
  \bibfield  {author} {\bibinfo {author} {\bibfnamefont {D.}~\bibnamefont {Bettoni}}, \bibinfo {author} {\bibfnamefont {G.}~\bibnamefont {Laverda}}, \bibinfo {author} {\bibfnamefont {A.~L.}\ \bibnamefont {Eiguren}}, \ and\ \bibinfo {author} {\bibfnamefont {J.}~\bibnamefont {Rubio}},\ }\href {\doibase 10.1088/1475-7516/2025/03/027} {\bibfield  {journal} {\bibinfo  {journal} {JCAP}\ }\textbf {\bibinfo {volume} {03}},\ \bibinfo {pages} {027} (\bibinfo {year} {2025})},\ \Eprint {http://arxiv.org/abs/2409.15450} {arXiv:2409.15450 [gr-qc]} \BibitemShut {NoStop}%
\bibitem [{\citenamefont {Laverda}\ and\ \citenamefont {Rubio}(2024)}]{Laverda:2023uqv}%
  \BibitemOpen
  \bibfield  {author} {\bibinfo {author} {\bibfnamefont {G.}~\bibnamefont {Laverda}}\ and\ \bibinfo {author} {\bibfnamefont {J.}~\bibnamefont {Rubio}},\ }\href {\doibase 10.1088/1475-7516/2024/03/033} {\bibfield  {journal} {\bibinfo  {journal} {JCAP}\ }\textbf {\bibinfo {volume} {03}},\ \bibinfo {pages} {033} (\bibinfo {year} {2024})},\ \bibinfo {note} {[Erratum: JCAP 06, E01 (2024)]},\ \Eprint {http://arxiv.org/abs/2307.03774} {arXiv:2307.03774 [astro-ph.CO]} \BibitemShut {NoStop}%
\bibitem [{\citenamefont {Cline}\ \emph {et~al.}(2020)\citenamefont {Cline}, \citenamefont {Puel},\ and\ \citenamefont {Toma}}]{Cline:2019fxx}%
  \BibitemOpen
  \bibfield  {author} {\bibinfo {author} {\bibfnamefont {J.~M.}\ \bibnamefont {Cline}}, \bibinfo {author} {\bibfnamefont {M.}~\bibnamefont {Puel}}, \ and\ \bibinfo {author} {\bibfnamefont {T.}~\bibnamefont {Toma}},\ }\href {\doibase 10.1103/PhysRevD.101.043014} {\bibfield  {journal} {\bibinfo  {journal} {Phys. Rev. D}\ }\textbf {\bibinfo {volume} {101}},\ \bibinfo {pages} {043014} (\bibinfo {year} {2020})},\ \Eprint {http://arxiv.org/abs/1909.12300} {arXiv:1909.12300 [hep-ph]} \BibitemShut {NoStop}%
\bibitem [{\citenamefont {Lloyd-Stubbs}\ and\ \citenamefont {McDonald}(2021)}]{Lloyd-Stubbs:2020sed}%
  \BibitemOpen
  \bibfield  {author} {\bibinfo {author} {\bibfnamefont {A.}~\bibnamefont {Lloyd-Stubbs}}\ and\ \bibinfo {author} {\bibfnamefont {J.}~\bibnamefont {McDonald}},\ }\href {\doibase 10.1103/PhysRevD.103.123514} {\bibfield  {journal} {\bibinfo  {journal} {Phys. Rev. D}\ }\textbf {\bibinfo {volume} {103}},\ \bibinfo {pages} {123514} (\bibinfo {year} {2021})},\ \Eprint {http://arxiv.org/abs/2008.04339} {arXiv:2008.04339 [hep-ph]} \BibitemShut {NoStop}%
\bibitem [{\citenamefont {Davoudiasl}\ \emph {et~al.}(2004)\citenamefont {Davoudiasl}, \citenamefont {Kitano}, \citenamefont {Kribs}, \citenamefont {Murayama},\ and\ \citenamefont {Steinhardt}}]{Davoudiasl:2004gf}%
  \BibitemOpen
  \bibfield  {author} {\bibinfo {author} {\bibfnamefont {H.}~\bibnamefont {Davoudiasl}}, \bibinfo {author} {\bibfnamefont {R.}~\bibnamefont {Kitano}}, \bibinfo {author} {\bibfnamefont {G.~D.}\ \bibnamefont {Kribs}}, \bibinfo {author} {\bibfnamefont {H.}~\bibnamefont {Murayama}}, \ and\ \bibinfo {author} {\bibfnamefont {P.~J.}\ \bibnamefont {Steinhardt}},\ }\href {\doibase 10.1103/PhysRevLett.93.201301} {\bibfield  {journal} {\bibinfo  {journal} {Phys. Rev. Lett.}\ }\textbf {\bibinfo {volume} {93}},\ \bibinfo {pages} {201301} (\bibinfo {year} {2004})},\ \Eprint {http://arxiv.org/abs/hep-ph/0403019} {arXiv:hep-ph/0403019} \BibitemShut {NoStop}%
\bibitem [{\citenamefont {Goodarzi}(2023)}]{Goodarzi:2023ltp}%
  \BibitemOpen
  \bibfield  {author} {\bibinfo {author} {\bibfnamefont {P.}~\bibnamefont {Goodarzi}},\ }\href {\doibase 10.1140/epjc/s10052-023-12182-7} {\bibfield  {journal} {\bibinfo  {journal} {Eur. Phys. J. C}\ }\textbf {\bibinfo {volume} {83}},\ \bibinfo {pages} {990} (\bibinfo {year} {2023})},\ \Eprint {http://arxiv.org/abs/2307.10709} {arXiv:2307.10709 [hep-th]} \BibitemShut {NoStop}%
\bibitem [{\citenamefont {Arbuzova}\ and\ \citenamefont {Dolgov}(2019)}]{Arbuzova:2017zby}%
  \BibitemOpen
  \bibfield  {author} {\bibinfo {author} {\bibfnamefont {E.~V.}\ \bibnamefont {Arbuzova}}\ and\ \bibinfo {author} {\bibfnamefont {A.~D.}\ \bibnamefont {Dolgov}},\ }in\ \href {\doibase 10.1142/9789811202339_0059} {\emph {\bibinfo {booktitle} {{18th Lomonosov Conference on Elementary Particle Physics}}}}\ (\bibinfo {year} {2019})\ pp.\ \bibinfo {pages} {309--313},\ \Eprint {http://arxiv.org/abs/1712.04627} {arXiv:1712.04627 [hep-ph]} \BibitemShut {NoStop}%
\bibitem [{\citenamefont {Mojahed}\ \emph {et~al.}(2025)\citenamefont {Mojahed}, \citenamefont {Schmitz},\ and\ \citenamefont {Xu}}]{Mojahed:2024mvb}%
  \BibitemOpen
  \bibfield  {author} {\bibinfo {author} {\bibfnamefont {M.~A.}\ \bibnamefont {Mojahed}}, \bibinfo {author} {\bibfnamefont {K.}~\bibnamefont {Schmitz}}, \ and\ \bibinfo {author} {\bibfnamefont {X.-J.}\ \bibnamefont {Xu}},\ }\href {\doibase 10.1103/PhysRevD.111.055005} {\bibfield  {journal} {\bibinfo  {journal} {Phys. Rev. D}\ }\textbf {\bibinfo {volume} {111}},\ \bibinfo {pages} {055005} (\bibinfo {year} {2025})},\ \Eprint {http://arxiv.org/abs/2409.10605} {arXiv:2409.10605 [hep-ph]} \BibitemShut {NoStop}%
\bibitem [{\citenamefont {Sadjadi}(2007)}]{Sadjadi:2007dx}%
  \BibitemOpen
  \bibfield  {author} {\bibinfo {author} {\bibfnamefont {H.~M.}\ \bibnamefont {Sadjadi}},\ }\href {\doibase 10.1103/PhysRevD.76.123507} {\bibfield  {journal} {\bibinfo  {journal} {Phys. Rev. D}\ }\textbf {\bibinfo {volume} {76}},\ \bibinfo {pages} {123507} (\bibinfo {year} {2007})},\ \Eprint {http://arxiv.org/abs/0709.0697} {arXiv:0709.0697 [gr-qc]} \BibitemShut {NoStop}%
\bibitem [{\citenamefont {Arbuzova}(2018)}]{Arbuzova:2018hcj}%
  \BibitemOpen
  \bibfield  {author} {\bibinfo {author} {\bibfnamefont {E.~V.}\ \bibnamefont {Arbuzova}},\ }\href {\doibase 10.1142/S0217751X18440232} {\bibfield  {journal} {\bibinfo  {journal} {Int. J. Mod. Phys. A}\ }\textbf {\bibinfo {volume} {33}},\ \bibinfo {pages} {1844023} (\bibinfo {year} {2018})},\ \Eprint {http://arxiv.org/abs/1808.08577} {arXiv:1808.08577 [gr-qc]} \BibitemShut {NoStop}%
\bibitem [{\citenamefont {Dolgov}\ and\ \citenamefont {Freese}(1995)}]{Dolgov:1994zq}%
  \BibitemOpen
  \bibfield  {author} {\bibinfo {author} {\bibfnamefont {A.}~\bibnamefont {Dolgov}}\ and\ \bibinfo {author} {\bibfnamefont {K.}~\bibnamefont {Freese}},\ }\href {\doibase 10.1103/PhysRevD.51.2693} {\bibfield  {journal} {\bibinfo  {journal} {Phys. Rev. D}\ }\textbf {\bibinfo {volume} {51}},\ \bibinfo {pages} {2693} (\bibinfo {year} {1995})},\ \Eprint {http://arxiv.org/abs/hep-ph/9410346} {arXiv:hep-ph/9410346} \BibitemShut {NoStop}%
\bibitem [{\citenamefont {Dolgov}\ \emph {et~al.}(1997)\citenamefont {Dolgov}, \citenamefont {Freese}, \citenamefont {Rangarajan},\ and\ \citenamefont {Srednicki}}]{Dolgov:1996qq}%
  \BibitemOpen
  \bibfield  {author} {\bibinfo {author} {\bibfnamefont {A.}~\bibnamefont {Dolgov}}, \bibinfo {author} {\bibfnamefont {K.}~\bibnamefont {Freese}}, \bibinfo {author} {\bibfnamefont {R.}~\bibnamefont {Rangarajan}}, \ and\ \bibinfo {author} {\bibfnamefont {M.}~\bibnamefont {Srednicki}},\ }\href {\doibase 10.1103/PhysRevD.56.6155} {\bibfield  {journal} {\bibinfo  {journal} {Phys. Rev. D}\ }\textbf {\bibinfo {volume} {56}},\ \bibinfo {pages} {6155} (\bibinfo {year} {1997})},\ \Eprint {http://arxiv.org/abs/hep-ph/9610405} {arXiv:hep-ph/9610405} \BibitemShut {NoStop}%
\bibitem [{\citenamefont {Luongo}\ \emph {et~al.}(2023)\citenamefont {Luongo}, \citenamefont {Marcantognini},\ and\ \citenamefont {Muccino}}]{Luongo:2021gho}%
  \BibitemOpen
  \bibfield  {author} {\bibinfo {author} {\bibfnamefont {O.}~\bibnamefont {Luongo}}, \bibinfo {author} {\bibfnamefont {N.}~\bibnamefont {Marcantognini}}, \ and\ \bibinfo {author} {\bibfnamefont {M.}~\bibnamefont {Muccino}},\ }\href {\doibase 10.1007/s10714-023-03079-7} {\bibfield  {journal} {\bibinfo  {journal} {Gen. Rel. Grav.}\ }\textbf {\bibinfo {volume} {55}},\ \bibinfo {pages} {33} (\bibinfo {year} {2023})},\ \Eprint {http://arxiv.org/abs/2112.05730} {arXiv:2112.05730 [hep-ph]} \BibitemShut {NoStop}%
\bibitem [{\citenamefont {De~Simone}\ and\ \citenamefont {Kobayashi}(2016)}]{DeSimone:2016ofp}%
  \BibitemOpen
  \bibfield  {author} {\bibinfo {author} {\bibfnamefont {A.}~\bibnamefont {De~Simone}}\ and\ \bibinfo {author} {\bibfnamefont {T.}~\bibnamefont {Kobayashi}},\ }\href {\doibase 10.1088/1475-7516/2016/08/052} {\bibfield  {journal} {\bibinfo  {journal} {JCAP}\ }\textbf {\bibinfo {volume} {08}},\ \bibinfo {pages} {052} (\bibinfo {year} {2016})},\ \Eprint {http://arxiv.org/abs/1605.00670} {arXiv:1605.00670 [hep-ph]} \BibitemShut {NoStop}%
\bibitem [{\citenamefont {Dubbini}\ \emph {et~al.}(2025{\natexlab{a}})\citenamefont {Dubbini}, \citenamefont {Luongo},\ and\ \citenamefont {Muccino}}]{Dubbini:2025jjz}%
  \BibitemOpen
  \bibfield  {author} {\bibinfo {author} {\bibfnamefont {M.}~\bibnamefont {Dubbini}}, \bibinfo {author} {\bibfnamefont {O.}~\bibnamefont {Luongo}}, \ and\ \bibinfo {author} {\bibfnamefont {M.}~\bibnamefont {Muccino}},\ }\href@noop {} {\  (\bibinfo {year} {2025}{\natexlab{a}})},\ \Eprint {http://arxiv.org/abs/2505.03644} {arXiv:2505.03644 [gr-qc]} \BibitemShut {NoStop}%
\bibitem [{\citenamefont {Dubbini}\ \emph {et~al.}(2025{\natexlab{b}})\citenamefont {Dubbini}, \citenamefont {Luongo},\ and\ \citenamefont {Muccino}}]{Dubbini:2025hcw}%
  \BibitemOpen
  \bibfield  {author} {\bibinfo {author} {\bibfnamefont {M.}~\bibnamefont {Dubbini}}, \bibinfo {author} {\bibfnamefont {O.}~\bibnamefont {Luongo}}, \ and\ \bibinfo {author} {\bibfnamefont {M.}~\bibnamefont {Muccino}},\ }\href@noop {} {\  (\bibinfo {year} {2025}{\natexlab{b}})},\ \Eprint {http://arxiv.org/abs/2507.01112} {arXiv:2507.01112 [gr-qc]} \BibitemShut {NoStop}%
\bibitem [{\citenamefont {Jejelava}\ and\ \citenamefont {Kepuladze}(2025)}]{Jejelava:2025ijk}%
  \BibitemOpen
  \bibfield  {author} {\bibinfo {author} {\bibfnamefont {J.}~\bibnamefont {Jejelava}}\ and\ \bibinfo {author} {\bibfnamefont {Z.}~\bibnamefont {Kepuladze}},\ }\href@noop {} {\  (\bibinfo {year} {2025})},\ \Eprint {http://arxiv.org/abs/2504.11248} {arXiv:2504.11248 [hep-ph]} \BibitemShut {NoStop}%
\bibitem [{\citenamefont {Collins}\ \emph {et~al.}(2004{\natexlab{a}})\citenamefont {Collins}, \citenamefont {Perez}, \citenamefont {Sudarsky}, \citenamefont {Urrutia},\ and\ \citenamefont {Vucetich}}]{Collins:2004bp}%
  \BibitemOpen
  \bibfield  {author} {\bibinfo {author} {\bibfnamefont {J.}~\bibnamefont {Collins}}, \bibinfo {author} {\bibfnamefont {A.}~\bibnamefont {Perez}}, \bibinfo {author} {\bibfnamefont {D.}~\bibnamefont {Sudarsky}}, \bibinfo {author} {\bibfnamefont {L.}~\bibnamefont {Urrutia}}, \ and\ \bibinfo {author} {\bibfnamefont {H.}~\bibnamefont {Vucetich}},\ }\href {\doibase 10.1103/PhysRevLett.93.191301} {\bibfield  {journal} {\bibinfo  {journal} {Phys. Rev. Lett.}\ }\textbf {\bibinfo {volume} {93}},\ \bibinfo {pages} {191301} (\bibinfo {year} {2004}{\natexlab{a}})},\ \Eprint {http://arxiv.org/abs/gr-qc/0403053} {arXiv:gr-qc/0403053} \BibitemShut {NoStop}%
\bibitem [{\citenamefont {Alfaro}(2005)}]{Alfaro:2004aa}%
  \BibitemOpen
  \bibfield  {author} {\bibinfo {author} {\bibfnamefont {J.}~\bibnamefont {Alfaro}},\ }\href {\doibase 10.1103/PhysRevLett.94.221302} {\bibfield  {journal} {\bibinfo  {journal} {Phys. Rev. Lett.}\ }\textbf {\bibinfo {volume} {94}},\ \bibinfo {pages} {221302} (\bibinfo {year} {2005})},\ \Eprint {http://arxiv.org/abs/hep-th/0412295} {arXiv:hep-th/0412295} \BibitemShut {NoStop}%
\bibitem [{\citenamefont {Mavromatos}(2007)}]{Mavromatos:2007xe}%
  \BibitemOpen
  \bibfield  {author} {\bibinfo {author} {\bibfnamefont {N.~E.}\ \bibnamefont {Mavromatos}},\ }\href {\doibase 10.22323/1.043.0027} {\bibfield  {journal} {\bibinfo  {journal} {PoS}\ }\textbf {\bibinfo {volume} {QG-PH}},\ \bibinfo {pages} {027} (\bibinfo {year} {2007})},\ \Eprint {http://arxiv.org/abs/0708.2250} {arXiv:0708.2250 [hep-th]} \BibitemShut {NoStop}%
\bibitem [{\citenamefont {Li}\ and\ \citenamefont {Ma}(2025)}]{Li:2025yvq}%
  \BibitemOpen
  \bibfield  {author} {\bibinfo {author} {\bibfnamefont {C.}~\bibnamefont {Li}}\ and\ \bibinfo {author} {\bibfnamefont {B.-Q.}\ \bibnamefont {Ma}},\ }\href {\doibase 10.3390/sym17060974} {\bibfield  {journal} {\bibinfo  {journal} {Symmetry}\ }\textbf {\bibinfo {volume} {17}},\ \bibinfo {pages} {974} (\bibinfo {year} {2025})},\ \Eprint {http://arxiv.org/abs/2508.11172} {arXiv:2508.11172 [hep-ph]} \BibitemShut {NoStop}%
\bibitem [{\citenamefont {Kepuladze}(2025)}]{Kepuladze:2025czd}%
  \BibitemOpen
  \bibfield  {author} {\bibinfo {author} {\bibfnamefont {Z.}~\bibnamefont {Kepuladze}},\ }\href@noop {} {\  (\bibinfo {year} {2025})},\ \Eprint {http://arxiv.org/abs/2504.15608} {arXiv:2504.15608 [hep-ph]} \BibitemShut {NoStop}%
\bibitem [{\citenamefont {Williams}(2024)}]{Williams:2024woi}%
  \BibitemOpen
  \bibfield  {author} {\bibinfo {author} {\bibfnamefont {H.}~\bibnamefont {Williams}},\ }\href {\doibase 10.5506/APhysPolB.55.11-A3} {\bibfield  {journal} {\bibinfo  {journal} {Acta Phys. Polon. B}\ }\textbf {\bibinfo {volume} {55}},\ \bibinfo {pages} {11} (\bibinfo {year} {2024})},\ \Eprint {http://arxiv.org/abs/2502.00890} {arXiv:2502.00890 [hep-ph]} \BibitemShut {NoStop}%
\bibitem [{\citenamefont {Piran}\ and\ \citenamefont {Ofengeim}(2024)}]{Piran:2023xfg}%
  \BibitemOpen
  \bibfield  {author} {\bibinfo {author} {\bibfnamefont {T.}~\bibnamefont {Piran}}\ and\ \bibinfo {author} {\bibfnamefont {D.~D.}\ \bibnamefont {Ofengeim}},\ }\href {\doibase 10.1103/PhysRevD.109.L081501} {\bibfield  {journal} {\bibinfo  {journal} {Phys. Rev. D}\ }\textbf {\bibinfo {volume} {109}},\ \bibinfo {pages} {L081501} (\bibinfo {year} {2024})},\ \Eprint {http://arxiv.org/abs/2308.03031} {arXiv:2308.03031 [astro-ph.HE]} \BibitemShut {NoStop}%
\bibitem [{\citenamefont {Xi}\ and\ \citenamefont {Shu}(2025)}]{Xi:2025ruv}%
  \BibitemOpen
  \bibfield  {author} {\bibinfo {author} {\bibfnamefont {Y.}~\bibnamefont {Xi}}\ and\ \bibinfo {author} {\bibfnamefont {F.-W.}\ \bibnamefont {Shu}},\ }\href@noop {} {\  (\bibinfo {year} {2025})},\ \Eprint {http://arxiv.org/abs/2508.00656} {arXiv:2508.00656 [astro-ph.HE]} \BibitemShut {NoStop}%
\bibitem [{\citenamefont {Yang}\ \emph {et~al.}(2025)\citenamefont {Yang}, \citenamefont {Lv}, \citenamefont {Bi},\ and\ \citenamefont {Yin}}]{Yang:2025kfr}%
  \BibitemOpen
  \bibfield  {author} {\bibinfo {author} {\bibfnamefont {Y.-M.}\ \bibnamefont {Yang}}, \bibinfo {author} {\bibfnamefont {X.-J.}\ \bibnamefont {Lv}}, \bibinfo {author} {\bibfnamefont {X.-J.}\ \bibnamefont {Bi}}, \ and\ \bibinfo {author} {\bibfnamefont {P.-F.}\ \bibnamefont {Yin}},\ }\href {\doibase 10.1103/6zzg-tv4s} {\bibfield  {journal} {\bibinfo  {journal} {Phys. Rev. D}\ }\textbf {\bibinfo {volume} {111}},\ \bibinfo {pages} {123037} (\bibinfo {year} {2025})},\ \Eprint {http://arxiv.org/abs/2502.18256} {arXiv:2502.18256 [hep-ph]} \BibitemShut {NoStop}%
\bibitem [{\citenamefont {Desai}(2024)}]{Desai:2023rkd}%
  \BibitemOpen
  \bibfield  {author} {\bibinfo {author} {\bibfnamefont {S.}~\bibnamefont {Desai}},\ }\enquote {\bibinfo {title} {{Astrophysical and Cosmological Searches for Lorentz Invariance Violation}},}\ \ (\bibinfo {year} {2024})\ \Eprint {http://arxiv.org/abs/2303.10643} {arXiv:2303.10643 [astro-ph.CO]} \BibitemShut {NoStop}%
\bibitem [{\citenamefont {Xiao}\ and\ \citenamefont {Ma}(2009)}]{Xiao:2009xe}%
  \BibitemOpen
  \bibfield  {author} {\bibinfo {author} {\bibfnamefont {Z.}~\bibnamefont {Xiao}}\ and\ \bibinfo {author} {\bibfnamefont {B.-Q.}\ \bibnamefont {Ma}},\ }\href {\doibase 10.1103/PhysRevD.80.116005} {\bibfield  {journal} {\bibinfo  {journal} {Phys. Rev. D}\ }\textbf {\bibinfo {volume} {80}},\ \bibinfo {pages} {116005} (\bibinfo {year} {2009})},\ \Eprint {http://arxiv.org/abs/0909.4927} {arXiv:0909.4927 [hep-ph]} \BibitemShut {NoStop}%
\bibitem [{\citenamefont {Carroll}\ and\ \citenamefont {Shu}(2006)}]{Carroll:2005dj}%
  \BibitemOpen
  \bibfield  {author} {\bibinfo {author} {\bibfnamefont {S.~M.}\ \bibnamefont {Carroll}}\ and\ \bibinfo {author} {\bibfnamefont {J.}~\bibnamefont {Shu}},\ }\href {\doibase 10.1103/PhysRevD.73.103515} {\bibfield  {journal} {\bibinfo  {journal} {Phys. Rev. D}\ }\textbf {\bibinfo {volume} {73}},\ \bibinfo {pages} {103515} (\bibinfo {year} {2006})},\ \Eprint {http://arxiv.org/abs/hep-ph/0510081} {arXiv:hep-ph/0510081} \BibitemShut {NoStop}%
\bibitem [{\citenamefont {Shu}(2008)}]{Shu:2007wi}%
  \BibitemOpen
  \bibfield  {author} {\bibinfo {author} {\bibfnamefont {J.}~\bibnamefont {Shu}},\ }in\ \href {\doibase 10.1142/9789812779519_0037} {\emph {\bibinfo {booktitle} {{4th Meeting on CPT and Lorentz Symmetry}}}}\ (\bibinfo {year} {2008})\ pp.\ \bibinfo {pages} {244--249},\ \Eprint {http://arxiv.org/abs/0711.2519} {arXiv:0711.2519 [hep-ph]} \BibitemShut {NoStop}%
\bibitem [{\citenamefont {Alfaro}\ and\ \citenamefont {Gonzalez}(2009)}]{Alfaro:2009xc}%
  \BibitemOpen
  \bibfield  {author} {\bibinfo {author} {\bibfnamefont {J.}~\bibnamefont {Alfaro}}\ and\ \bibinfo {author} {\bibfnamefont {P.}~\bibnamefont {Gonzalez}},\ }\href@noop {} {\  (\bibinfo {year} {2009})},\ \Eprint {http://arxiv.org/abs/0909.3883} {arXiv:0909.3883 [hep-ph]} \BibitemShut {NoStop}%
\bibitem [{\citenamefont {Adams}\ \emph {et~al.}(1993)\citenamefont {Adams}, \citenamefont {Bond}, \citenamefont {Freese}, \citenamefont {Frieman},\ and\ \citenamefont {Olinto}}]{Adams:1992bn}%
  \BibitemOpen
  \bibfield  {author} {\bibinfo {author} {\bibfnamefont {F.~C.}\ \bibnamefont {Adams}}, \bibinfo {author} {\bibfnamefont {J.~R.}\ \bibnamefont {Bond}}, \bibinfo {author} {\bibfnamefont {K.}~\bibnamefont {Freese}}, \bibinfo {author} {\bibfnamefont {J.~A.}\ \bibnamefont {Frieman}}, \ and\ \bibinfo {author} {\bibfnamefont {A.~V.}\ \bibnamefont {Olinto}},\ }\href {\doibase 10.1103/PhysRevD.47.426} {\bibfield  {journal} {\bibinfo  {journal} {Phys. Rev. D}\ }\textbf {\bibinfo {volume} {47}},\ \bibinfo {pages} {426} (\bibinfo {year} {1993})},\ \Eprint {http://arxiv.org/abs/hep-ph/9207245} {arXiv:hep-ph/9207245} \BibitemShut {NoStop}%
\bibitem [{\citenamefont {Colladay}\ and\ \citenamefont {Kosteleck\'y}(1998)}]{Kost98}%
  \BibitemOpen
  \bibfield  {author} {\bibinfo {author} {\bibfnamefont {D.}~\bibnamefont {Colladay}}\ and\ \bibinfo {author} {\bibfnamefont {V.~A.}\ \bibnamefont {Kosteleck\'y}},\ }\href {\doibase 10.1103/PhysRevD.58.116002} {\bibfield  {journal} {\bibinfo  {journal} {Phys. Rev. D}\ }\textbf {\bibinfo {volume} {58}},\ \bibinfo {pages} {116002} (\bibinfo {year} {1998})}\BibitemShut {NoStop}%
\bibitem [{\citenamefont {Abbott}\ \emph {et~al.}(2017)\citenamefont {Abbott} \emph {et~al.}}]{LIGOScientific:2017zic}%
  \BibitemOpen
  \bibfield  {author} {\bibinfo {author} {\bibfnamefont {B.~P.}\ \bibnamefont {Abbott}} \emph {et~al.} (\bibinfo {collaboration} {LIGO Scientific, Virgo, Fermi-GBM, INTEGRAL}),\ }\href {\doibase 10.3847/2041-8213/aa920c} {\bibfield  {journal} {\bibinfo  {journal} {Astrophys. J. Lett.}\ }\textbf {\bibinfo {volume} {848}},\ \bibinfo {pages} {L13} (\bibinfo {year} {2017})},\ \Eprint {http://arxiv.org/abs/1710.05834} {arXiv:1710.05834 [astro-ph.HE]} \BibitemShut {NoStop}%
\bibitem [{\citenamefont {Ellis}\ \emph {et~al.}(2019)\citenamefont {Ellis}, \citenamefont {Mavromatos}, \citenamefont {Sakharov},\ and\ \citenamefont {Sarkisyan-Grinbaum}}]{Ellis:2018ogq}%
  \BibitemOpen
  \bibfield  {author} {\bibinfo {author} {\bibfnamefont {J.}~\bibnamefont {Ellis}}, \bibinfo {author} {\bibfnamefont {N.~E.}\ \bibnamefont {Mavromatos}}, \bibinfo {author} {\bibfnamefont {A.~S.}\ \bibnamefont {Sakharov}}, \ and\ \bibinfo {author} {\bibfnamefont {E.~K.}\ \bibnamefont {Sarkisyan-Grinbaum}},\ }\href {\doibase 10.1016/j.physletb.2018.11.062} {\bibfield  {journal} {\bibinfo  {journal} {Phys. Lett. B}\ }\textbf {\bibinfo {volume} {789}},\ \bibinfo {pages} {352} (\bibinfo {year} {2019})},\ \Eprint {http://arxiv.org/abs/1807.05155} {arXiv:1807.05155 [astro-ph.HE]} \BibitemShut {NoStop}%
\bibitem [{\citenamefont {Maccione}\ \emph {et~al.}(2009)\citenamefont {Maccione}, \citenamefont {Taylor}, \citenamefont {Mattingly},\ and\ \citenamefont {Liberati}}]{Maccione:2009ju}%
  \BibitemOpen
  \bibfield  {author} {\bibinfo {author} {\bibfnamefont {L.}~\bibnamefont {Maccione}}, \bibinfo {author} {\bibfnamefont {A.~M.}\ \bibnamefont {Taylor}}, \bibinfo {author} {\bibfnamefont {D.~M.}\ \bibnamefont {Mattingly}}, \ and\ \bibinfo {author} {\bibfnamefont {S.}~\bibnamefont {Liberati}},\ }\href {\doibase 10.1088/1475-7516/2009/04/022} {\bibfield  {journal} {\bibinfo  {journal} {JCAP}\ }\textbf {\bibinfo {volume} {04}},\ \bibinfo {pages} {022} (\bibinfo {year} {2009})},\ \Eprint {http://arxiv.org/abs/0902.1756} {arXiv:0902.1756 [astro-ph.HE]} \BibitemShut {NoStop}%
\bibitem [{\citenamefont {Bluhm}\ and\ \citenamefont {Kosteleck\'y}(2005)}]{SPLV1}%
  \BibitemOpen
  \bibfield  {author} {\bibinfo {author} {\bibfnamefont {R.}~\bibnamefont {Bluhm}}\ and\ \bibinfo {author} {\bibfnamefont {V.~A.}\ \bibnamefont {Kosteleck\'y}},\ }\href {\doibase 10.1103/PhysRevD.71.065008} {\bibfield  {journal} {\bibinfo  {journal} {Phys. Rev. D}\ }\textbf {\bibinfo {volume} {71}},\ \bibinfo {pages} {065008} (\bibinfo {year} {2005})}\BibitemShut {NoStop}%
\bibitem [{\citenamefont {Seifert}(2009)}]{SPLV2}%
  \BibitemOpen
  \bibfield  {author} {\bibinfo {author} {\bibfnamefont {M.~D.}\ \bibnamefont {Seifert}},\ }\href {\doibase 10.1103/PhysRevD.79.124012} {\bibfield  {journal} {\bibinfo  {journal} {Phys. Rev. D}\ }\textbf {\bibinfo {volume} {79}},\ \bibinfo {pages} {124012} (\bibinfo {year} {2009})}\BibitemShut {NoStop}%
\bibitem [{\citenamefont {Bluhm}\ \emph {et~al.}(2008)\citenamefont {Bluhm}, \citenamefont {Fung},\ and\ \citenamefont {Kosteleck\'y}}]{SPLV3}%
  \BibitemOpen
  \bibfield  {author} {\bibinfo {author} {\bibfnamefont {R.}~\bibnamefont {Bluhm}}, \bibinfo {author} {\bibfnamefont {S.-H.}\ \bibnamefont {Fung}}, \ and\ \bibinfo {author} {\bibfnamefont {V.~A.}\ \bibnamefont {Kosteleck\'y}},\ }\href {\doibase 10.1103/PhysRevD.77.065020} {\bibfield  {journal} {\bibinfo  {journal} {Phys. Rev. D}\ }\textbf {\bibinfo {volume} {77}},\ \bibinfo {pages} {065020} (\bibinfo {year} {2008})}\BibitemShut {NoStop}%
\bibitem [{\citenamefont {Weinberg}(1974)}]{WeinbergISB}%
  \BibitemOpen
  \bibfield  {author} {\bibinfo {author} {\bibfnamefont {S.}~\bibnamefont {Weinberg}},\ }\href {\doibase 10.1103/PhysRevD.9.3357} {\bibfield  {journal} {\bibinfo  {journal} {Phys. Rev. D}\ }\textbf {\bibinfo {volume} {9}},\ \bibinfo {pages} {3357} (\bibinfo {year} {1974})}\BibitemShut {NoStop}%
\bibitem [{\citenamefont {Jansen}\ and\ \citenamefont {Laine}(1998)}]{ISB0}%
  \BibitemOpen
  \bibfield  {author} {\bibinfo {author} {\bibfnamefont {K.}~\bibnamefont {Jansen}}\ and\ \bibinfo {author} {\bibfnamefont {M.}~\bibnamefont {Laine}},\ }\href {\doibase https://doi.org/10.1016/S0370-2693(98)00775-8} {\bibfield  {journal} {\bibinfo  {journal} {Physics Letters B}\ }\textbf {\bibinfo {volume} {435}},\ \bibinfo {pages} {166} (\bibinfo {year} {1998})}\BibitemShut {NoStop}%
\bibitem [{\citenamefont {Pietroni}\ \emph {et~al.}(1997)\citenamefont {Pietroni}, \citenamefont {Rius},\ and\ \citenamefont {Tetradis}}]{ISB1}%
  \BibitemOpen
  \bibfield  {author} {\bibinfo {author} {\bibfnamefont {M.}~\bibnamefont {Pietroni}}, \bibinfo {author} {\bibfnamefont {N.}~\bibnamefont {Rius}}, \ and\ \bibinfo {author} {\bibfnamefont {N.}~\bibnamefont {Tetradis}},\ }\href {\doibase https://doi.org/10.1016/S0370-2693(97)00150-0} {\bibfield  {journal} {\bibinfo  {journal} {Physics Letters B}\ }\textbf {\bibinfo {volume} {397}},\ \bibinfo {pages} {119} (\bibinfo {year} {1997})}\BibitemShut {NoStop}%
\bibitem [{\citenamefont {Ireland}\ and\ \citenamefont {Koren}(2024)}]{ISB2}%
  \BibitemOpen
  \bibfield  {author} {\bibinfo {author} {\bibfnamefont {A.}~\bibnamefont {Ireland}}\ and\ \bibinfo {author} {\bibfnamefont {S.}~\bibnamefont {Koren}},\ }\href {\doibase 10.1103/PhysRevD.109.103537} {\bibfield  {journal} {\bibinfo  {journal} {Phys. Rev. D}\ }\textbf {\bibinfo {volume} {109}},\ \bibinfo {pages} {103537} (\bibinfo {year} {2024})}\BibitemShut {NoStop}%
\bibitem [{\citenamefont {Pinto}\ and\ \citenamefont {Ramos}(2000)}]{ISBX}%
  \BibitemOpen
  \bibfield  {author} {\bibinfo {author} {\bibfnamefont {M.~B.}\ \bibnamefont {Pinto}}\ and\ \bibinfo {author} {\bibfnamefont {R.~O.}\ \bibnamefont {Ramos}},\ }\href {\doibase 10.1103/PhysRevD.61.125016} {\bibfield  {journal} {\bibinfo  {journal} {Phys. Rev. D}\ }\textbf {\bibinfo {volume} {61}},\ \bibinfo {pages} {125016} (\bibinfo {year} {2000})}\BibitemShut {NoStop}%
\bibitem [{\citenamefont {Ara{\'u}jo}\ \emph {et~al.}(2024)\citenamefont {Ara{\'u}jo}, \citenamefont {Mariz}, \citenamefont {Nascimento},\ and\ \citenamefont {Petrov}}]{ISBB1}%
  \BibitemOpen
  \bibfield  {author} {\bibinfo {author} {\bibfnamefont {R.}~\bibnamefont {Ara{\'u}jo}}, \bibinfo {author} {\bibfnamefont {T.}~\bibnamefont {Mariz}}, \bibinfo {author} {\bibfnamefont {J.~R.}\ \bibnamefont {Nascimento}}, \ and\ \bibinfo {author} {\bibfnamefont {A.~Y.}\ \bibnamefont {Petrov}},\ }\href {\doibase 10.1140/epjc/s10052-024-13391-4} {\bibfield  {journal} {\bibinfo  {journal} {The European Physical Journal C}\ }\textbf {\bibinfo {volume} {84}},\ \bibinfo {pages} {1034} (\bibinfo {year} {2024})}\BibitemShut {NoStop}%
\bibitem [{\citenamefont {Assun\ifmmode \mbox{\c{c}}\else~\c{c}\fi{}\ ao}\ \emph {et~al.}(2017)\citenamefont {Assun\ifmmode \mbox{\c{c}}\else~\c{c}\fi{}\ ao}, \citenamefont {Mariz}, \citenamefont {Nascimento},\ and\ \citenamefont {Petrov}}]{ISBB2}%
  \BibitemOpen
  \bibfield  {author} {\bibinfo {author} {\bibfnamefont {J.~F.}\ \bibnamefont {Assun\ifmmode \mbox{\c{c}}\else~\c{c}\fi{}\ ao}}, \bibinfo {author} {\bibfnamefont {T.}~\bibnamefont {Mariz}}, \bibinfo {author} {\bibfnamefont {J.~R.}\ \bibnamefont {Nascimento}}, \ and\ \bibinfo {author} {\bibfnamefont {A.~Y.}\ \bibnamefont {Petrov}},\ }\href {\doibase 10.1103/PhysRevD.96.065021} {\bibfield  {journal} {\bibinfo  {journal} {Phys. Rev. D}\ }\textbf {\bibinfo {volume} {96}},\ \bibinfo {pages} {065021} (\bibinfo {year} {2017})}\BibitemShut {NoStop}%
\bibitem [{\citenamefont {Gomes}\ \emph {et~al.}(2008)\citenamefont {Gomes}, \citenamefont {Mariz}, \citenamefont {Nascimento},\ and\ \citenamefont {da~Silva}}]{ISBB3}%
  \BibitemOpen
  \bibfield  {author} {\bibinfo {author} {\bibfnamefont {M.}~\bibnamefont {Gomes}}, \bibinfo {author} {\bibfnamefont {T.}~\bibnamefont {Mariz}}, \bibinfo {author} {\bibfnamefont {J.~R.}\ \bibnamefont {Nascimento}}, \ and\ \bibinfo {author} {\bibfnamefont {A.~J.}\ \bibnamefont {da~Silva}},\ }\href {\doibase 10.1103/PhysRevD.77.105002} {\bibfield  {journal} {\bibinfo  {journal} {Phys. Rev. D}\ }\textbf {\bibinfo {volume} {77}},\ \bibinfo {pages} {105002} (\bibinfo {year} {2008})}\BibitemShut {NoStop}%
\bibitem [{\citenamefont {Allahverdi}\ \emph {et~al.}(2000)\citenamefont {Allahverdi}, \citenamefont {Shaw},\ and\ \citenamefont {Campbell}}]{ISBPR}%
  \BibitemOpen
  \bibfield  {author} {\bibinfo {author} {\bibfnamefont {R.}~\bibnamefont {Allahverdi}}, \bibinfo {author} {\bibfnamefont {R.~H. A.~D.}\ \bibnamefont {Shaw}}, \ and\ \bibinfo {author} {\bibfnamefont {B.~A.}\ \bibnamefont {Campbell}},\ }\href {\doibase 10.1016/S0370-2693(99)01302-7} {\bibfield  {journal} {\bibinfo  {journal} {Phys. Lett. B}\ }\textbf {\bibinfo {volume} {473}},\ \bibinfo {pages} {246} (\bibinfo {year} {2000})},\ \Eprint {http://arxiv.org/abs/hep-ph/9909256} {arXiv:hep-ph/9909256} \BibitemShut {NoStop}%
\bibitem [{\citenamefont {Lee}\ and\ \citenamefont {Koh}(1996)}]{PhysRevD.54.7153}%
  \BibitemOpen
  \bibfield  {author} {\bibinfo {author} {\bibfnamefont {J.-w.}\ \bibnamefont {Lee}}\ and\ \bibinfo {author} {\bibfnamefont {I.-g.}\ \bibnamefont {Koh}},\ }\href {\doibase 10.1103/PhysRevD.54.7153} {\bibfield  {journal} {\bibinfo  {journal} {Phys. Rev. D}\ }\textbf {\bibinfo {volume} {54}},\ \bibinfo {pages} {7153} (\bibinfo {year} {1996})}\BibitemShut {NoStop}%
\bibitem [{\citenamefont {Ho\ifmmode~\check{r}\else \v{r}\fi{}ava}(2009)}]{Hor1}%
  \BibitemOpen
  \bibfield  {author} {\bibinfo {author} {\bibfnamefont {P.}~\bibnamefont {Ho\ifmmode~\check{r}\else \v{r}\fi{}ava}},\ }\href {\doibase 10.1103/PhysRevD.79.084008} {\bibfield  {journal} {\bibinfo  {journal} {Phys. Rev. D}\ }\textbf {\bibinfo {volume} {79}},\ \bibinfo {pages} {084008} (\bibinfo {year} {2009})}\BibitemShut {NoStop}%
\bibitem [{\citenamefont {Myers}\ and\ \citenamefont {Pospelov}(2003)}]{ELV1}%
  \BibitemOpen
  \bibfield  {author} {\bibinfo {author} {\bibfnamefont {R.~C.}\ \bibnamefont {Myers}}\ and\ \bibinfo {author} {\bibfnamefont {M.}~\bibnamefont {Pospelov}},\ }\href {\doibase 10.1103/PhysRevLett.90.211601} {\bibfield  {journal} {\bibinfo  {journal} {Phys. Rev. Lett.}\ }\textbf {\bibinfo {volume} {90}},\ \bibinfo {pages} {211601} (\bibinfo {year} {2003})},\ \Eprint {http://arxiv.org/abs/hep-ph/0301124} {arXiv:hep-ph/0301124} \BibitemShut {NoStop}%
\bibitem [{\citenamefont {Collins}\ \emph {et~al.}(2004{\natexlab{b}})\citenamefont {Collins}, \citenamefont {Perez}, \citenamefont {Sudarsky}, \citenamefont {Urrutia},\ and\ \citenamefont {Vucetich}}]{ELV2}%
  \BibitemOpen
  \bibfield  {author} {\bibinfo {author} {\bibfnamefont {J.}~\bibnamefont {Collins}}, \bibinfo {author} {\bibfnamefont {A.}~\bibnamefont {Perez}}, \bibinfo {author} {\bibfnamefont {D.}~\bibnamefont {Sudarsky}}, \bibinfo {author} {\bibfnamefont {L.}~\bibnamefont {Urrutia}}, \ and\ \bibinfo {author} {\bibfnamefont {H.}~\bibnamefont {Vucetich}},\ }\href {\doibase 10.1103/PhysRevLett.93.191301} {\bibfield  {journal} {\bibinfo  {journal} {Phys. Rev. Lett.}\ }\textbf {\bibinfo {volume} {93}},\ \bibinfo {pages} {191301} (\bibinfo {year} {2004}{\natexlab{b}})}\BibitemShut {NoStop}%
\bibitem [{\citenamefont {Jacobson}\ \emph {et~al.}(2006)\citenamefont {Jacobson}, \citenamefont {Liberati},\ and\ \citenamefont {Mattingly}}]{ELV3}%
  \BibitemOpen
  \bibfield  {author} {\bibinfo {author} {\bibfnamefont {T.}~\bibnamefont {Jacobson}}, \bibinfo {author} {\bibfnamefont {S.}~\bibnamefont {Liberati}}, \ and\ \bibinfo {author} {\bibfnamefont {D.}~\bibnamefont {Mattingly}},\ }\href {\doibase https://doi.org/10.1016/j.aop.2005.06.004} {\bibfield  {journal} {\bibinfo  {journal} {Annals of Physics}\ }\textbf {\bibinfo {volume} {321}},\ \bibinfo {pages} {150} (\bibinfo {year} {2006})},\ \bibinfo {note} {january Special Issue}\BibitemShut {NoStop}%
\bibitem [{\citenamefont {Gambini}\ \emph {et~al.}(2011)\citenamefont {Gambini}, \citenamefont {Rastgoo},\ and\ \citenamefont {Pullin}}]{ELV4}%
  \BibitemOpen
  \bibfield  {author} {\bibinfo {author} {\bibfnamefont {R.}~\bibnamefont {Gambini}}, \bibinfo {author} {\bibfnamefont {S.}~\bibnamefont {Rastgoo}}, \ and\ \bibinfo {author} {\bibfnamefont {J.}~\bibnamefont {Pullin}},\ }\href {\doibase 10.1088/0264-9381/28/15/155005} {\bibfield  {journal} {\bibinfo  {journal} {Classical and Quantum Gravity}\ }\textbf {\bibinfo {volume} {28}},\ \bibinfo {pages} {155005} (\bibinfo {year} {2011})}\BibitemShut {NoStop}%
\bibitem [{\citenamefont {Coates}\ \emph {et~al.}(2019)\citenamefont {Coates}, \citenamefont {Melby-Thompson},\ and\ \citenamefont {Mukohyama}}]{Hor2}%
  \BibitemOpen
  \bibfield  {author} {\bibinfo {author} {\bibfnamefont {A.}~\bibnamefont {Coates}}, \bibinfo {author} {\bibfnamefont {C.}~\bibnamefont {Melby-Thompson}}, \ and\ \bibinfo {author} {\bibfnamefont {S.}~\bibnamefont {Mukohyama}},\ }\href {\doibase 10.1103/PhysRevD.100.064046} {\bibfield  {journal} {\bibinfo  {journal} {Phys. Rev. D}\ }\textbf {\bibinfo {volume} {100}},\ \bibinfo {pages} {064046} (\bibinfo {year} {2019})}\BibitemShut {NoStop}%
\bibitem [{\citenamefont {Gasperini}(1985)}]{Gasp85}%
  \BibitemOpen
  \bibfield  {author} {\bibinfo {author} {\bibfnamefont {M.}~\bibnamefont {Gasperini}},\ }\href {\doibase https://doi.org/10.1016/0370-2693(85)90197-2} {\bibfield  {journal} {\bibinfo  {journal} {Physics Letters B}\ }\textbf {\bibinfo {volume} {163}},\ \bibinfo {pages} {84} (\bibinfo {year} {1985})}\BibitemShut {NoStop}%
\bibitem [{\citenamefont {Kosteleck\'y}\ and\ \citenamefont {Samuel}(1989)}]{Kost89}%
  \BibitemOpen
  \bibfield  {author} {\bibinfo {author} {\bibfnamefont {V.~A.}\ \bibnamefont {Kosteleck\'y}}\ and\ \bibinfo {author} {\bibfnamefont {S.}~\bibnamefont {Samuel}},\ }\href {\doibase 10.1103/PhysRevD.39.683} {\bibfield  {journal} {\bibinfo  {journal} {Phys. Rev. D}\ }\textbf {\bibinfo {volume} {39}},\ \bibinfo {pages} {683} (\bibinfo {year} {1989})}\BibitemShut {NoStop}%
\bibitem [{\citenamefont {Kosteleck\'y}(2004)}]{Kost04}%
  \BibitemOpen
  \bibfield  {author} {\bibinfo {author} {\bibfnamefont {V.~A.}\ \bibnamefont {Kosteleck\'y}},\ }\href {\doibase 10.1103/PhysRevD.69.105009} {\bibfield  {journal} {\bibinfo  {journal} {Phys. Rev. D}\ }\textbf {\bibinfo {volume} {69}},\ \bibinfo {pages} {105009} (\bibinfo {year} {2004})}\BibitemShut {NoStop}%
\bibitem [{\citenamefont {Kosteleck\'y}\ and\ \citenamefont {Russell}(2011)}]{Kost11}%
  \BibitemOpen
  \bibfield  {author} {\bibinfo {author} {\bibfnamefont {V.~A.}\ \bibnamefont {Kosteleck\'y}}\ and\ \bibinfo {author} {\bibfnamefont {N.}~\bibnamefont {Russell}},\ }\href {\doibase 10.1103/RevModPhys.83.11} {\bibfield  {journal} {\bibinfo  {journal} {Rev. Mod. Phys.}\ }\textbf {\bibinfo {volume} {83}},\ \bibinfo {pages} {11} (\bibinfo {year} {2011})}\BibitemShut {NoStop}%
\bibitem [{\citenamefont {Seifert}(2010)}]{BB2}%
  \BibitemOpen
  \bibfield  {author} {\bibinfo {author} {\bibfnamefont {M.~D.}\ \bibnamefont {Seifert}},\ }\href {\doibase 10.1103/PhysRevD.81.065010} {\bibfield  {journal} {\bibinfo  {journal} {Phys. Rev. D}\ }\textbf {\bibinfo {volume} {81}},\ \bibinfo {pages} {065010} (\bibinfo {year} {2010})}\BibitemShut {NoStop}%
\bibitem [{\citenamefont {Bertolami}\ and\ \citenamefont {P\'aramos}(2005)}]{BB3}%
  \BibitemOpen
  \bibfield  {author} {\bibinfo {author} {\bibfnamefont {O.}~\bibnamefont {Bertolami}}\ and\ \bibinfo {author} {\bibfnamefont {J.}~\bibnamefont {P\'aramos}},\ }\href {\doibase 10.1103/PhysRevD.72.044001} {\bibfield  {journal} {\bibinfo  {journal} {Phys. Rev. D}\ }\textbf {\bibinfo {volume} {72}},\ \bibinfo {pages} {044001} (\bibinfo {year} {2005})}\BibitemShut {NoStop}%
\bibitem [{\citenamefont {Guiomar}\ and\ \citenamefont {P\'aramos}(2014)}]{BB4}%
  \BibitemOpen
  \bibfield  {author} {\bibinfo {author} {\bibfnamefont {G.~m.~c.}\ \bibnamefont {Guiomar}}\ and\ \bibinfo {author} {\bibfnamefont {J.}~\bibnamefont {P\'aramos}},\ }\href {\doibase 10.1103/PhysRevD.90.082002} {\bibfield  {journal} {\bibinfo  {journal} {Phys. Rev. D}\ }\textbf {\bibinfo {volume} {90}},\ \bibinfo {pages} {082002} (\bibinfo {year} {2014})}\BibitemShut {NoStop}%
\bibitem [{\citenamefont {Capelo}\ and\ \citenamefont {P\'aramos}(2015)}]{BB5}%
  \BibitemOpen
  \bibfield  {author} {\bibinfo {author} {\bibfnamefont {D.}~\bibnamefont {Capelo}}\ and\ \bibinfo {author} {\bibfnamefont {J.}~\bibnamefont {P\'aramos}},\ }\href {\doibase 10.1103/PhysRevD.91.104007} {\bibfield  {journal} {\bibinfo  {journal} {Phys. Rev. D}\ }\textbf {\bibinfo {volume} {91}},\ \bibinfo {pages} {104007} (\bibinfo {year} {2015})}\BibitemShut {NoStop}%
\bibitem [{\citenamefont {Maluf}\ and\ \citenamefont {Neves}(2021)}]{BB6}%
  \BibitemOpen
  \bibfield  {author} {\bibinfo {author} {\bibfnamefont {R.~V.}\ \bibnamefont {Maluf}}\ and\ \bibinfo {author} {\bibfnamefont {J.~C.~S.}\ \bibnamefont {Neves}},\ }\href {\doibase 10.1088/1475-7516/2021/10/038} {\bibfield  {journal} {\bibinfo  {journal} {JCAP}\ }\textbf {\bibinfo {volume} {10}},\ \bibinfo {pages} {038} (\bibinfo {year} {2021})},\ \Eprint {http://arxiv.org/abs/2105.08659} {arXiv:2105.08659 [gr-qc]} \BibitemShut {NoStop}%
\bibitem [{\citenamefont {Luongo}\ \emph {et~al.}(2019)\citenamefont {Luongo}, \citenamefont {Muccino},\ and\ \citenamefont {Quevedo}}]{Luongo:2018oil}%
  \BibitemOpen
  \bibfield  {author} {\bibinfo {author} {\bibfnamefont {O.}~\bibnamefont {Luongo}}, \bibinfo {author} {\bibfnamefont {M.}~\bibnamefont {Muccino}}, \ and\ \bibinfo {author} {\bibfnamefont {H.}~\bibnamefont {Quevedo}},\ }\href {\doibase 10.1016/j.dark.2019.100313} {\bibfield  {journal} {\bibinfo  {journal} {Phys. Dark Univ.}\ }\textbf {\bibinfo {volume} {25}},\ \bibinfo {pages} {100313} (\bibinfo {year} {2019})},\ \Eprint {http://arxiv.org/abs/1811.05227} {arXiv:1811.05227 [gr-qc]} \BibitemShut {NoStop}%
\bibitem [{\citenamefont {Biesiada}\ and\ \citenamefont {Piorkowska}(2007)}]{Biesiada:2007zzb}%
  \BibitemOpen
  \bibfield  {author} {\bibinfo {author} {\bibfnamefont {M.}~\bibnamefont {Biesiada}}\ and\ \bibinfo {author} {\bibfnamefont {A.}~\bibnamefont {Piorkowska}},\ }\href {\doibase 10.1088/1475-7516/2007/05/011} {\bibfield  {journal} {\bibinfo  {journal} {JCAP}\ }\textbf {\bibinfo {volume} {05}},\ \bibinfo {pages} {011} (\bibinfo {year} {2007})},\ \Eprint {http://arxiv.org/abs/0712.0937} {arXiv:0712.0937 [astro-ph]} \BibitemShut {NoStop}%
\bibitem [{\citenamefont {Mattingly}(2005)}]{Mattingly:2005re}%
  \BibitemOpen
  \bibfield  {author} {\bibinfo {author} {\bibfnamefont {D.}~\bibnamefont {Mattingly}},\ }\href {\doibase 10.12942/lrr-2005-5} {\bibfield  {journal} {\bibinfo  {journal} {Living Rev. Rel.}\ }\textbf {\bibinfo {volume} {8}},\ \bibinfo {pages} {5} (\bibinfo {year} {2005})},\ \Eprint {http://arxiv.org/abs/gr-qc/0502097} {arXiv:gr-qc/0502097} \BibitemShut {NoStop}%
\bibitem [{\citenamefont {Liberati}(2013)}]{Liberati:2013xla}%
  \BibitemOpen
  \bibfield  {author} {\bibinfo {author} {\bibfnamefont {S.}~\bibnamefont {Liberati}},\ }\href {\doibase 10.1088/0264-9381/30/13/133001} {\bibfield  {journal} {\bibinfo  {journal} {Class. Quant. Grav.}\ }\textbf {\bibinfo {volume} {30}},\ \bibinfo {pages} {133001} (\bibinfo {year} {2013})},\ \Eprint {http://arxiv.org/abs/1304.5795} {arXiv:1304.5795 [gr-qc]} \BibitemShut {NoStop}%
\bibitem [{\citenamefont {Minami}\ and\ \citenamefont {Komatsu}(2020)}]{Minami:2020odp}%
  \BibitemOpen
  \bibfield  {author} {\bibinfo {author} {\bibfnamefont {Y.}~\bibnamefont {Minami}}\ and\ \bibinfo {author} {\bibfnamefont {E.}~\bibnamefont {Komatsu}},\ }\href {\doibase 10.1103/PhysRevLett.125.221301} {\bibfield  {journal} {\bibinfo  {journal} {Phys. Rev. Lett.}\ }\textbf {\bibinfo {volume} {125}},\ \bibinfo {pages} {221301} (\bibinfo {year} {2020})},\ \Eprint {http://arxiv.org/abs/2011.11254} {arXiv:2011.11254 [astro-ph.CO]} \BibitemShut {NoStop}%
\end{thebibliography}
\end{document}